\documentclass[aps,twocolumn,pre,floatfix,superscriptaddress]{revtex4-2}

\usepackage{graphicx}
\usepackage{amsmath,amsfonts,amssymb,bm}
\usepackage{lmodern}
\usepackage{xcolor}
\usepackage[normalem]{ulem}
\usepackage[shortlabels]{enumitem}
\usepackage{float}
\usepackage[caption=false]{subfig}
\usepackage{natbib}
\usepackage{upgreek}

\newcommand{\figcaption}[2]{\caption{\textsf{\textbf{#1.} #2}}}
\renewcommand{\figurename}{\textsf{\textbf{{Figure}}}}
\newcommand{\tabcaption}[2]{\caption{{\textbf{#1.} #2}}}

\newcommand{\fs}[1]{\textit{\textsf{#1}}}
\newcommand{\bfs}[1]{\textbf{\textit{\textsf{#1}}}}

\newcommand{\cto}[1]{CoTiO$_3$ }
\newcommand{\beq}{\begin{equation}}
\newcommand{\eeq}{\end{equation}}

\begin{document}

\title{Order-by-Disorder from Bond-Dependent Exchange and Intensity Signature of \\ Nodal Quasiparticles in a Honeycomb Cobaltate}

\author{M.~Elliot}
\affiliation{Clarendon Laboratory, University of Oxford, Parks Road, Oxford, OX1 3PU, UK}

\author{P.~A.~McClarty}
\affiliation{Max Planck Institute for the Physics of Complex Systems, N\"{o}thnitzer Str. 38, 01187 Dresden, Germany}

\author{D.~Prabhakaran}
\affiliation{Clarendon Laboratory, University of Oxford, Parks Road, Oxford, OX1 3PU, UK}

\author{R.~D.~Johnson}
\affiliation{Department of Physics and Astronomy, University College London, Gower Street, London WC1E 6BT, UK}

\author{H.~C.~Walker}
\affiliation{ISIS Facility, Rutherford Appleton Laboratory-STFC, Chilton, Didcot, OX11 0QX, UK}

\author{P.~Manuel}
\affiliation{ISIS Facility, Rutherford Appleton Laboratory-STFC, Chilton, Didcot, OX11 0QX, UK}

\author{R.~Coldea}
\affiliation{Clarendon Laboratory, University of Oxford, Parks Road, Oxford, OX1 3PU, UK}

\maketitle
\section*{Abstract}
Recent theoretical proposals have argued that cobaltates with edge-sharing octahedral coordination can have significant bond-dependent exchange couplings thus offering a platform in 3$d$ ions for such physics beyond the much-explored realizations in 4$d$ and 5$d$ materials. Here we present high-resolution inelastic neutron scattering data within the magnetically ordered phase of the stacked honeycomb magnet CoTiO$_3$ revealing the presence of a finite energy gap and demonstrate that this implies the presence of bond-dependent anisotropic couplings. We also show through an extensive theoretical analysis that the gap further implies the existence of a quantum order-by-disorder mechanism that, in this material, crucially involves virtual crystal field fluctuations. Our data also provide an experimental observation of a universal winding of the scattering intensity in angular scans around linear band-touching points for both magnons and dispersive spin-orbit excitons, which is directly related to the non-trivial topology of the quasiparticle wavefunction in momentum space near nodal points.

\section*{Introduction}
Spin-orbit coupling is at the origin of many remarkable properties of condensed matter uncovered in recent years \cite{RevModPhys.82.3045,RevModPhys.90.015001,bernevig2013topological,SOReview,SOReview2}. It is central to the appearance of nontrivial topological invariants in electronic band structures and underlies the existence of bond-dependent exchange couplings that have been shown to bring about exotic features in many quantum magnets \cite{Winter_review,hermanns2018,Takagi_review}. In the latter case much of the effort in materials discovery has focussed on heavy 5$d$ and 4$d$ ions in which the spin-orbit coupling is one of the dominant energy scales. Notable are the honeycomb iridates $A_2$IrO$_3$ ($A$=Na,Li) and related materials, and $\alpha$-RuCl$_3$, which displayed a range of many novel exotic magnetic properties including spin-momentum locking \cite{Chun}, incommensurate orders with counter-rotating spin spirals \cite{Winter_review}, broad scattering continua in the spectrum of spin excitations \cite{banerjee2017neutron} or unconventional field-dependent thermal Hall effect \cite{Matsuda}. The origin of these exotic forms of behaviour is the presence of significant anisotropic, bond-dependent exchange, which in extreme cases has been predicted to stabilize quantum spin liquids, such as the celebrated Kitaev honeycomb model with Ising exchanges along orthogonal directions for the three bonds that meet at each site \cite{Kitaev}. The path to the discovery of the unusual magnetic properties of those materials has been a fruitful one starting with theoretical proposals that bond-dependent exchange couplings can arise in certain iridates and ruthenates with edge-sharing octahedra \cite{PhysRevLett.102.017205,PhysRevLett.105.027204}. The octahedra supply a crystal field environment that leads to an effective low-energy spin one-half degree of freedom for the magnetic ions and the edge-sharing provides the local exchange pathway that, in conjunction with the spin-orbit coupling, produces anisotropic bond-dependent exchange. There is now evidence for significant such exchanges in honeycomb iridates and ruthenates \cite{Winter_review,hermanns2018,Takagi_review}.

More recent theoretical work has argued that significant bond-dependent exchange in the form of Kitaev and related couplings may also arise between Co$^{2+}$ ions in edge-sharing octahedral coordination  \cite{PhysRevB.97.014407,PhysRevB.97.014408,PhysRevLett.125.047201} thus extending the original proposals into a surprising new setting. To investigate such effects we report here inelastic neutron scattering (INS) measurements of the spin dynamics in the stacked honeycomb magnet CoTiO$_3$. Our data show propagating spin wave excitations with a clear low energy spectral gap, which was inferred but could not be resolved by previous studies \cite{Yuan}. We show that the spin wave spectrum is not merely compatible with the presence of bond-dependent exchange, but that such couplings must be present in the low energy pseudo-spin one-half theory in order to explain the origin of the gap. Moreover, we show that the gap opening must occur via a quantum order-by-disorder mechanism \cite{villain:jpa-00208953,PhysRevLett.62.2056,PhysRevB.58.12049,PhysRevB.46.11137,PhysRevB.68.020401,ObD,PhysRevLett.109.077204} as a consequence of unusually strong constraints on the possible mechanisms that can open the spectral gap. In view of the low-lying crystal field excitations in this material compared to the exchange coupling, we provide compelling evidence that virtual crystal field excitations are the driving mechanism for order-by-disorder \cite{McClarty_2009,PhysRevB.93.184408} assisted by spin-orbital exchange and supply a calculation of the spin wave spectrum including this effect that captures the principal features of the data.

\begin{figure*}[!htbp]
\centering
\includegraphics[width=\linewidth,keepaspectratio]{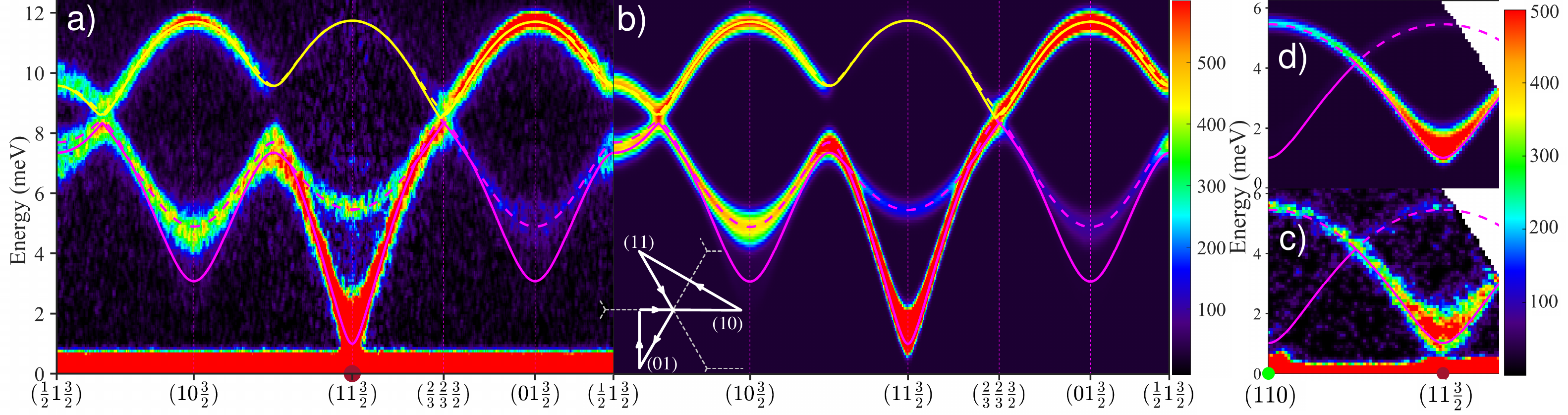}
\figcaption{Magnon dispersions}{INS data at 8~K observing the magnon dispersions along high-symmetry directions (a) in-plane and (c) out-of-plane, compared in (b) and (d) with the XXZ$\Delta$ model. Lines are the model dispersions $\tilde{\omega}(\bf{k})$, green/brown dots on the elastic line indicate location of structural/magnetic Bragg peaks. Lower left inset in (b) shows the wavevector path in a)-b) (arrowed solid white lines) projected onto the ($hk$) plane, gray dashed lines are the 2D Brillouin zone boundaries. Intensities are averaged for a transverse wavevector range of $\pm0.1$~\AA$^{-1}$. The incident neutron energy was $E_{\rm i}=18$~meV in (a) and $9.6$~meV in (c). The colour bars indicate scattering intensity in arbitrary units on a linear scale. \label{fig:dispersions}}
\end{figure*}

CoTiO$_3$ is part of a growing list of materials \cite{Bao2018,yao2018topological,Yuan} explored as candidates displaying Dirac magnons. Earlier studies established the presence of Dirac nodal lines \cite{Yuan}, which make this material ideal for the exploration of a recently predicted \cite{shivam2017neutron} fingerprint of a topologically non-trivial magnon band structure, namely a universal azimuthal modulation in the dynamical structure factor around linear band touching points, not probed experimentally before and which  originates from the special topological features in the wavefunction of nodal quasiparticles. We indeed observe clear evidence for the predicted intensity winding around the nodal points, thus providing a direct measurement of the non-trivial topology of the Dirac magnon wavefunctions and establishing that there are meaningful features in the momentum-and-energy dependent dynamical structure factor beyond simply revealing the quasiparticle dispersion relations. Furthermore, we observe analogous features in the dispersive spin-orbital excitations at higher energy, highlighting the universal properties of Dirac bosonic quasiparticles. Finally, we investigate the effect of the bond-dependent exchange on the Dirac nodal lines arguing that they are robust to gap opening and likely appear as `double helices' winding around each zone corner. We show that  the same type of bond-dependent anisotropic exchange that opens up the spectral gap provides a natural explanation for a `double-peak' structure in energy scans near the nodal points.

\section*{Results}
{\bf Magnon dispersions} - The magnon dispersions along high-symmetry directions in the honeycomb plane obtained using inelastic neutron scattering (INS) measurements on single crystals of CoTiO$_3$ (for details see Supplementary Note 6A) are summarized in Fig.~\ref{fig:dispersions}a). Wavevectors are indexed in reciprocal lattice units of the hexagonal structural unit cell. Near the (1,1,3/2) magnetic Bragg peak the lowest mode has a near-linear in-plane dispersion. As the honeycomb layers are ferromagnetically ordered with moments confined to the crystallographic $ab$ plane, the linear dispersion indicates predominant easy-plane-type exchange couplings for in-plane neighbors. Fig.~\ref{fig:dispersions}c) observes a finite dispersion at low energies in the direction normal to the layers, indicating finite inter-layer couplings, and a small but finite spectral gap $\Delta=1.0(1)$~meV, clearly resolved above the magnetic Bragg peak. Ref. \cite{Yuan} proposed that a finite spin gap would be needed to account for the observed non-linear magnetization curve in small in-plane fields \cite{Balbashov}, but it was not possible to directly resolve the gap excitation in the earlier lower-resolution INS data \cite{Yuan}.
Apart from the finite gap, the main features of the magnon spectrum can be accounted for by a minimal exchange Hamiltonian ${\cal H}_{\rm XXZ}$ for the stacked honeycomb geometry in CoTiO$_3$, allowing for each bond a different exchange coupling between the moment components along the $c$-axis, and between the components in the $ab$ plane. For a single ferromagnetic honeycomb layer, two magnon bands (acoustic/optic) would be expected with linear crossings at the corners (K-points) of the hexagonal Brillouin zone. For finite interlayer couplings that stabilize antiferromagnetic stacking of layers, the number of bands doubles and inter-layer resolved lower bands are expected with almost degenerate higher bands, as observed in Figs.~\ref{fig:dispersions}a,c).
${\cal H}_{\rm XXZ}$ has a gapless (Goldstone) mode corresponding to moments rotating freely in the $ab$ plane, so to capture the observed gap we assume that the physical mechanism responsible for gap generation only modifies the dispersion relations $\omega(\mathbf{k})$ of ${\cal H}_{\rm XXZ}$ by adding a gap in quadrature, i.e. experimental dispersion points are compared with $\tilde{\omega}(\mathbf{k})=\sqrt{\omega^2(\mathbf{k})+\Delta^2}$. We call this parameterization the  XXZ$\Delta$ model to emphasize that the gap $\Delta$ is not intrinsic, but is an additional, empirical fitting parameter. We find that exchanges up to $6$th nearest-neighbor (nn) are important and obtain a very good level of agreement for both the dispersions and intensities as shown by comparing Figs.~\ref{fig:dispersions}a) with b), and c) with d) (for more details see Supplementary Notes 5B, 5C, and 6C).

\begin{figure}[htbp!]
\centering
\includegraphics[width=0.97\linewidth,keepaspectratio]{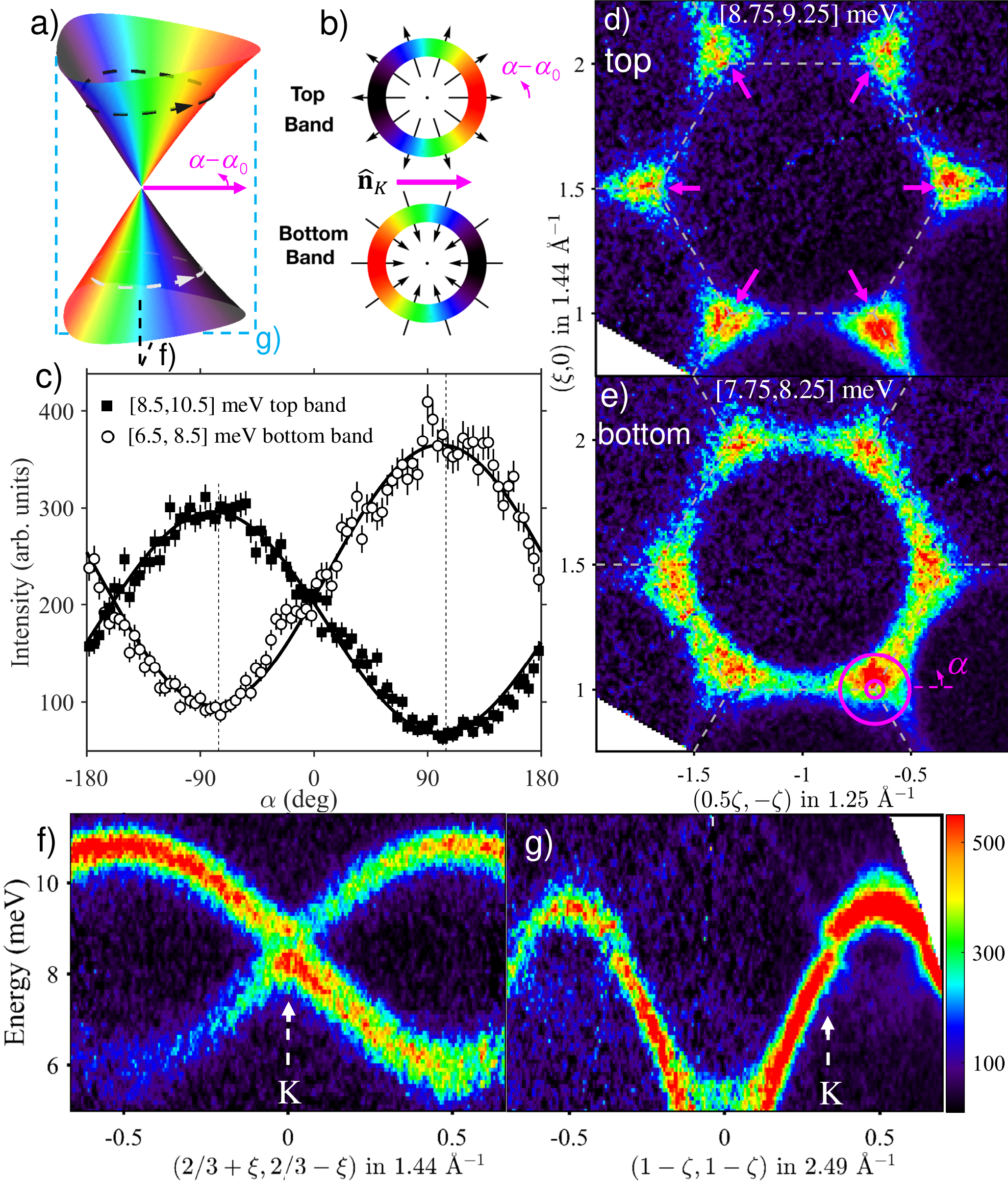}
\figcaption{Intensity winding for Dirac magnons: theory and experiment}{a) Conical dispersion surfaces meeting at a Dirac node for a honeycomb ferromagnet. b) Winding pattern of the isospin polarization $\bm{\sigma}$ (radial arrows) out/in from the nodal point for the top/bottom bands. In both a-b) colour is the dynamical structure factor $1+\bm{\sigma}\cdot\hat{\mathbf{n}}_{\rm K}$, which winds around the node in antiphase between the top and bottom bands. c) Constant-energy INS intensity in CoTiO$_3$ as a function of azimuthal angle $\alpha$ around the (2/3,2/3) Dirac node, showing expected two-fold winding periodicity in anti-phase between the top/bottom bands (filled/open symbols) in agreement with b). The black squares/white circles denote the inelastic neutron scattering intensity for the top/bottom bands, with error bars representing one standard deviation. Solid lines are fits to cosine dependencies described in the text with dotted vertical lines showing the extreme points $180^\circ$ apart. d/e) Momentum intensity maps above/below the Dirac node energy, highlighting dramatic changes in the angular intensity dependence around the Dirac nodes. Dashed gray lines outline the edges of the 2D Brillouin zones and radial magenta arrows in d) indicate the direction of the vectors $\hat{\mathbf{n}}$ at the zone corners at  $L=0$. Magenta annular region in e) shows the radial in-plane wavevector range $[0.05,0.2]$~\AA$^{-1}$ in the angular scans in c). f/g) INS data through a nodal point (vertical dashed arrow) along orthogonal in-plane directions that maximise the intensity asymmetry effect (slices shown in a) by dashed black/cyan rectangles, respectively): in f) both crossing modes are visible, in g) only one mode carries weight. All data were collected with $E_{\rm i}=18$~meV. In panels c-g) intensities are averaged for $L=[0,2.4]$, and in f) and g) for an in-plane transverse momentum range of $\pm0.026$ and $\pm0.028$~\AA$^{-1}$, respectively. The colour bar in g) applies also to panels d-f), indicating scattering intensity in arbitrary units on a linear scale.
\label{fig:node}}
\end{figure}

{\bf Quantum Order-by-Disorder} - The presence of the finite magnon spectral gap $\Delta$ is important as it indicates preferential moment orientations inside the easy plane. The magnetic ground state of Co$^{2+}$ ($3d^7$) ions in the local crystal field environment is a Kramers doublet with pseudospin-1/2, for which there is no local anisotropy, so any preferential orientation must be selected by interactions beyond the minimal ${\cal H}_{\rm XXZ}$ Hamiltonian. We focus our attention on bilinear couplings in the pseudospin as higher order two-site couplings project down to such couplings. As outlined in Supplementary Note 10, multi-site couplings will be suppressed by the large charge gap. As there is no detectable distortion of the crystal lattice following the onset of the magnetic order, we perform the analysis of bi-linear couplings between cobalt moments that are symmetry-allowed by the crystal structure space group. We find that whilst various bond-dependent exchange couplings can be present in principle, at the classical level, surprisingly, the ground state energy remains independent of the moment orientation in the $ab$ plane - see Supplementary Notes 7 and 8.

This degeneracy must however be an artefact of the mean-field approximation, as the real material Hamiltonian has only discrete, rather than continuous rotational symmetry around the $c$-axis. Such degeneracies would in general be expected to be lifted by quantum fluctuations via an order-by-disorder mechanism \cite{villain:jpa-00208953,PhysRevLett.62.2056,PhysRevB.46.11137,PhysRevB.58.12049,PhysRevB.68.020401,ObD}, when the ground state energy (per site) acquires a contribution from zero-point fluctuations of the form $\epsilon_{\rm qu}(\phi)=\frac{1}{2}\sum_m\langle\omega_m(\mathbf{k})\rangle$, where $\phi$ defines the moments' orientation in the $ab$-plane relative to the $a$-axis and $\langle\omega_m(\mathbf{k})\rangle$ is the average energy of dispersive branch $m=1$ to 4 over the Brillouin zone. The possibility that an order-by-disorder mechanism might be relevant for the ground state selection in CoTiO$_3$ was mentioned in \cite{Yuan}, but no quantitative model was proposed. We show by direct calculations in Supplementary Note 8 that the semi-classical degeneracy is indeed lifted by zero-point fluctuations from bond-dependent anisotropic couplings such as $\eta \equiv J^{yy}-J^{xx}$ on the 1st neighbor bond where $y$ defines the local bond direction and $x$ is in-plane transverse to $y$, and we find an induced gap that scales as $\Delta \sim \vert \eta \vert^{3/2}$ at leading order. At the level of the low energy pseudospin-1/2 moments this provides a natural qualitative mechanism for the observed gap. One can also place this finding in the context of a theory that operates within the full set of $12$ single-ion spin and orbital states. In fact, working within the pseudospin-1/2 picture suggests an unphysically large coupling $\eta$ calculated in Supplementary Note 8 compared to the coupling $\eta$ fitted in Supplementary Note 7. Since the crystal field excitations are comparable to the exchange scale, an entirely natural mechanism for order-by-disorder to arise is through virtual crystal field fluctuations in a model that includes small spin-orbital exchange. The virtual crystal field mechanism has been discussed in the context of Er$_2$Ti$_2$O$_7$ \cite{McClarty_2009,PhysRevB.93.184408} $-$ essentially the only other well-characterized example of order-by-disorder $-$ where the linear spin wave mechanism and virtual crystal field mechanism are complementary. However, in CoTiO$_3$ virtual crystal field excitations are the leading cause of the discrete symmetry breaking. A so-called flavour-wave expansion \cite{PAPANICOLAOU1984281,PAPANICOLAOU1988367,Joshi1999,Chubukov1990,Dong2018} incorporating this effect captures the magnon dispersions including the spectral gap and the dispersing crystal field excitations, as shown in Supplementary Note 10.

{\bf Neutron Intensity Fingerprint of Magnon Isospin Winding} - Having established the presence of bond-dependent exchange in this material, we now focus on the Dirac points in the magnon spectrum which provide an ideal setting to explore predicted intensity modulations associated with the isospin winding around nodal points. To explain this physics we use the simple example of a two-dimensional (2D) honeycomb Heisenberg ferromagnet ${\cal{H}}=-J\sum_{\langle i,j\rangle} \mathbf{S}_i\cdot\mathbf{S}_j$ ($J>0$) taken from Ref.~\cite{shivam2017neutron} to which we refer for further details, generalizations to band structures in 3D, and different types of touching points. The magnon band structure for this model computed within linear spin wave theory around the collinear ferromagnetic ground state has Dirac points at finite frequency at the corners (K-points) of the 2D hexagonal Brillouin zone (dashed outline in Fig.~\ref{fig:node}d). For a small momentum $\delta \mathbf{k}$ measured from a Dirac node, the effective spin wave Hamiltonian takes the famous form ${\cal{H}}_{\rm eff}=v~\delta\mathbf{k}\cdot \bm{\upsigma}$ where $v=3JSa_0/2$ is the Dirac velocity ($a_0$ is the nearest-neighbor distance) and the isospin encoded in the Pauli matrices $\bm{\upsigma}$ originates from the two sublattice honeycomb structure. By analogy with the Zeeman Hamiltonian, it follows that magnon wavefunctions carry an isospin polarization that is locked to the offset momentum $\delta \mathbf{k}$ thus  winding around each Dirac point, see Fig.~\ref{fig:node}b). This feature is directly observable via INS because, in the vicinity of these points, the intensity is, up to a constant, the projection of the isospin polarization onto some direction $\hat{\mathbf{n}}$ characteristic of each Dirac point \cite{shivam2017neutron}, illustrated by the pink radial arrows in Fig.~\ref{fig:node}d). Explicitly, the intensity takes the form  $1\pm \cos(\alpha-\alpha_0)$ where $\alpha$ is the polar angle around the K point and $\alpha_0$ defines the direction of $\hat{\mathbf{n}}$, with the upper/lower sign for the top/bottom band, respectively. Therefore, the intensity winds smoothly around the Dirac point (as illustrated by the colour shading on the two conical bands in Fig.~\ref{fig:node}a).

{\bf Isospin of Dirac Magnons} - CoTiO$_3$ provides a nearly ideal experimental platform to see the theoretically predicted winding of neutron intensity in the vicinity of the Dirac points.  Fig.~\ref{fig:node}f) shows the INS data along the (1,$\bar{1}$) in-plane direction through the nominal Dirac point at (2/3,2/3) where a clear near-linear band crossing is observed. In contrast, Fig.~\ref{fig:node}g) shows that the INS data through the same K point, but along the orthogonal (1,1) direction, has vanishingly small intensity in one of the two crossing bands. This strong intensity asymmetry in orthogonal scans is precisely what is expected based on the predicted isospin winding around a Dirac node in Fig.~\ref{fig:node}a). This can be seen more directly in Fig.~\ref{fig:node}c), which plots the intensity dependence as a function of angle $\alpha$ winding around the Dirac node in the top/bottom bands (filled/open symbols), the maxima and minima in each band are 180$^{\circ}$ apart and in anti-phase between the two bands, the solid lines show fits to the generic form $A_{\pm} \pm B_{\pm} \cos(\alpha-\alpha_0)$ with the upper/lower sign for top/bottom band. The fits give $\alpha_0=-80(3)^{\circ}$, in good agreement with the XXZ$\Delta$ model for the same scan $-81(1)^{\circ}$, the offset from $-60^{\circ}$ is due to the buckling of the honeycomb layers, which rotate the $\hat{\mathbf{n}}$ vectors in plane upon varying $L$, for more details see Supplementary Note 5. The observed two-fold angular dependence is precisely the fingerprint of the predicted isospin winding for the near-nodal quasiparticles.

\begin{figure}[tbph!]
\includegraphics[width=0.96\linewidth,keepaspectratio]{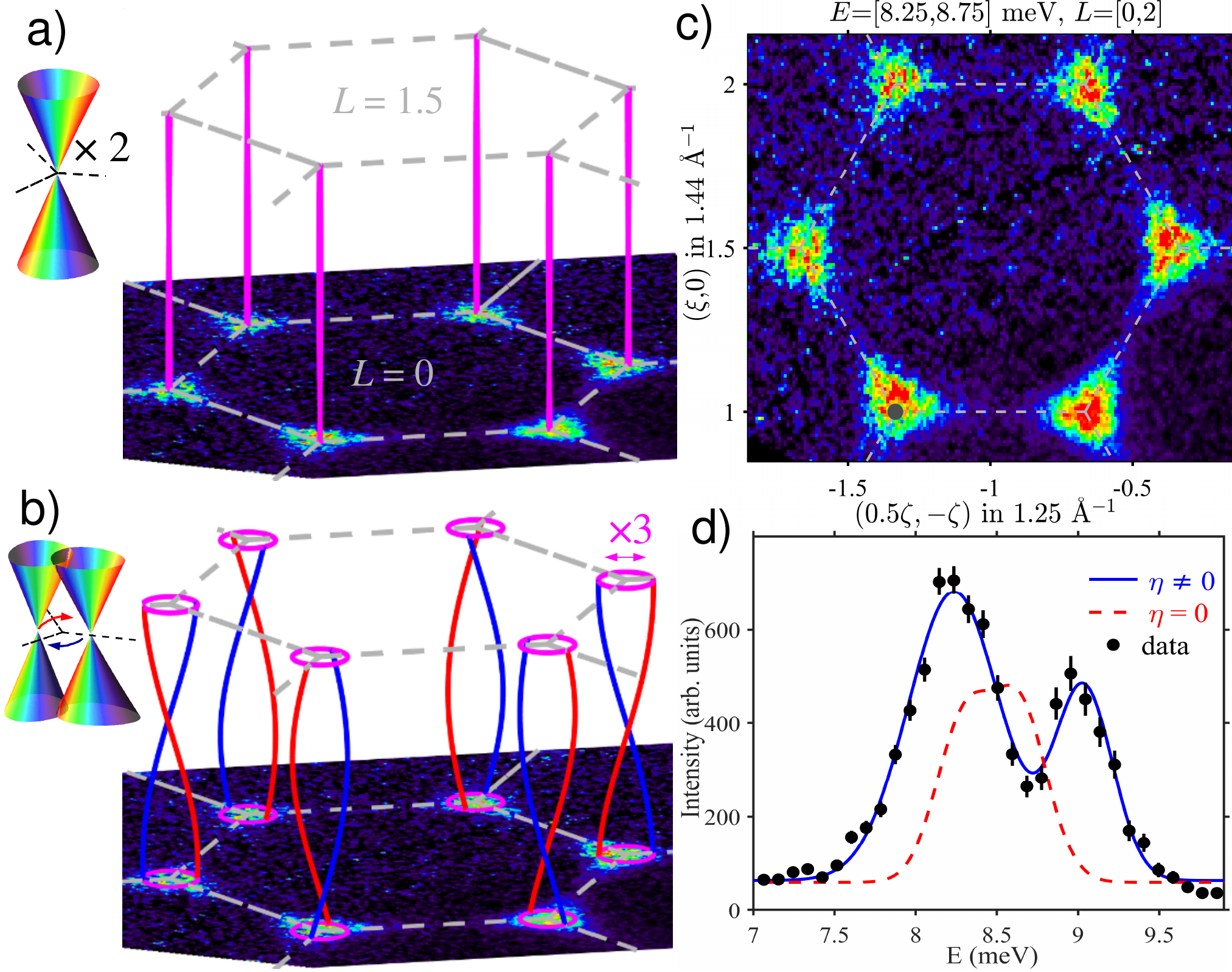}
\figcaption{Dirac magnon nodal lines}{Nodal lines along $L$ for a) Heisenberg and b) XXZ interlayer couplings (blue/red lines correspond to in-/out-of-plane polarization). Top left insets show the band structure near the nodal points, two doubly-degenerate touching cones in a) and momentum-offset pairs of touching cones in b), curly arrows indicate precession of the nodal points along $L$. In b) the diameter of the `double helix' nodal lines is amplified for visibility by $\times 3$ compared to the XXZ$\Delta$ model. c) Momentum INS intensity map as in Fig.~\ref{fig:node}d-e), but centred at the nominal nodal energy. Dashed lines show 2D Brillouin zone edges. The intensities are on a linear scale as per the colour bar in Fig.{~\ref{fig:node}g)}. d) Energy scan averaged between all six K-points in c) as well as (1/3,1/3), for a cylindrical wavevector range of in-plane radius 0.03~\AA$^{-1}$ (dark gray dot at (1/3,4/3) in c) bottom left) and $L=[0,2.3]$. Error bars represent one standard deviation, the dashed line is the calculated lineshape for the XXZ$\Delta$ model ($\eta=0$), and the solid line is a fit that includes an additional exchange anisotropy $\eta=-1.7$~meV, both calculations include instrumental resolution effects.\label{fig:helix}}
\end{figure}

{\bf Fine Structure of Dirac Magnons} - The bond-dependent exchange that is responsible for the spectral gap also affects the Dirac nodal lines. For antiferromagnetic Heisenberg interlayer couplings the nodal points form lines along $L$, each 4-fold degenerate (the top and bottom cones in Fig.~\ref{fig:node}a) are each doubly degenerate due to the antiferromagnetic doubling of the number of magnetic sublattices). For an XXZ Hamiltonian two cases can occur depending on the anisotropy of the interlayer coupling $J_2$: i) for Heisenberg $J_2$ the nodal lines are degenerate and are straight  along $L$ [see Fig.~\ref{fig:helix}a)], ii) for XXZ $J_2$ they are separated in momentum, but remain at the same energy and wind along $L$ in a `double-helix' [see Fig.~\ref{fig:helix}b)], in opposite senses between adjacent K-type points due to the $\bar{3}$ point group symmetry of the crystal lattice. However, neither of those cases can explain the fine structure observed by the energy scan in Fig.~\ref{fig:helix}d) centred at K-points, where two peaks are clearly resolved, 0.75(5)~meV apart, instead of a single peak (XXZ$\Delta$ model, dashed red line, case ii) above; for case i) the single peak would be even sharper). This fine structure was not detected by earlier lower-resolution studies \cite{Yuan} and accounting for it requires anisotropic coupling terms beyond ${\cal{H}}_{\rm XXZ}$. We have already argued that such terms must be present in order to account for the spectral gap. As shown in Supplementary Note 9, these terms all preserve the Dirac nodal lines while shifting their position in momentum space along in-plane directions related to the moment orientation in the ground state. To make quantitative contact with the experiment, we demonstrate that adding a finite nearest neighbor bond-dependent exchange $\eta$ leaves the dispersions largely unaffected relative to the $\eta=0$ case in the magnetic Brillouin zone interior while leading to the observed double peak structure in Fig.~\ref{fig:helix}d)(solid line).
\begin{figure}[htbp!]
\includegraphics[width=0.97\linewidth,keepaspectratio]{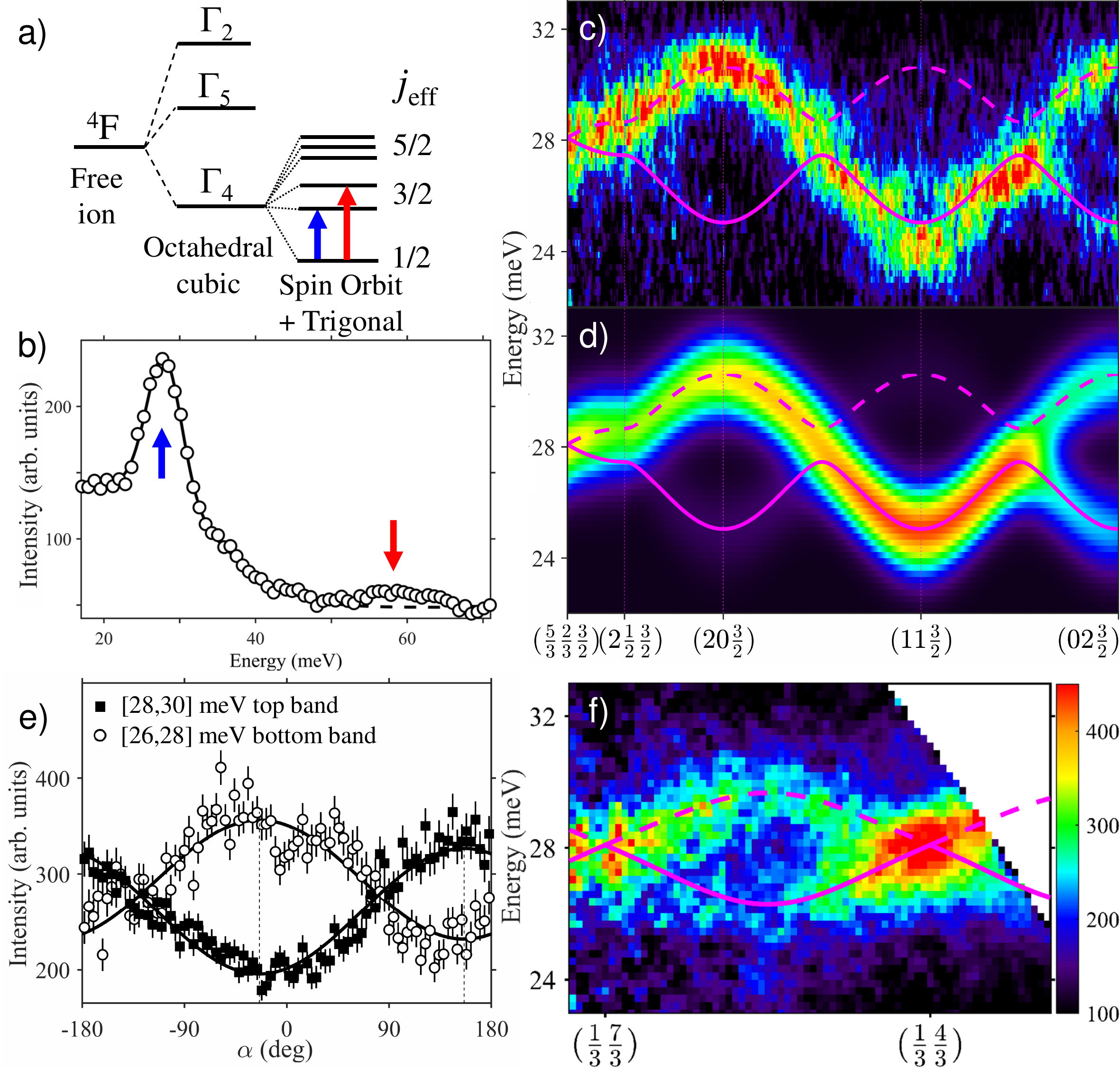}
\figcaption{Spin-orbit excitons: dispersions and Dirac node}{a) Schematic level splitting for a Co$^{2+}$ ion in an octahedral crystal field of trigonal symmetry including spin-orbit coupling. b) INS energy scan observing transitions to the first two excited crystal field levels (the blue/red arrows above the peaks show the transitions indicated by matching colour vertical arrows in a), the solid line is a guide to the eye. c) INS data probing the dispersions of the first crystal level along high-symmetry directions, compared in d) with a tight-binding model (thick solid/dashed lines through both graphs show best fit dispersions). e) Angular intensity dependence around the nodal point (2/3,5/3) for the top/bottom exciton bands fitted to an ${\cal A}_{\pm} \pm {\cal B}_{\pm}\cos(\alpha-\tilde{\alpha}_0)$ form (solid lines, $\tilde{\alpha}_0=155(3)^{\circ}$, calculated 153(1)$^{\circ}$, in-plane radial wavevector range $[0.075,0.3]$~\AA$^{-1}$). The black squares/white circles denote the inelastic neutron scattering intensity for the top/bottom exciton bands, respectively, with error bars representing one standard deviation. Note the analogous behaviour to the intensity dependence in azimuthal scans for Dirac magnons in Fig.~\ref{fig:node}c). f) Exciton bands crossing at the two labelled nodal Dirac points, analogous to the magnon bands crossing in Fig.~\ref{fig:node}f). In e-f) intensities are averaged for $L=[0,3.5]$, in c) for a transverse wavevector range $\pm0.1$~\AA$^{-1}$, and in f) for a transverse in-plane wavevector range $\pm0.025$~\AA$^{-1}$. Data were collected at 8~K with $E_{\rm i}=83$~meV in b) and 45~meV in c,e,f). The colour bar in f) also applies to c) and d), and indicate scattering intensity in arbitrary units on a linear scale. \label{fig:cf}}
\end{figure}

{\bf Dirac Excitons} - We now describe high-energy excitations, which we attribute to transitions to higher crystal field levels, where we also observe propagating excitations with linear band touching points and intensity winding around nodal points. The local spin-orbit coupled and trigonally distorted octahedral crystal field scheme for a Co$^{2+}$ ($3d^7$) ion ($L=3$ and $\fs{S}=3/2$) is shown in Fig.~\ref{fig:cf}a). Fig.~\ref{fig:cf}b) shows INS measurements observing two peaks centred near 28 and 58~meV, which we identify with the (exciton) transitions to the two trigonally-split doublets of the $j_{\rm eff}=3/2$ excited quadruplet (blue and red vertical thick arrow in Fig.~\ref{fig:cf}a).

Fig.~\ref{fig:cf}c) shows higher resolution INS measurements observing clear in-plane dispersions for the lower exciton modes near 28~meV, attributed to hopping due to spin and orbital exchange. Two modes are expected due to the two sublattices of the honeycomb structure and Fig.~\ref{fig:cf}f) shows clear evidence for mode crossing at the two labelled nodal positions. Angular intensity maps around a nodal point in Fig.~\ref{fig:cf}e) show a clear two-fold angular dependence, in anti-phase between the top/bottom bands (filled/open symbols), as expected from the intensity winding picture, again in complete analogy with the spectroscopic signature seen for the Dirac magnon wavefunctions in Fig.~\ref{fig:node}c). The observed dispersions and relative intensities of the two exciton modes can be well captured by a tight-binding model, detailed in Supplementary Note 4. The experimental and modelled exciton dispersions are compared in Figs.~\ref{fig:cf}c) and d). We note that after this work was completed, {Ref.~\cite{Yuan_exciton}} appeared, also reporting INS measurements of the exciton dispersion in CoTiO$_3$.

\section*{Discussion}
To summarise, we have reported INS measurements of the magnon dispersions in the stacked honeycomb CoTiO$_3$, which reveal the presence of a spectral gap and Dirac nodal lines. We have shown that the gap implies the presence of significant bond-dependent anisotropic exchange originating from spin-orbit coupling and we have proposed a minimal model compatible with the experimental data to explain the discrete symmetry breaking via a quantum order-by-disorder mechanism. We have also observed key signatures of proximity to Dirac magnon physics through near-linear band touching and characteristic two-fold intensity periodicity in azimuthal scans attributed to the isospin winding around the Dirac node. The similar features seen also at the nodal band crossing in the spin-orbit excitons show that neutron scattering provides a window into the universal properties of highly constrained wavefunctions around linear band-touching points in bosonic systems in the solid state.

\section*{Data Availability}
The experimental data in this study is available from Ref.~\cite{data_archive}.
\newpage

\bibliography{CoTiO3_paper.bib}

\begin{thebibliography}{59}%
\makeatletter
\providecommand \@ifxundefined [1]{%
 \@ifx{#1\undefined}
}%
\providecommand \@ifnum [1]{%
 \ifnum #1\expandafter \@firstoftwo
 \else \expandafter \@secondoftwo
 \fi
}%
\providecommand \@ifx [1]{%
 \ifx #1\expandafter \@firstoftwo
 \else \expandafter \@secondoftwo
 \fi
}%
\providecommand \natexlab [1]{#1}%
\providecommand \enquote  [1]{``#1''}%
\providecommand \bibnamefont  [1]{#1}%
\providecommand \bibfnamefont [1]{#1}%
\providecommand \citenamefont [1]{#1}%
\providecommand \href@noop [0]{\@secondoftwo}%
\providecommand \href [0]{\begingroup \@sanitize@url \@href}%
\providecommand \@href[1]{\@@startlink{#1}\@@href}%
\providecommand \@@href[1]{\endgroup#1\@@endlink}%
\providecommand \@sanitize@url [0]{\catcode `\\12\catcode `\$12\catcode
  `\&12\catcode `\#12\catcode `\^12\catcode `\_12\catcode `\%12\relax}%
\providecommand \@@startlink[1]{}%
\providecommand \@@endlink[0]{}%
\providecommand \url  [0]{\begingroup\@sanitize@url \@url }%
\providecommand \@url [1]{\endgroup\@href {#1}{\urlprefix }}%
\providecommand \urlprefix  [0]{URL }%
\providecommand \Eprint [0]{\href }%
\providecommand \doibase [0]{https://doi.org/}%
\providecommand \selectlanguage [0]{\@gobble}%
\providecommand \bibinfo  [0]{\@secondoftwo}%
\providecommand \bibfield  [0]{\@secondoftwo}%
\providecommand \translation [1]{[#1]}%
\providecommand \BibitemOpen [0]{}%
\providecommand \bibitemStop [0]{}%
\providecommand \bibitemNoStop [0]{.\EOS\space}%
\providecommand \EOS [0]{\spacefactor3000\relax}%
\providecommand \BibitemShut  [1]{\csname bibitem#1\endcsname}%
\let\auto@bib@innerbib\@empty
\bibitem [{\citenamefont {Hasan}\ and\ \citenamefont
  {Kane}(2010)}]{RevModPhys.82.3045}%
  \BibitemOpen
  \bibfield  {author} {\bibinfo {author} {\bibfnamefont {M.~Z.}\ \bibnamefont
  {Hasan}}\ and\ \bibinfo {author} {\bibfnamefont {C.~L.}\ \bibnamefont
  {Kane}},\ }\bibfield  {title} {\bibinfo {title} {Colloquium: Topological
  insulators},\ }\href {https://doi.org/10.1103/RevModPhys.82.3045} {\bibfield
  {journal} {\bibinfo  {journal} {Rev. Mod. Phys.}\ }\textbf {\bibinfo {volume}
  {82}},\ \bibinfo {pages} {3045} (\bibinfo {year} {2010})}\BibitemShut
  {NoStop}%
\bibitem [{\citenamefont {Armitage}\ \emph {et~al.}(2018)\citenamefont
  {Armitage}, \citenamefont {Mele},\ and\ \citenamefont
  {Vishwanath}}]{RevModPhys.90.015001}%
  \BibitemOpen
  \bibfield  {author} {\bibinfo {author} {\bibfnamefont {N.~P.}\ \bibnamefont
  {Armitage}}, \bibinfo {author} {\bibfnamefont {E.~J.}\ \bibnamefont {Mele}},\
  and\ \bibinfo {author} {\bibfnamefont {A.}~\bibnamefont {Vishwanath}},\
  }\bibfield  {title} {\bibinfo {title} {Weyl and {Dirac} semimetals in
  three-dimensional solids},\ }\href
  {https://doi.org/10.1103/RevModPhys.90.015001} {\bibfield  {journal}
  {\bibinfo  {journal} {Rev. Mod. Phys.}\ }\textbf {\bibinfo {volume} {90}},\
  \bibinfo {pages} {015001} (\bibinfo {year} {2018})}\BibitemShut {NoStop}%
\bibitem [{\citenamefont {Bernevig}\ and\ \citenamefont
  {Hughes}(2013)}]{bernevig2013topological}%
  \BibitemOpen
  \bibfield  {author} {\bibinfo {author} {\bibfnamefont {B.~A.}\ \bibnamefont
  {Bernevig}}\ and\ \bibinfo {author} {\bibfnamefont {T.~L.}\ \bibnamefont
  {Hughes}},\ }\href@noop {} {\emph {\bibinfo {title} {Topological insulators
  and topological superconductors}}}\ (\bibinfo  {publisher} {Princeton
  University Press},\ \bibinfo {year} {2013})\BibitemShut {NoStop}%
\bibitem [{\citenamefont {Witczak-Krempa}\ \emph {et~al.}(2014)\citenamefont
  {Witczak-Krempa}, \citenamefont {Chen}, \citenamefont {Kim},\ and\
  \citenamefont {Balents}}]{SOReview}%
  \BibitemOpen
  \bibfield  {author} {\bibinfo {author} {\bibfnamefont {W.}~\bibnamefont
  {Witczak-Krempa}}, \bibinfo {author} {\bibfnamefont {G.}~\bibnamefont
  {Chen}}, \bibinfo {author} {\bibfnamefont {Y.~B.}\ \bibnamefont {Kim}},\ and\
  \bibinfo {author} {\bibfnamefont {L.}~\bibnamefont {Balents}},\ }\bibfield
  {title} {\bibinfo {title} {Correlated quantum phenomena in the strong
  spin-orbit regime},\ }\href
  {https://doi.org/10.1146/annurev-conmatphys-020911-125138} {\bibfield
  {journal} {\bibinfo  {journal} {Ann. Rev. Cond. Matt. Phys.}\ }\textbf
  {\bibinfo {volume} {5}},\ \bibinfo {pages} {57} (\bibinfo {year}
  {2014})}\BibitemShut {NoStop}%
\bibitem [{\citenamefont {Rau}\ \emph {et~al.}(2016{\natexlab{a}})\citenamefont
  {Rau}, \citenamefont {Lee},\ and\ \citenamefont {Kee}}]{SOReview2}%
  \BibitemOpen
  \bibfield  {author} {\bibinfo {author} {\bibfnamefont {J.~G.}\ \bibnamefont
  {Rau}}, \bibinfo {author} {\bibfnamefont {E.~K.-H.}\ \bibnamefont {Lee}},\
  and\ \bibinfo {author} {\bibfnamefont {H.-Y.}\ \bibnamefont {Kee}},\
  }\bibfield  {title} {\bibinfo {title} {Spin-orbit physics giving rise to
  novel phases in correlated systems: Iridates and related materials},\ }\href
  {https://doi.org/10.1146/annurev-conmatphys-031115-011319} {\bibfield
  {journal} {\bibinfo  {journal} {Ann. Rev. Cond. Matt. Phys.}\ }\textbf
  {\bibinfo {volume} {7}},\ \bibinfo {pages} {195} (\bibinfo {year}
  {2016}{\natexlab{a}})}\BibitemShut {NoStop}%
\bibitem [{\citenamefont {Winter}\ \emph {et~al.}(2016)\citenamefont {Winter},
  \citenamefont {Li}, \citenamefont {Jeschke},\ and\ \citenamefont
  {Valent\'{\i}}}]{Winter_review}%
  \BibitemOpen
  \bibfield  {author} {\bibinfo {author} {\bibfnamefont {S.~M.}\ \bibnamefont
  {Winter}}, \bibinfo {author} {\bibfnamefont {Y.}~\bibnamefont {Li}}, \bibinfo
  {author} {\bibfnamefont {H.~O.}\ \bibnamefont {Jeschke}},\ and\ \bibinfo
  {author} {\bibfnamefont {R.}~\bibnamefont {Valent\'{\i}}},\ }\bibfield
  {title} {\bibinfo {title} {Challenges in design of {Kitaev} materials:
  Magnetic interactions from competing energy scales},\ }\href
  {https://doi.org/10.1103/PhysRevB.93.214431} {\bibfield  {journal} {\bibinfo
  {journal} {Phys. Rev. B}\ }\textbf {\bibinfo {volume} {93}},\ \bibinfo
  {pages} {214431} (\bibinfo {year} {2016})}\BibitemShut {NoStop}%
\bibitem [{\citenamefont {Hermanns}\ \emph {et~al.}(2018)\citenamefont
  {Hermanns}, \citenamefont {Kimchi},\ and\ \citenamefont
  {Knolle}}]{hermanns2018}%
  \BibitemOpen
  \bibfield  {author} {\bibinfo {author} {\bibfnamefont {M.}~\bibnamefont
  {Hermanns}}, \bibinfo {author} {\bibfnamefont {I.}~\bibnamefont {Kimchi}},\
  and\ \bibinfo {author} {\bibfnamefont {J.}~\bibnamefont {Knolle}},\
  }\bibfield  {title} {\bibinfo {title} {Physics of the {Kitaev} model:
  Fractionalization, dynamic correlations, and material connections},\ }\href
  {https://doi.org/10.1146/annurev-conmatphys-033117-053934} {\bibfield
  {journal} {\bibinfo  {journal} {Ann. Rev. Cond. Matt. Phys.}\ }\textbf
  {\bibinfo {volume} {9}},\ \bibinfo {pages} {17} (\bibinfo {year}
  {2018})}\BibitemShut {NoStop}%
\bibitem [{\citenamefont {Takagi}\ \emph {et~al.}(2019)\citenamefont {Takagi},
  \citenamefont {Takayama}, \citenamefont {Jackeli}, \citenamefont
  {Khaliullin},\ and\ \citenamefont {Nagler}}]{Takagi_review}%
  \BibitemOpen
  \bibfield  {author} {\bibinfo {author} {\bibfnamefont {H.}~\bibnamefont
  {Takagi}}, \bibinfo {author} {\bibfnamefont {T.}~\bibnamefont {Takayama}},
  \bibinfo {author} {\bibfnamefont {G.}~\bibnamefont {Jackeli}}, \bibinfo
  {author} {\bibfnamefont {G.}~\bibnamefont {Khaliullin}},\ and\ \bibinfo
  {author} {\bibfnamefont {S.~E.}\ \bibnamefont {Nagler}},\ }\bibfield  {title}
  {\bibinfo {title} {Concept and realization of {Kitaev} quantum spin
  liquids},\ }\href {https://doi.org/10.1038/s42254-019-0038-2} {\bibfield
  {journal} {\bibinfo  {journal} {Nat. Rev. Phys.}\ }\textbf {\bibinfo {volume}
  {1}},\ \bibinfo {pages} {264} (\bibinfo {year} {2019})}\BibitemShut {NoStop}%
\bibitem [{\citenamefont {Hwan~Chun}\ \emph {et~al.}(2015)\citenamefont
  {Hwan~Chun}, \citenamefont {Kim}, \citenamefont {Kim}, \citenamefont {Zheng},
  \citenamefont {Stoumpos}, \citenamefont {Malliakas}, \citenamefont
  {Mitchell}, \citenamefont {Mehlawat}, \citenamefont {Singh}, \citenamefont
  {Choi}, \citenamefont {Gog}, \citenamefont {Al-Zein}, \citenamefont {Sala},
  \citenamefont {Krisch}, \citenamefont {Chaloupka}, \citenamefont {Jackeli},
  \citenamefont {Khaliullin},\ and\ \citenamefont {Kim}}]{Chun}%
  \BibitemOpen
  \bibfield  {author} {\bibinfo {author} {\bibfnamefont {S.}~\bibnamefont
  {Hwan~Chun}}, \bibinfo {author} {\bibfnamefont {J.-W.}\ \bibnamefont {Kim}},
  \bibinfo {author} {\bibfnamefont {J.}~\bibnamefont {Kim}}, \bibinfo {author}
  {\bibfnamefont {H.}~\bibnamefont {Zheng}}, \bibinfo {author} {\bibfnamefont
  {C.~C.}\ \bibnamefont {Stoumpos}}, \bibinfo {author} {\bibfnamefont {C.~D.}\
  \bibnamefont {Malliakas}}, \bibinfo {author} {\bibfnamefont {J.~F.}\
  \bibnamefont {Mitchell}}, \bibinfo {author} {\bibfnamefont {K.}~\bibnamefont
  {Mehlawat}}, \bibinfo {author} {\bibfnamefont {Y.}~\bibnamefont {Singh}},
  \bibinfo {author} {\bibfnamefont {Y.}~\bibnamefont {Choi}}, \bibinfo {author}
  {\bibfnamefont {T.}~\bibnamefont {Gog}}, \bibinfo {author} {\bibfnamefont
  {A.}~\bibnamefont {Al-Zein}}, \bibinfo {author} {\bibfnamefont {M.~M.}\
  \bibnamefont {Sala}}, \bibinfo {author} {\bibfnamefont {M.}~\bibnamefont
  {Krisch}}, \bibinfo {author} {\bibfnamefont {J.}~\bibnamefont {Chaloupka}},
  \bibinfo {author} {\bibfnamefont {G.}~\bibnamefont {Jackeli}}, \bibinfo
  {author} {\bibfnamefont {G.}~\bibnamefont {Khaliullin}},\ and\ \bibinfo
  {author} {\bibfnamefont {B.~J.}\ \bibnamefont {Kim}},\ }\bibfield  {title}
  {\bibinfo {title} {Direct evidence for dominant bond-directional interactions
  in a honeycomb lattice iridate {$\mathrm{Na}_{2}{\mathrm{IrO}}_{3}$}},\
  }\href {https://doi.org/10.1038/nphys3322} {\bibfield  {journal} {\bibinfo
  {journal} {Nature Phys.}\ }\textbf {\bibinfo {volume} {11}},\ \bibinfo
  {pages} {462} (\bibinfo {year} {2015})}\BibitemShut {NoStop}%
\bibitem [{\citenamefont {Banerjee}\ \emph {et~al.}(2017)\citenamefont
  {Banerjee}, \citenamefont {Yan}, \citenamefont {Knolle}, \citenamefont
  {Bridges}, \citenamefont {Stone}, \citenamefont {Lumsden}, \citenamefont
  {Mandrus}, \citenamefont {Tennant}, \citenamefont {Moessner},\ and\
  \citenamefont {Nagler}}]{banerjee2017neutron}%
  \BibitemOpen
  \bibfield  {author} {\bibinfo {author} {\bibfnamefont {A.}~\bibnamefont
  {Banerjee}}, \bibinfo {author} {\bibfnamefont {J.}~\bibnamefont {Yan}},
  \bibinfo {author} {\bibfnamefont {J.}~\bibnamefont {Knolle}}, \bibinfo
  {author} {\bibfnamefont {C.~A.}\ \bibnamefont {Bridges}}, \bibinfo {author}
  {\bibfnamefont {M.~B.}\ \bibnamefont {Stone}}, \bibinfo {author}
  {\bibfnamefont {M.~D.}\ \bibnamefont {Lumsden}}, \bibinfo {author}
  {\bibfnamefont {D.~G.}\ \bibnamefont {Mandrus}}, \bibinfo {author}
  {\bibfnamefont {D.~A.}\ \bibnamefont {Tennant}}, \bibinfo {author}
  {\bibfnamefont {R.}~\bibnamefont {Moessner}},\ and\ \bibinfo {author}
  {\bibfnamefont {S.~E.}\ \bibnamefont {Nagler}},\ }\bibfield  {title}
  {\bibinfo {title} {Neutron scattering in the proximate quantum spin liquid
  {$\alpha$-${\mathrm{RuCl}}_3$}},\ }\href@noop {} {\bibfield  {journal}
  {\bibinfo  {journal} {Science}\ }\textbf {\bibinfo {volume} {356}},\ \bibinfo
  {pages} {1055} (\bibinfo {year} {2017})}\BibitemShut {NoStop}%
\bibitem [{\citenamefont {Kasahara}\ \emph {et~al.}(2018)\citenamefont
  {Kasahara}, \citenamefont {Ohnishi}, \citenamefont {Mizukami}, \citenamefont
  {Tanaka}, \citenamefont {Ma}, \citenamefont {Sugii}, \citenamefont {Kurita},
  \citenamefont {Tanaka}, \citenamefont {Nasu}, \citenamefont {Motome},
  \citenamefont {Shibauchi},\ and\ \citenamefont {Matsuda}}]{Matsuda}%
  \BibitemOpen
  \bibfield  {author} {\bibinfo {author} {\bibfnamefont {Y.}~\bibnamefont
  {Kasahara}}, \bibinfo {author} {\bibfnamefont {T.}~\bibnamefont {Ohnishi}},
  \bibinfo {author} {\bibfnamefont {Y.}~\bibnamefont {Mizukami}}, \bibinfo
  {author} {\bibfnamefont {O.}~\bibnamefont {Tanaka}}, \bibinfo {author}
  {\bibfnamefont {S.}~\bibnamefont {Ma}}, \bibinfo {author} {\bibfnamefont
  {K.}~\bibnamefont {Sugii}}, \bibinfo {author} {\bibfnamefont
  {N.}~\bibnamefont {Kurita}}, \bibinfo {author} {\bibfnamefont
  {H.}~\bibnamefont {Tanaka}}, \bibinfo {author} {\bibfnamefont
  {J.}~\bibnamefont {Nasu}}, \bibinfo {author} {\bibfnamefont {Y.}~\bibnamefont
  {Motome}}, \bibinfo {author} {\bibfnamefont {T.}~\bibnamefont {Shibauchi}},\
  and\ \bibinfo {author} {\bibfnamefont {Y.}~\bibnamefont {Matsuda}},\
  }\bibfield  {title} {\bibinfo {title} {Majorana quantization and half-integer
  thermal quantum hall effect in a {Kitaev} spin liquid},\ }\href
  {https://doi.org/10.1038/s41586-018-0274-0} {\bibfield  {journal} {\bibinfo
  {journal} {Nature}\ }\textbf {\bibinfo {volume} {559}},\ \bibinfo {pages}
  {227} (\bibinfo {year} {2018})}\BibitemShut {NoStop}%
\bibitem [{\citenamefont {Kitaev}(2006)}]{Kitaev}%
  \BibitemOpen
  \bibfield  {author} {\bibinfo {author} {\bibfnamefont {A.}~\bibnamefont
  {Kitaev}},\ }\bibfield  {title} {\bibinfo {title} {Anyons in an exactly
  solved model and beyond},\ }\href {https://doi.org/10.1016/j.aop.2005.10.005}
  {\bibfield  {journal} {\bibinfo  {journal} {Ann. Phys.}\ }\textbf {\bibinfo
  {volume} {321}},\ \bibinfo {pages} {2} (\bibinfo {year} {2006})}\BibitemShut
  {NoStop}%
\bibitem [{\citenamefont {Jackeli}\ and\ \citenamefont
  {Khaliullin}(2009)}]{PhysRevLett.102.017205}%
  \BibitemOpen
  \bibfield  {author} {\bibinfo {author} {\bibfnamefont {G.}~\bibnamefont
  {Jackeli}}\ and\ \bibinfo {author} {\bibfnamefont {G.}~\bibnamefont
  {Khaliullin}},\ }\bibfield  {title} {\bibinfo {title} {Mott insulators in the
  strong spin-orbit coupling limit: From {Heisenberg} to a quantum compass and
  {Kitaev} models},\ }\href {https://doi.org/10.1103/PhysRevLett.102.017205}
  {\bibfield  {journal} {\bibinfo  {journal} {Phys. Rev. Lett.}\ }\textbf
  {\bibinfo {volume} {102}},\ \bibinfo {pages} {017205} (\bibinfo {year}
  {2009})}\BibitemShut {NoStop}%
\bibitem [{\citenamefont {Chaloupka}\ \emph {et~al.}(2010)\citenamefont
  {Chaloupka}, \citenamefont {Jackeli},\ and\ \citenamefont
  {Khaliullin}}]{PhysRevLett.105.027204}%
  \BibitemOpen
  \bibfield  {author} {\bibinfo {author} {\bibfnamefont {J.}~\bibnamefont
  {Chaloupka}}, \bibinfo {author} {\bibfnamefont {G.}~\bibnamefont {Jackeli}},\
  and\ \bibinfo {author} {\bibfnamefont {G.}~\bibnamefont {Khaliullin}},\
  }\bibfield  {title} {\bibinfo {title} {Kitaev-{Heisenberg} model on a
  honeycomb lattice: Possible exotic phases in iridium oxides
  {${A}_{2}{\mathrm{IrO}}_{3}$}},\ }\href
  {https://doi.org/10.1103/PhysRevLett.105.027204} {\bibfield  {journal}
  {\bibinfo  {journal} {Phys. Rev. Lett.}\ }\textbf {\bibinfo {volume} {105}},\
  \bibinfo {pages} {027204} (\bibinfo {year} {2010})}\BibitemShut {NoStop}%
\bibitem [{\citenamefont {Liu}\ and\ \citenamefont
  {Khaliullin}(2018)}]{PhysRevB.97.014407}%
  \BibitemOpen
  \bibfield  {author} {\bibinfo {author} {\bibfnamefont {H.}~\bibnamefont
  {Liu}}\ and\ \bibinfo {author} {\bibfnamefont {G.}~\bibnamefont
  {Khaliullin}},\ }\bibfield  {title} {\bibinfo {title} {Pseudospin exchange
  interactions in ${d}^{7}$ cobalt compounds: Possible realization of the
  {Kitaev} model},\ }\href {https://doi.org/10.1103/PhysRevB.97.014407}
  {\bibfield  {journal} {\bibinfo  {journal} {Phys. Rev. B}\ }\textbf {\bibinfo
  {volume} {97}},\ \bibinfo {pages} {014407} (\bibinfo {year}
  {2018})}\BibitemShut {NoStop}%
\bibitem [{\citenamefont {Sano}\ \emph {et~al.}(2018)\citenamefont {Sano},
  \citenamefont {Kato},\ and\ \citenamefont {Motome}}]{PhysRevB.97.014408}%
  \BibitemOpen
  \bibfield  {author} {\bibinfo {author} {\bibfnamefont {R.}~\bibnamefont
  {Sano}}, \bibinfo {author} {\bibfnamefont {Y.}~\bibnamefont {Kato}},\ and\
  \bibinfo {author} {\bibfnamefont {Y.}~\bibnamefont {Motome}},\ }\bibfield
  {title} {\bibinfo {title} {Kitaev-{Heisenberg} {Hamiltonian} for high-spin
  ${d}^{7}$ {Mott} insulators},\ }\href
  {https://doi.org/10.1103/PhysRevB.97.014408} {\bibfield  {journal} {\bibinfo
  {journal} {Phys. Rev. B}\ }\textbf {\bibinfo {volume} {97}},\ \bibinfo
  {pages} {014408} (\bibinfo {year} {2018})}\BibitemShut {NoStop}%
\bibitem [{\citenamefont {Liu}\ \emph {et~al.}(2020)\citenamefont {Liu},
  \citenamefont {Chaloupka},\ and\ \citenamefont
  {Khaliullin}}]{PhysRevLett.125.047201}%
  \BibitemOpen
  \bibfield  {author} {\bibinfo {author} {\bibfnamefont {H.}~\bibnamefont
  {Liu}}, \bibinfo {author} {\bibfnamefont {J.}~\bibnamefont {Chaloupka}},\
  and\ \bibinfo {author} {\bibfnamefont {G.}~\bibnamefont {Khaliullin}},\
  }\bibfield  {title} {\bibinfo {title} {Kitaev spin liquid in $3d$ transition
  metal compounds},\ }\href {https://doi.org/10.1103/PhysRevLett.125.047201}
  {\bibfield  {journal} {\bibinfo  {journal} {Phys. Rev. Lett.}\ }\textbf
  {\bibinfo {volume} {125}},\ \bibinfo {pages} {047201} (\bibinfo {year}
  {2020})}\BibitemShut {NoStop}%
\bibitem [{\citenamefont {Yuan}\ \emph
  {et~al.}(2020{\natexlab{a}})\citenamefont {Yuan}, \citenamefont {Khait},
  \citenamefont {Shu}, \citenamefont {Chou}, \citenamefont {Stone},
  \citenamefont {Clancy}, \citenamefont {Paramekanti},\ and\ \citenamefont
  {Kim}}]{Yuan}%
  \BibitemOpen
  \bibfield  {author} {\bibinfo {author} {\bibfnamefont {B.}~\bibnamefont
  {Yuan}}, \bibinfo {author} {\bibfnamefont {I.}~\bibnamefont {Khait}},
  \bibinfo {author} {\bibfnamefont {G.-J.}\ \bibnamefont {Shu}}, \bibinfo
  {author} {\bibfnamefont {F.~C.}\ \bibnamefont {Chou}}, \bibinfo {author}
  {\bibfnamefont {M.~B.}\ \bibnamefont {Stone}}, \bibinfo {author}
  {\bibfnamefont {J.~P.}\ \bibnamefont {Clancy}}, \bibinfo {author}
  {\bibfnamefont {A.}~\bibnamefont {Paramekanti}},\ and\ \bibinfo {author}
  {\bibfnamefont {Y.-J.}\ \bibnamefont {Kim}},\ }\bibfield  {title} {\bibinfo
  {title} {Dirac magnons in a honeycomb lattice quantum {$\mathit{XY}$} magnet
  {${\mathrm{CoTiO}}_{3}$}},\ }\href
  {https://doi.org/10.1103/PhysRevX.10.011062} {\bibfield  {journal} {\bibinfo
  {journal} {Phys. Rev. X}\ }\textbf {\bibinfo {volume} {10}},\ \bibinfo
  {pages} {011062} (\bibinfo {year} {2020}{\natexlab{a}})}\BibitemShut
  {NoStop}%
\bibitem [{\citenamefont {Villain}\ \emph {et~al.}(1980)\citenamefont
  {Villain}, \citenamefont {Bidaux}, \citenamefont {Carton},\ and\
  \citenamefont {Conte}}]{villain:jpa-00208953}%
  \BibitemOpen
  \bibfield  {author} {\bibinfo {author} {\bibfnamefont {J.}~\bibnamefont
  {Villain}}, \bibinfo {author} {\bibfnamefont {R.}~\bibnamefont {Bidaux}},
  \bibinfo {author} {\bibfnamefont {J.-P.}\ \bibnamefont {Carton}},\ and\
  \bibinfo {author} {\bibfnamefont {R.}~\bibnamefont {Conte}},\ }\bibfield
  {title} {\bibinfo {title} {{Order as an effect of disorder}},\ }\href
  {https://doi.org/10.1051/jphys:0198000410110126300} {\bibfield  {journal}
  {\bibinfo  {journal} {{Journal de Physique}}\ }\textbf {\bibinfo {volume}
  {41}},\ \bibinfo {pages} {1263} (\bibinfo {year} {1980})}\BibitemShut
  {NoStop}%
\bibitem [{\citenamefont {Henley}(1989)}]{PhysRevLett.62.2056}%
  \BibitemOpen
  \bibfield  {author} {\bibinfo {author} {\bibfnamefont {C.~L.}\ \bibnamefont
  {Henley}},\ }\bibfield  {title} {\bibinfo {title} {Ordering due to disorder
  in a frustrated vector antiferromagnet},\ }\href
  {https://doi.org/10.1103/PhysRevLett.62.2056} {\bibfield  {journal} {\bibinfo
   {journal} {Phys. Rev. Lett.}\ }\textbf {\bibinfo {volume} {62}},\ \bibinfo
  {pages} {2056} (\bibinfo {year} {1989})}\BibitemShut {NoStop}%
\bibitem [{\citenamefont {Moessner}\ and\ \citenamefont
  {Chalker}(1998)}]{PhysRevB.58.12049}%
  \BibitemOpen
  \bibfield  {author} {\bibinfo {author} {\bibfnamefont {R.}~\bibnamefont
  {Moessner}}\ and\ \bibinfo {author} {\bibfnamefont {J.~T.}\ \bibnamefont
  {Chalker}},\ }\bibfield  {title} {\bibinfo {title} {Low-temperature
  properties of classical geometrically frustrated antiferromagnets},\ }\href
  {https://doi.org/10.1103/PhysRevB.58.12049} {\bibfield  {journal} {\bibinfo
  {journal} {Phys. Rev. B}\ }\textbf {\bibinfo {volume} {58}},\ \bibinfo
  {pages} {12049} (\bibinfo {year} {1998})}\BibitemShut {NoStop}%
\bibitem [{\citenamefont {Chubukov}\ and\ \citenamefont
  {Jolicoeur}(1992)}]{PhysRevB.46.11137}%
  \BibitemOpen
  \bibfield  {author} {\bibinfo {author} {\bibfnamefont {A.~V.}\ \bibnamefont
  {Chubukov}}\ and\ \bibinfo {author} {\bibfnamefont {T.}~\bibnamefont
  {Jolicoeur}},\ }\bibfield  {title} {\bibinfo {title} {Order-from-disorder
  phenomena in heisenberg antiferromagnets on a triangular lattice},\ }\href
  {https://doi.org/10.1103/PhysRevB.46.11137} {\bibfield  {journal} {\bibinfo
  {journal} {Phys. Rev. B}\ }\textbf {\bibinfo {volume} {46}},\ \bibinfo
  {pages} {11137} (\bibinfo {year} {1992})}\BibitemShut {NoStop}%
\bibitem [{\citenamefont {Champion}\ \emph {et~al.}(2003)\citenamefont
  {Champion}, \citenamefont {Harris}, \citenamefont {Holdsworth}, \citenamefont
  {Wills}, \citenamefont {Balakrishnan}, \citenamefont {Bramwell},
  \citenamefont {\ifmmode \check{C}\else \v{C}\fi{}i\ifmmode~\check{z}\else
  \v{z}\fi{}m\'ar}, \citenamefont {Fennell}, \citenamefont {Gardner},
  \citenamefont {Lago}, \citenamefont {McMorrow}, \citenamefont
  {Orend\'a\ifmmode~\check{c}\else \v{c}\fi{}}, \citenamefont
  {Orend\'a\ifmmode~\check{c}\else \v{c}\fi{}ov\'a}, \citenamefont {Paul},
  \citenamefont {Smith}, \citenamefont {Telling},\ and\ \citenamefont
  {Wildes}}]{PhysRevB.68.020401}%
  \BibitemOpen
  \bibfield  {author} {\bibinfo {author} {\bibfnamefont {J.~D.~M.}\
  \bibnamefont {Champion}}, \bibinfo {author} {\bibfnamefont {M.~J.}\
  \bibnamefont {Harris}}, \bibinfo {author} {\bibfnamefont {P.~C.~W.}\
  \bibnamefont {Holdsworth}}, \bibinfo {author} {\bibfnamefont {A.~S.}\
  \bibnamefont {Wills}}, \bibinfo {author} {\bibfnamefont {G.}~\bibnamefont
  {Balakrishnan}}, \bibinfo {author} {\bibfnamefont {S.~T.}\ \bibnamefont
  {Bramwell}}, \bibinfo {author} {\bibfnamefont {E.}~\bibnamefont {\ifmmode
  \check{C}\else \v{C}\fi{}i\ifmmode~\check{z}\else \v{z}\fi{}m\'ar}}, \bibinfo
  {author} {\bibfnamefont {T.}~\bibnamefont {Fennell}}, \bibinfo {author}
  {\bibfnamefont {J.~S.}\ \bibnamefont {Gardner}}, \bibinfo {author}
  {\bibfnamefont {J.}~\bibnamefont {Lago}}, \bibinfo {author} {\bibfnamefont
  {D.~F.}\ \bibnamefont {McMorrow}}, \bibinfo {author} {\bibfnamefont
  {M.}~\bibnamefont {Orend\'a\ifmmode~\check{c}\else \v{c}\fi{}}}, \bibinfo
  {author} {\bibfnamefont {A.}~\bibnamefont {Orend\'a\ifmmode~\check{c}\else
  \v{c}\fi{}ov\'a}}, \bibinfo {author} {\bibfnamefont {D.~M.}\ \bibnamefont
  {Paul}}, \bibinfo {author} {\bibfnamefont {R.~I.}\ \bibnamefont {Smith}},
  \bibinfo {author} {\bibfnamefont {M.~T.~F.}\ \bibnamefont {Telling}},\ and\
  \bibinfo {author} {\bibfnamefont {A.}~\bibnamefont {Wildes}},\ }\bibfield
  {title} {\bibinfo {title}
  {{${\mathrm{Er}}_{2}{\mathrm{Ti}}_{2}{\mathrm{O}}_{7}$}: {Evidence} of
  quantum order by disorder in a frustrated antiferromagnet},\ }\href
  {https://doi.org/10.1103/PhysRevB.68.020401} {\bibfield  {journal} {\bibinfo
  {journal} {Phys. Rev. B}\ }\textbf {\bibinfo {volume} {68}},\ \bibinfo
  {pages} {020401} (\bibinfo {year} {2003})}\BibitemShut {NoStop}%
\bibitem [{\citenamefont {Savary}\ \emph {et~al.}(2012)\citenamefont {Savary},
  \citenamefont {Ross}, \citenamefont {Gaulin}, \citenamefont {Ruff},\ and\
  \citenamefont {Balents}}]{ObD}%
  \BibitemOpen
  \bibfield  {author} {\bibinfo {author} {\bibfnamefont {L.}~\bibnamefont
  {Savary}}, \bibinfo {author} {\bibfnamefont {K.~A.}\ \bibnamefont {Ross}},
  \bibinfo {author} {\bibfnamefont {B.~D.}\ \bibnamefont {Gaulin}}, \bibinfo
  {author} {\bibfnamefont {J.~P.~C.}\ \bibnamefont {Ruff}},\ and\ \bibinfo
  {author} {\bibfnamefont {L.}~\bibnamefont {Balents}},\ }\bibfield  {title}
  {\bibinfo {title} {Order by quantum disorder in
  {${\mathrm{Er}}_{2}{\mathrm{Ti}}_{2}{\mathrm{O}}_{7}$}},\ }\href
  {https://doi.org/10.1103/PhysRevLett.109.167201} {\bibfield  {journal}
  {\bibinfo  {journal} {Phys. Rev. Lett.}\ }\textbf {\bibinfo {volume} {109}},\
  \bibinfo {pages} {167201} (\bibinfo {year} {2012})}\BibitemShut {NoStop}%
\bibitem [{\citenamefont {Zhitomirsky}\ \emph {et~al.}(2012)\citenamefont
  {Zhitomirsky}, \citenamefont {Gvozdikova}, \citenamefont {Holdsworth},\ and\
  \citenamefont {Moessner}}]{PhysRevLett.109.077204}%
  \BibitemOpen
  \bibfield  {author} {\bibinfo {author} {\bibfnamefont {M.~E.}\ \bibnamefont
  {Zhitomirsky}}, \bibinfo {author} {\bibfnamefont {M.~V.}\ \bibnamefont
  {Gvozdikova}}, \bibinfo {author} {\bibfnamefont {P.~C.~W.}\ \bibnamefont
  {Holdsworth}},\ and\ \bibinfo {author} {\bibfnamefont {R.}~\bibnamefont
  {Moessner}},\ }\bibfield  {title} {\bibinfo {title} {Quantum order by
  disorder and accidental soft mode in
  {${\mathrm{Er}}_{2}{\mathrm{Ti}}_{2}{\mathrm{O}}_{7}$}},\ }\href
  {https://doi.org/10.1103/PhysRevLett.109.077204} {\bibfield  {journal}
  {\bibinfo  {journal} {Phys. Rev. Lett.}\ }\textbf {\bibinfo {volume} {109}},\
  \bibinfo {pages} {077204} (\bibinfo {year} {2012})}\BibitemShut {NoStop}%
\bibitem [{\citenamefont {McClarty}\ \emph {et~al.}(2009)\citenamefont
  {McClarty}, \citenamefont {Curnoe},\ and\ \citenamefont
  {Gingras}}]{McClarty_2009}%
  \BibitemOpen
  \bibfield  {author} {\bibinfo {author} {\bibfnamefont {P.~A.}\ \bibnamefont
  {McClarty}}, \bibinfo {author} {\bibfnamefont {S.~H.}\ \bibnamefont
  {Curnoe}},\ and\ \bibinfo {author} {\bibfnamefont {M.~J.~P.}\ \bibnamefont
  {Gingras}},\ }\bibfield  {title} {\bibinfo {title} {Energetic selection of
  ordered states in a model of the
  {${\mathrm{Er}}_{2}{\mathrm{Ti}}_{2}{\mathrm{O}}_{7}$} frustrated pyrochlore
  {XY} antiferromagnet},\ }\href
  {https://doi.org/10.1088/1742-6596/145/1/012032} {\bibfield  {journal}
  {\bibinfo  {journal} {J. Phys. Conf. Ser.}\ }\textbf {\bibinfo {volume}
  {145}},\ \bibinfo {pages} {012032} (\bibinfo {year} {2009})}\BibitemShut
  {NoStop}%
\bibitem [{\citenamefont {Rau}\ \emph {et~al.}(2016{\natexlab{b}})\citenamefont
  {Rau}, \citenamefont {Petit},\ and\ \citenamefont
  {Gingras}}]{PhysRevB.93.184408}%
  \BibitemOpen
  \bibfield  {author} {\bibinfo {author} {\bibfnamefont {J.~G.}\ \bibnamefont
  {Rau}}, \bibinfo {author} {\bibfnamefont {S.}~\bibnamefont {Petit}},\ and\
  \bibinfo {author} {\bibfnamefont {M.~J.~P.}\ \bibnamefont {Gingras}},\
  }\bibfield  {title} {\bibinfo {title} {Order by virtual crystal field
  fluctuations in pyrochlore {XY} antiferromagnets},\ }\href
  {https://doi.org/10.1103/PhysRevB.93.184408} {\bibfield  {journal} {\bibinfo
  {journal} {Phys. Rev. B}\ }\textbf {\bibinfo {volume} {93}},\ \bibinfo
  {pages} {184408} (\bibinfo {year} {2016}{\natexlab{b}})}\BibitemShut
  {NoStop}%
\bibitem [{\citenamefont {{Bao}}\ \emph {et~al.}(2018)\citenamefont {{Bao}},
  \citenamefont {{Wang}}, \citenamefont {{Wang}}, \citenamefont {{Cai}},
  \citenamefont {{Li}}, \citenamefont {{Ma}}, \citenamefont {{Wang}},
  \citenamefont {{Ran}}, \citenamefont {{Dong}}, \citenamefont {{Abernathy}},
  \citenamefont {{Yu}}, \citenamefont {{Wan}}, \citenamefont {{Li}},\ and\
  \citenamefont {{Wen}}}]{Bao2018}%
  \BibitemOpen
  \bibfield  {author} {\bibinfo {author} {\bibfnamefont {S.}~\bibnamefont
  {{Bao}}}, \bibinfo {author} {\bibfnamefont {J.}~\bibnamefont {{Wang}}},
  \bibinfo {author} {\bibfnamefont {W.}~\bibnamefont {{Wang}}}, \bibinfo
  {author} {\bibfnamefont {Z.}~\bibnamefont {{Cai}}}, \bibinfo {author}
  {\bibfnamefont {S.}~\bibnamefont {{Li}}}, \bibinfo {author} {\bibfnamefont
  {Z.}~\bibnamefont {{Ma}}}, \bibinfo {author} {\bibfnamefont {D.}~\bibnamefont
  {{Wang}}}, \bibinfo {author} {\bibfnamefont {K.}~\bibnamefont {{Ran}}},
  \bibinfo {author} {\bibfnamefont {Z.-Y.}\ \bibnamefont {{Dong}}}, \bibinfo
  {author} {\bibfnamefont {D.~L.}\ \bibnamefont {{Abernathy}}}, \bibinfo
  {author} {\bibfnamefont {S.-L.}\ \bibnamefont {{Yu}}}, \bibinfo {author}
  {\bibfnamefont {X.}~\bibnamefont {{Wan}}}, \bibinfo {author} {\bibfnamefont
  {J.-X.}\ \bibnamefont {{Li}}},\ and\ \bibinfo {author} {\bibfnamefont
  {J.}~\bibnamefont {{Wen}}},\ }\bibfield  {title} {\bibinfo {title}
  {{Discovery of coexisting Dirac and triply degenerate magnons in a
  three-dimensional antiferromagnet}},\ }\href
  {https://doi.org/10.1038/s41467-018-05054-2} {\bibfield  {journal} {\bibinfo
  {journal} {Nature Communications}\ }\textbf {\bibinfo {volume} {9}},\
  \bibinfo {eid} {2591} (\bibinfo {year} {2018})},\ \Eprint
  {https://arxiv.org/abs/1711.02960} {arXiv:1711.02960 [cond-mat.str-el]}
  \BibitemShut {NoStop}%
\bibitem [{\citenamefont {Yao}\ \emph {et~al.}(2018)\citenamefont {Yao},
  \citenamefont {Li}, \citenamefont {Wang}, \citenamefont {Xue}, \citenamefont
  {Dan}, \citenamefont {Iida}, \citenamefont {Kamazawa}, \citenamefont {Li},
  \citenamefont {Fang},\ and\ \citenamefont {Li}}]{yao2018topological}%
  \BibitemOpen
  \bibfield  {author} {\bibinfo {author} {\bibfnamefont {W.}~\bibnamefont
  {Yao}}, \bibinfo {author} {\bibfnamefont {C.}~\bibnamefont {Li}}, \bibinfo
  {author} {\bibfnamefont {L.}~\bibnamefont {Wang}}, \bibinfo {author}
  {\bibfnamefont {S.}~\bibnamefont {Xue}}, \bibinfo {author} {\bibfnamefont
  {Y.}~\bibnamefont {Dan}}, \bibinfo {author} {\bibfnamefont {K.}~\bibnamefont
  {Iida}}, \bibinfo {author} {\bibfnamefont {K.}~\bibnamefont {Kamazawa}},
  \bibinfo {author} {\bibfnamefont {K.}~\bibnamefont {Li}}, \bibinfo {author}
  {\bibfnamefont {C.}~\bibnamefont {Fang}},\ and\ \bibinfo {author}
  {\bibfnamefont {Y.}~\bibnamefont {Li}},\ }\bibfield  {title} {\bibinfo
  {title} {Topological spin excitations in a three-dimensional
  antiferromagnet},\ }\href {https://doi.org/10.1038/s41567-018-0213-x}
  {\bibfield  {journal} {\bibinfo  {journal} {Nature Physics}\ }\textbf
  {\bibinfo {volume} {14}},\ \bibinfo {pages} {1011} (\bibinfo {year}
  {2018})}\BibitemShut {NoStop}%
\bibitem [{\citenamefont {Shivam}\ \emph {et~al.}(2017)\citenamefont {Shivam},
  \citenamefont {Coldea}, \citenamefont {Moessner},\ and\ \citenamefont
  {McClarty}}]{shivam2017neutron}%
  \BibitemOpen
  \bibfield  {author} {\bibinfo {author} {\bibfnamefont {S.}~\bibnamefont
  {Shivam}}, \bibinfo {author} {\bibfnamefont {R.}~\bibnamefont {Coldea}},
  \bibinfo {author} {\bibfnamefont {R.}~\bibnamefont {Moessner}},\ and\
  \bibinfo {author} {\bibfnamefont {P.}~\bibnamefont {McClarty}},\ }\bibfield
  {title} {\bibinfo {title} {Neutron scattering signatures of magnon {Weyl}
  points},\ }\href@noop {} {\bibfield  {journal} {\bibinfo  {journal}
  {arXiv:1712.08535}\ } (\bibinfo {year} {2017})}\BibitemShut {NoStop}%
\bibitem [{\citenamefont {Balbashov}\ \emph {et~al.}(2017)\citenamefont
  {Balbashov}, \citenamefont {Mukhin}, \citenamefont {Ivanov}, \citenamefont
  {Iskhakova},\ and\ \citenamefont {Voronchikhina}}]{Balbashov}%
  \BibitemOpen
  \bibfield  {author} {\bibinfo {author} {\bibfnamefont {A.~M.}\ \bibnamefont
  {Balbashov}}, \bibinfo {author} {\bibfnamefont {A.~A.}\ \bibnamefont
  {Mukhin}}, \bibinfo {author} {\bibfnamefont {V.~Y.}\ \bibnamefont {Ivanov}},
  \bibinfo {author} {\bibfnamefont {L.~D.}\ \bibnamefont {Iskhakova}},\ and\
  \bibinfo {author} {\bibfnamefont {M.~E.}\ \bibnamefont {Voronchikhina}},\
  }\bibfield  {title} {\bibinfo {title} {Electric and magnetic properties of
  titanium-cobalt-oxide single crystals produced by floating zone melting with
  light heating},\ }\href {https://doi.org/10.1063/1.5001297} {\bibfield
  {journal} {\bibinfo  {journal} {Low Temp. Phys.}\ }\textbf {\bibinfo {volume}
  {43}},\ \bibinfo {pages} {965} (\bibinfo {year} {2017})}\BibitemShut
  {NoStop}%
\bibitem [{\citenamefont {Papanicolaou}(1984)}]{PAPANICOLAOU1984281}%
  \BibitemOpen
  \bibfield  {author} {\bibinfo {author} {\bibfnamefont {N.}~\bibnamefont
  {Papanicolaou}},\ }\bibfield  {title} {\bibinfo {title} {Pseudospin approach
  for planar ferromagnets},\ }\href
  {https://doi.org/https://doi.org/10.1016/0550-3213(84)90268-2} {\bibfield
  {journal} {\bibinfo  {journal} {Nuclear Physics B}\ }\textbf {\bibinfo
  {volume} {240}},\ \bibinfo {pages} {281} (\bibinfo {year}
  {1984})}\BibitemShut {NoStop}%
\bibitem [{\citenamefont
  {Papanicolaou}(1988{\natexlab{a}})}]{PAPANICOLAOU1988367}%
  \BibitemOpen
  \bibfield  {author} {\bibinfo {author} {\bibfnamefont {N.}~\bibnamefont
  {Papanicolaou}},\ }\bibfield  {title} {\bibinfo {title} {Unusual phases in
  quantum spin-1 systems},\ }\href
  {https://doi.org/https://doi.org/10.1016/0550-3213(88)90073-9} {\bibfield
  {journal} {\bibinfo  {journal} {Nuclear Physics B}\ }\textbf {\bibinfo
  {volume} {305}},\ \bibinfo {pages} {367} (\bibinfo {year}
  {1988}{\natexlab{a}})}\BibitemShut {NoStop}%
\bibitem [{\citenamefont {Joshi}\ \emph {et~al.}(1999)\citenamefont {Joshi},
  \citenamefont {Ma}, \citenamefont {Mila}, \citenamefont {Shi},\ and\
  \citenamefont {Zhang}}]{Joshi1999}%
  \BibitemOpen
  \bibfield  {author} {\bibinfo {author} {\bibfnamefont {A.}~\bibnamefont
  {Joshi}}, \bibinfo {author} {\bibfnamefont {M.}~\bibnamefont {Ma}}, \bibinfo
  {author} {\bibfnamefont {F.}~\bibnamefont {Mila}}, \bibinfo {author}
  {\bibfnamefont {D.~N.}\ \bibnamefont {Shi}},\ and\ \bibinfo {author}
  {\bibfnamefont {F.~C.}\ \bibnamefont {Zhang}},\ }\bibfield  {title} {\bibinfo
  {title} {Elementary excitations in magnetically ordered systems with orbital
  degeneracy},\ }\href {https://doi.org/10.1103/PhysRevB.60.6584} {\bibfield
  {journal} {\bibinfo  {journal} {Phys. Rev. B}\ }\textbf {\bibinfo {volume}
  {60}},\ \bibinfo {pages} {6584} (\bibinfo {year} {1999})}\BibitemShut
  {NoStop}%
\bibitem [{\citenamefont {{Chubukov}}(1990)}]{Chubukov1990}%
  \BibitemOpen
  \bibfield  {author} {\bibinfo {author} {\bibfnamefont {A.~V.}\ \bibnamefont
  {{Chubukov}}},\ }\bibfield  {title} {\bibinfo {title} {{Fluctuations in spin
  nematics}},\ }\href {https://doi.org/10.1088/0953-8984/2/6/018} {\bibfield
  {journal} {\bibinfo  {journal} {Journal of Physics Condensed Matter}\
  }\textbf {\bibinfo {volume} {2}},\ \bibinfo {pages} {1593} (\bibinfo {year}
  {1990})}\BibitemShut {NoStop}%
\bibitem [{\citenamefont {Dong}\ \emph {et~al.}(2018)\citenamefont {Dong},
  \citenamefont {Wang},\ and\ \citenamefont {Li}}]{Dong2018}%
  \BibitemOpen
  \bibfield  {author} {\bibinfo {author} {\bibfnamefont {Z.-Y.}\ \bibnamefont
  {Dong}}, \bibinfo {author} {\bibfnamefont {W.}~\bibnamefont {Wang}},\ and\
  \bibinfo {author} {\bibfnamefont {J.-X.}\ \bibnamefont {Li}},\ }\bibfield
  {title} {\bibinfo {title} {{$\mathrm{SU}(\mathrm{N})$} spin-wave theory:
  Application to spin-orbital mott insulators},\ }\href
  {https://doi.org/10.1103/PhysRevB.97.205106} {\bibfield  {journal} {\bibinfo
  {journal} {Phys. Rev. B}\ }\textbf {\bibinfo {volume} {97}},\ \bibinfo
  {pages} {205106} (\bibinfo {year} {2018})}\BibitemShut {NoStop}%
\bibitem [{\citenamefont {Yuan}\ \emph
  {et~al.}(2020{\natexlab{b}})\citenamefont {Yuan}, \citenamefont {Stone},
  \citenamefont {Shu}, \citenamefont {Chou}, \citenamefont {Rao}, \citenamefont
  {Clancy},\ and\ \citenamefont {Kim}}]{Yuan_exciton}%
  \BibitemOpen
  \bibfield  {author} {\bibinfo {author} {\bibfnamefont {B.}~\bibnamefont
  {Yuan}}, \bibinfo {author} {\bibfnamefont {M.~B.}\ \bibnamefont {Stone}},
  \bibinfo {author} {\bibfnamefont {G.-J.}\ \bibnamefont {Shu}}, \bibinfo
  {author} {\bibfnamefont {F.~C.}\ \bibnamefont {Chou}}, \bibinfo {author}
  {\bibfnamefont {X.}~\bibnamefont {Rao}}, \bibinfo {author} {\bibfnamefont
  {J.~P.}\ \bibnamefont {Clancy}},\ and\ \bibinfo {author} {\bibfnamefont
  {Y.-J.}\ \bibnamefont {Kim}},\ }\bibfield  {title} {\bibinfo {title}
  {Spin-orbit exciton in a honeycomb lattice magnet {${\mathrm{CoTiO}}_{3}$}:
  Revealing a link between magnetism in $d$- and $f$-electron systems},\ }\href
  {https://doi.org/10.1103/PhysRevB.102.134404} {\bibfield  {journal} {\bibinfo
   {journal} {Phys. Rev. B}\ }\textbf {\bibinfo {volume} {102}},\ \bibinfo
  {pages} {134404} (\bibinfo {year} {2020}{\natexlab{b}})}\BibitemShut
  {NoStop}%
\bibitem [{\citenamefont {Elliot}\ \emph {et~al.}(2021)\citenamefont {Elliot}
  \emph {et~al.}}]{data_archive}%
  \BibitemOpen
  \bibfield  {author} {\bibinfo {author} {\bibfnamefont {M.}~\bibnamefont
  {Elliot}} \emph {et~al.},\ }\href@noop {} {}\bibinfo {howpublished} {ORA data
  deposit} (\bibinfo {year} {2021}),\ \bibinfo {note}
  {https://doi.org/10.5287/bodleian:OR1BRxw0R}\BibitemShut {NoStop}%
\bibitem [{\citenamefont {Coldea}\ \emph {et~al.}(2019)\citenamefont {Coldea}
  \emph {et~al.}}]{exp_doi}%
  \BibitemOpen
  \bibfield  {author} {\bibinfo {author} {\bibfnamefont {R.}~\bibnamefont
  {Coldea}} \emph {et~al.},\ }\href {https://doi.org/10.5286/ISIS.E.RB1820500}
  {}\bibinfo {howpublished} {ISIS Pulsed Neutron and Muon Source} (\bibinfo
  {year} {2019}),\ \bibinfo {note} {doi:
  \href{https://doi.org/10.5286/ISIS.E.RB1820500}{10.5286/ISIS.E.RB1820500}}\BibitemShut
  {NoStop}%
\bibitem [{\citenamefont {Chapon}\ \emph {et~al.}(2011)\citenamefont {Chapon},
  \citenamefont {Manuel}, \citenamefont {Radaelli}, \citenamefont {Benson},
  \citenamefont {Perrott}, \citenamefont {Ansell}, \citenamefont {Rhodes},
  \citenamefont {Raspino}, \citenamefont {Duxbury}, \citenamefont {Spill},\
  and\ \citenamefont {Norris}}]{Chapon11}%
  \BibitemOpen
  \bibfield  {author} {\bibinfo {author} {\bibfnamefont {L.~C.}\ \bibnamefont
  {Chapon}}, \bibinfo {author} {\bibfnamefont {P.}~\bibnamefont {Manuel}},
  \bibinfo {author} {\bibfnamefont {P.~G.}\ \bibnamefont {Radaelli}}, \bibinfo
  {author} {\bibfnamefont {C.}~\bibnamefont {Benson}}, \bibinfo {author}
  {\bibfnamefont {L.}~\bibnamefont {Perrott}}, \bibinfo {author} {\bibfnamefont
  {S.}~\bibnamefont {Ansell}}, \bibinfo {author} {\bibfnamefont {N.~J.}\
  \bibnamefont {Rhodes}}, \bibinfo {author} {\bibfnamefont {D.}~\bibnamefont
  {Raspino}}, \bibinfo {author} {\bibfnamefont {D.}~\bibnamefont {Duxbury}},
  \bibinfo {author} {\bibfnamefont {E.}~\bibnamefont {Spill}},\ and\ \bibinfo
  {author} {\bibfnamefont {J.}~\bibnamefont {Norris}},\ }\bibfield  {title}
  {\bibinfo {title} {Wish: The new powder and single crystal magnetic
  diffractometer on the second target station},\ }\href@noop {} {\bibfield
  {journal} {\bibinfo  {journal} {Neutron News}\ }\textbf {\bibinfo {volume}
  {22}},\ \bibinfo {pages} {22} (\bibinfo {year} {2011})}\BibitemShut {NoStop}%
\bibitem [{\citenamefont {Rodr{\'\i}guez-Carvajal}(1993)}]{rodriguezcarvaja93}%
  \BibitemOpen
  \bibfield  {author} {\bibinfo {author} {\bibfnamefont {J.}~\bibnamefont
  {Rodr{\'\i}guez-Carvajal}},\ }\bibfield  {title} {\bibinfo {title} {Recent
  advances in magnetic structure determination by neutron powder diffraction},\
  }\href@noop {} {\bibfield  {journal} {\bibinfo  {journal} {Physica B}\
  }\textbf {\bibinfo {volume} {192}},\ \bibinfo {pages} {55} (\bibinfo {year}
  {1993})}\BibitemShut {NoStop}%
\bibitem [{\citenamefont {Newnham}\ \emph {et~al.}(1964)\citenamefont
  {Newnham}, \citenamefont {Fang},\ and\ \citenamefont {Santoro}}]{Newnham64}%
  \BibitemOpen
  \bibfield  {author} {\bibinfo {author} {\bibfnamefont {R.~E.}\ \bibnamefont
  {Newnham}}, \bibinfo {author} {\bibfnamefont {J.~H.}\ \bibnamefont {Fang}},\
  and\ \bibinfo {author} {\bibfnamefont {R.~P.}\ \bibnamefont {Santoro}},\
  }\bibfield  {title} {\bibinfo {title} {Crystal structure and magnetic
  properties of {${\mathrm{CoTiO}}_{3}$}},\ }\href@noop {} {\bibfield
  {journal} {\bibinfo  {journal} {Acta. Cryst.}\ }\textbf {\bibinfo {volume}
  {17}},\ \bibinfo {pages} {240} (\bibinfo {year} {1964})}\BibitemShut
  {NoStop}%
\bibitem [{\citenamefont {Campbell}\ \emph {et~al.}(2006)\citenamefont
  {Campbell}, \citenamefont {Stokes}, \citenamefont {Tanner},\ and\
  \citenamefont {Hatch}}]{Campbell06}%
  \BibitemOpen
  \bibfield  {author} {\bibinfo {author} {\bibfnamefont {B.~J.}\ \bibnamefont
  {Campbell}}, \bibinfo {author} {\bibfnamefont {H.~T.}\ \bibnamefont
  {Stokes}}, \bibinfo {author} {\bibfnamefont {D.~E.}\ \bibnamefont {Tanner}},\
  and\ \bibinfo {author} {\bibfnamefont {D.~M.}\ \bibnamefont {Hatch}},\
  }\bibfield  {title} {\bibinfo {title} {Isodisplace: a web-based tool for
  exploring structural distortions},\ }\href@noop {} {\bibfield  {journal}
  {\bibinfo  {journal} {J. Appl. Crystallogr.}\ }\textbf {\bibinfo {volume}
  {39}},\ \bibinfo {pages} {607} (\bibinfo {year} {2006})}\BibitemShut
  {NoStop}%
\bibitem [{\citenamefont {Stokes}\ \emph {et~al.}(2007)\citenamefont {Stokes},
  \citenamefont {Hatch},\ and\ \citenamefont {Campbell}}]{Stokes07}%
  \BibitemOpen
  \bibfield  {author} {\bibinfo {author} {\bibfnamefont {H.~T.}\ \bibnamefont
  {Stokes}}, \bibinfo {author} {\bibfnamefont {D.~M.}\ \bibnamefont {Hatch}},\
  and\ \bibinfo {author} {\bibfnamefont {B.~J.}\ \bibnamefont {Campbell}},\
  }\href {http://stokes.byu.edu/isotropy.html} {\bibinfo {title} {Isotropy}}
  (\bibinfo {year} {2007})\BibitemShut {NoStop}%
\bibitem [{\citenamefont {Abragam}\ and\ \citenamefont
  {Bleaney}(1970)}]{Abragam}%
  \BibitemOpen
  \bibfield  {author} {\bibinfo {author} {\bibfnamefont {A.}~\bibnamefont
  {Abragam}}\ and\ \bibinfo {author} {\bibfnamefont {B.}~\bibnamefont
  {Bleaney}},\ }\href@noop {} {\emph {\bibinfo {title} {Electron Paramagnetic
  Resonance of Transition Ions}}}\ (\bibinfo  {publisher} {Oxford University
  Press},\ \bibinfo {year} {1970})\BibitemShut {NoStop}%
\bibitem [{\citenamefont {Castro~Neto}\ \emph {et~al.}(2009)\citenamefont
  {Castro~Neto}, \citenamefont {Guinea}, \citenamefont {Peres}, \citenamefont
  {Novoselov},\ and\ \citenamefont {Geim}}]{graphene1}%
  \BibitemOpen
  \bibfield  {author} {\bibinfo {author} {\bibfnamefont {A.~H.}\ \bibnamefont
  {Castro~Neto}}, \bibinfo {author} {\bibfnamefont {F.}~\bibnamefont {Guinea}},
  \bibinfo {author} {\bibfnamefont {N.~M.~R.}\ \bibnamefont {Peres}}, \bibinfo
  {author} {\bibfnamefont {K.~S.}\ \bibnamefont {Novoselov}},\ and\ \bibinfo
  {author} {\bibfnamefont {A.~K.}\ \bibnamefont {Geim}},\ }\bibfield  {title}
  {\bibinfo {title} {The electronic properties of graphene},\ }\href
  {https://doi.org/10.1103/RevModPhys.81.109} {\bibfield  {journal} {\bibinfo
  {journal} {Rev. Mod. Phys.}\ }\textbf {\bibinfo {volume} {81}},\ \bibinfo
  {pages} {109} (\bibinfo {year} {2009})}\BibitemShut {NoStop}%
\bibitem [{\citenamefont {Mucha-Kruczy\ifmmode~\acute{n}\else \'{n}\fi{}ski}\
  \emph {et~al.}(2008)\citenamefont {Mucha-Kruczy\ifmmode~\acute{n}\else
  \'{n}\fi{}ski}, \citenamefont {Tsyplyatyev}, \citenamefont {Grishin},
  \citenamefont {McCann}, \citenamefont {Fal'ko}, \citenamefont {Bostwick},\
  and\ \citenamefont {Rotenberg}}]{graphene2}%
  \BibitemOpen
  \bibfield  {author} {\bibinfo {author} {\bibfnamefont {M.}~\bibnamefont
  {Mucha-Kruczy\ifmmode~\acute{n}\else \'{n}\fi{}ski}}, \bibinfo {author}
  {\bibfnamefont {O.}~\bibnamefont {Tsyplyatyev}}, \bibinfo {author}
  {\bibfnamefont {A.}~\bibnamefont {Grishin}}, \bibinfo {author} {\bibfnamefont
  {E.}~\bibnamefont {McCann}}, \bibinfo {author} {\bibfnamefont {V.~I.}\
  \bibnamefont {Fal'ko}}, \bibinfo {author} {\bibfnamefont {A.}~\bibnamefont
  {Bostwick}},\ and\ \bibinfo {author} {\bibfnamefont {E.}~\bibnamefont
  {Rotenberg}},\ }\bibfield  {title} {\bibinfo {title} {Characterization of
  graphene through anisotropy of constant-energy maps in angle-resolved
  photoemission},\ }\href {https://doi.org/10.1103/PhysRevB.77.195403}
  {\bibfield  {journal} {\bibinfo  {journal} {Phys. Rev. B}\ }\textbf {\bibinfo
  {volume} {77}},\ \bibinfo {pages} {195403} (\bibinfo {year}
  {2008})}\BibitemShut {NoStop}%
\bibitem [{\citenamefont {Bradley}\ and\ \citenamefont
  {Cracknell}(1972)}]{Bradley}%
  \BibitemOpen
  \bibfield  {author} {\bibinfo {author} {\bibfnamefont {C.}~\bibnamefont
  {Bradley}}\ and\ \bibinfo {author} {\bibfnamefont {A.}~\bibnamefont
  {Cracknell}},\ }\href@noop {} {\emph {\bibinfo {title} {The Mathematical
  Theory of Symmetry in Solids}}}\ (\bibinfo  {publisher} {Clarendon Press
  Oxford},\ \bibinfo {year} {1972})\BibitemShut {NoStop}%
\bibitem [{\citenamefont {Toth}\ and\ \citenamefont {Lake}(2015)}]{SpinW}%
  \BibitemOpen
  \bibfield  {author} {\bibinfo {author} {\bibfnamefont {S.}~\bibnamefont
  {Toth}}\ and\ \bibinfo {author} {\bibfnamefont {B.}~\bibnamefont {Lake}},\
  }\bibfield  {title} {\bibinfo {title} {{Linear spin wave theory for single-Q
  incommensurate magnetic structures}},\ }\href
  {https://doi.org/10.1088/0953-8984/27/16/166002} {\bibfield  {journal}
  {\bibinfo  {journal} {J. Phys. Condens. Matter}\ }\textbf {\bibinfo {volume}
  {27}},\ \bibinfo {pages} {166002} (\bibinfo {year} {2015})}\BibitemShut
  {NoStop}%
\bibitem [{\citenamefont {Bewley}\ \emph {et~al.}(2006)\citenamefont {Bewley},
  \citenamefont {Eccleston}, \citenamefont {McEwen}, \citenamefont {Hayden},
  \citenamefont {Dove}, \citenamefont {Bennington}, \citenamefont {Treadgold},\
  and\ \citenamefont {Coleman}}]{MERLIN}%
  \BibitemOpen
  \bibfield  {author} {\bibinfo {author} {\bibfnamefont {R.}~\bibnamefont
  {Bewley}}, \bibinfo {author} {\bibfnamefont {R.}~\bibnamefont {Eccleston}},
  \bibinfo {author} {\bibfnamefont {K.}~\bibnamefont {McEwen}}, \bibinfo
  {author} {\bibfnamefont {S.}~\bibnamefont {Hayden}}, \bibinfo {author}
  {\bibfnamefont {M.}~\bibnamefont {Dove}}, \bibinfo {author} {\bibfnamefont
  {S.}~\bibnamefont {Bennington}}, \bibinfo {author} {\bibfnamefont
  {J.}~\bibnamefont {Treadgold}},\ and\ \bibinfo {author} {\bibfnamefont
  {R.}~\bibnamefont {Coleman}},\ }\bibfield  {title} {\bibinfo {title}
  {{MERLIN, a new high count rate spectrometer at ISIS}},\ }\href
  {https://doi.org/10.1016/J.PHYSB.2006.05.328} {\bibfield  {journal} {\bibinfo
   {journal} {Phys. B Condens. Matter}\ }\textbf {\bibinfo {volume}
  {385-386}},\ \bibinfo {pages} {1029} (\bibinfo {year} {2006})}\BibitemShut
  {NoStop}%
\bibitem [{\citenamefont {Arnold}\ \emph {et~al.}(2014)\citenamefont {Arnold},
  \citenamefont {Bilheux}, \citenamefont {Borreguero}, \citenamefont {Buts},
  \citenamefont {Campbell}, \citenamefont {Chapon}, \citenamefont {Doucet},
  \citenamefont {Draper}, \citenamefont {{Ferraz Leal}}, \citenamefont {Gigg},
  \citenamefont {Lynch}, \citenamefont {Markvardsen}, \citenamefont
  {Mikkelson}, \citenamefont {Mikkelson}, \citenamefont {Miller}, \citenamefont
  {Palmen}, \citenamefont {Parker}, \citenamefont {Passos}, \citenamefont
  {Perring}, \citenamefont {Peterson}, \citenamefont {Ren}, \citenamefont
  {Reuter}, \citenamefont {Savici}, \citenamefont {Taylor}, \citenamefont
  {Taylor}, \citenamefont {Tolchenov}, \citenamefont {Zhou},\ and\
  \citenamefont {Zikovsky}}]{Arnold2014}%
  \BibitemOpen
  \bibfield  {author} {\bibinfo {author} {\bibfnamefont {O.}~\bibnamefont
  {Arnold}}, \bibinfo {author} {\bibfnamefont {J.~C.}\ \bibnamefont {Bilheux}},
  \bibinfo {author} {\bibfnamefont {J.~M.}\ \bibnamefont {Borreguero}},
  \bibinfo {author} {\bibfnamefont {A.}~\bibnamefont {Buts}}, \bibinfo {author}
  {\bibfnamefont {S.~I.}\ \bibnamefont {Campbell}}, \bibinfo {author}
  {\bibfnamefont {L.}~\bibnamefont {Chapon}}, \bibinfo {author} {\bibfnamefont
  {M.}~\bibnamefont {Doucet}}, \bibinfo {author} {\bibfnamefont
  {N.}~\bibnamefont {Draper}}, \bibinfo {author} {\bibfnamefont
  {R.}~\bibnamefont {{Ferraz Leal}}}, \bibinfo {author} {\bibfnamefont {M.~A.}\
  \bibnamefont {Gigg}}, \bibinfo {author} {\bibfnamefont {V.~E.}\ \bibnamefont
  {Lynch}}, \bibinfo {author} {\bibfnamefont {A.}~\bibnamefont {Markvardsen}},
  \bibinfo {author} {\bibfnamefont {D.~J.}\ \bibnamefont {Mikkelson}}, \bibinfo
  {author} {\bibfnamefont {R.~L.}\ \bibnamefont {Mikkelson}}, \bibinfo {author}
  {\bibfnamefont {R.}~\bibnamefont {Miller}}, \bibinfo {author} {\bibfnamefont
  {K.}~\bibnamefont {Palmen}}, \bibinfo {author} {\bibfnamefont
  {P.}~\bibnamefont {Parker}}, \bibinfo {author} {\bibfnamefont
  {G.}~\bibnamefont {Passos}}, \bibinfo {author} {\bibfnamefont {T.~G.}\
  \bibnamefont {Perring}}, \bibinfo {author} {\bibfnamefont {P.~F.}\
  \bibnamefont {Peterson}}, \bibinfo {author} {\bibfnamefont {S.}~\bibnamefont
  {Ren}}, \bibinfo {author} {\bibfnamefont {M.~A.}\ \bibnamefont {Reuter}},
  \bibinfo {author} {\bibfnamefont {A.~T.}\ \bibnamefont {Savici}}, \bibinfo
  {author} {\bibfnamefont {J.~W.}\ \bibnamefont {Taylor}}, \bibinfo {author}
  {\bibfnamefont {R.~J.}\ \bibnamefont {Taylor}}, \bibinfo {author}
  {\bibfnamefont {R.}~\bibnamefont {Tolchenov}}, \bibinfo {author}
  {\bibfnamefont {W.}~\bibnamefont {Zhou}},\ and\ \bibinfo {author}
  {\bibfnamefont {J.}~\bibnamefont {Zikovsky}},\ }\bibfield  {title} {\bibinfo
  {title} {{Mantid---Data analysis and visualization package for neutron
  scattering and $\mu$SR experiments}},\ }\href
  {https://doi.org/10.1016/j.nima.2014.07.029} {\bibfield  {journal} {\bibinfo
  {journal} {Nucl. Instruments Methods Phys. Res. Sect. A Accel. Spectrometers,
  Detect. Assoc. Equip.}\ }\textbf {\bibinfo {volume} {764}},\ \bibinfo {pages}
  {156} (\bibinfo {year} {2014})}\BibitemShut {NoStop}%
\bibitem [{\citenamefont {Ewings}\ \emph {et~al.}(2016)\citenamefont {Ewings},
  \citenamefont {Buts}, \citenamefont {Le}, \citenamefont {van Duijn},
  \citenamefont {Bustinduy},\ and\ \citenamefont {Perring}}]{Ewings2016}%
  \BibitemOpen
  \bibfield  {author} {\bibinfo {author} {\bibfnamefont {R.}~\bibnamefont
  {Ewings}}, \bibinfo {author} {\bibfnamefont {A.}~\bibnamefont {Buts}},
  \bibinfo {author} {\bibfnamefont {M.}~\bibnamefont {Le}}, \bibinfo {author}
  {\bibfnamefont {J.}~\bibnamefont {van Duijn}}, \bibinfo {author}
  {\bibfnamefont {I.}~\bibnamefont {Bustinduy}},\ and\ \bibinfo {author}
  {\bibfnamefont {T.}~\bibnamefont {Perring}},\ }\bibfield  {title} {\bibinfo
  {title} {{Horace: Software for the analysis of data from single crystal
  spectroscopy experiments at time-of-flight neutron instruments}},\ }\href
  {https://doi.org/10.1016/J.NIMA.2016.07.036} {\bibfield  {journal} {\bibinfo
  {journal} {Nucl. Instruments Methods Phys. Res. Sect. A}\ }\textbf {\bibinfo
  {volume} {834}},\ \bibinfo {pages} {132} (\bibinfo {year}
  {2016})}\BibitemShut {NoStop}%
\bibitem [{\citenamefont {Rau}\ \emph {et~al.}(2018)\citenamefont {Rau},
  \citenamefont {McClarty},\ and\ \citenamefont
  {Moessner}}]{PhysRevLett.121.237201}%
  \BibitemOpen
  \bibfield  {author} {\bibinfo {author} {\bibfnamefont {J.~G.}\ \bibnamefont
  {Rau}}, \bibinfo {author} {\bibfnamefont {P.~A.}\ \bibnamefont {McClarty}},\
  and\ \bibinfo {author} {\bibfnamefont {R.}~\bibnamefont {Moessner}},\
  }\bibfield  {title} {\bibinfo {title} {Pseudo-goldstone gaps and
  order-by-quantum disorder in frustrated magnets},\ }\href
  {https://doi.org/10.1103/PhysRevLett.121.237201} {\bibfield  {journal}
  {\bibinfo  {journal} {Phys. Rev. Lett.}\ }\textbf {\bibinfo {volume} {121}},\
  \bibinfo {pages} {237201} (\bibinfo {year} {2018})}\BibitemShut {NoStop}%
\bibitem [{\citenamefont {Bauer}\ and\ \citenamefont {Rotter}()}]{Bauer2010}%
  \BibitemOpen
  \bibfield  {author} {\bibinfo {author} {\bibfnamefont {E.}~\bibnamefont
  {Bauer}}\ and\ \bibinfo {author} {\bibfnamefont {M.}~\bibnamefont {Rotter}},\
  }\bibinfo {title} {Magnetism of complex metallic alloys: Crystalline electric
  field effects},\ in\ \href {https://doi.org/10.1142/9789814261647_0005}
  {\emph {\bibinfo {booktitle} {Properties and Applications of Complex
  Intermetallics}}},\ pp.\ \bibinfo {pages} {183--248}\BibitemShut {NoStop}%
\bibitem [{\citenamefont {Pershoguba}\ \emph {et~al.}(2018)\citenamefont
  {Pershoguba}, \citenamefont {Banerjee}, \citenamefont {Lashley},
  \citenamefont {Park}, \citenamefont {\AA{}gren}, \citenamefont {Aeppli},\
  and\ \citenamefont {Balatsky}}]{pershoguba2018}%
  \BibitemOpen
  \bibfield  {author} {\bibinfo {author} {\bibfnamefont {S.~S.}\ \bibnamefont
  {Pershoguba}}, \bibinfo {author} {\bibfnamefont {S.}~\bibnamefont
  {Banerjee}}, \bibinfo {author} {\bibfnamefont {J.~C.}\ \bibnamefont
  {Lashley}}, \bibinfo {author} {\bibfnamefont {J.}~\bibnamefont {Park}},
  \bibinfo {author} {\bibfnamefont {H.}~\bibnamefont {\AA{}gren}}, \bibinfo
  {author} {\bibfnamefont {G.}~\bibnamefont {Aeppli}},\ and\ \bibinfo {author}
  {\bibfnamefont {A.~V.}\ \bibnamefont {Balatsky}},\ }\bibfield  {title}
  {\bibinfo {title} {Dirac magnons in honeycomb ferromagnets},\ }\href
  {https://doi.org/10.1103/PhysRevX.8.011010} {\bibfield  {journal} {\bibinfo
  {journal} {Phys. Rev. X}\ }\textbf {\bibinfo {volume} {8}},\ \bibinfo {pages}
  {011010} (\bibinfo {year} {2018})}\BibitemShut {NoStop}%
\bibitem [{\citenamefont
  {Papanicolaou}(1988{\natexlab{b}})}]{papanicolaou1988unusual}%
  \BibitemOpen
  \bibfield  {author} {\bibinfo {author} {\bibfnamefont {N.}~\bibnamefont
  {Papanicolaou}},\ }\bibfield  {title} {\bibinfo {title} {Unusual phases in
  quantum spin-1 systems},\ }\href@noop {} {\bibfield  {journal} {\bibinfo
  {journal} {Nuclear Physics B}\ }\textbf {\bibinfo {volume} {305}},\ \bibinfo
  {pages} {367} (\bibinfo {year} {1988}{\natexlab{b}})}\BibitemShut {NoStop}%
\bibitem [{\citenamefont {Romh\'anyi}\ and\ \citenamefont
  {Penc}(2012)}]{PhysRevB.86.174428}%
  \BibitemOpen
  \bibfield  {author} {\bibinfo {author} {\bibfnamefont {J.}~\bibnamefont
  {Romh\'anyi}}\ and\ \bibinfo {author} {\bibfnamefont {K.}~\bibnamefont
  {Penc}},\ }\bibfield  {title} {\bibinfo {title} {Multiboson spin-wave theory
  for {Ba}${}_{2}${Co}{Ge}${}_{2}${O}${}_{7}$: A spin-3/2 easy-plane {N\'eel}
  antiferromagnet with strong single-ion anisotropy},\ }\href
  {https://doi.org/10.1103/PhysRevB.86.174428} {\bibfield  {journal} {\bibinfo
  {journal} {Phys. Rev. B}\ }\textbf {\bibinfo {volume} {86}},\ \bibinfo
  {pages} {174428} (\bibinfo {year} {2012})}\BibitemShut {NoStop}%
\bibitem [{\citenamefont {Coldea}\ \emph {et~al.}(2001)\citenamefont {Coldea},
  \citenamefont {Hayden}, \citenamefont {Aeppli}, \citenamefont {Perring},
  \citenamefont {Frost}, \citenamefont {Mason}, \citenamefont {Cheong},\ and\
  \citenamefont {Fisk}}]{cyclicexchange}%
  \BibitemOpen
  \bibfield  {author} {\bibinfo {author} {\bibfnamefont {R.}~\bibnamefont
  {Coldea}}, \bibinfo {author} {\bibfnamefont {S.~M.}\ \bibnamefont {Hayden}},
  \bibinfo {author} {\bibfnamefont {G.}~\bibnamefont {Aeppli}}, \bibinfo
  {author} {\bibfnamefont {T.~G.}\ \bibnamefont {Perring}}, \bibinfo {author}
  {\bibfnamefont {C.~D.}\ \bibnamefont {Frost}}, \bibinfo {author}
  {\bibfnamefont {T.~E.}\ \bibnamefont {Mason}}, \bibinfo {author}
  {\bibfnamefont {S.-W.}\ \bibnamefont {Cheong}},\ and\ \bibinfo {author}
  {\bibfnamefont {Z.}~\bibnamefont {Fisk}},\ }\bibfield  {title} {\bibinfo
  {title} {Spin waves and electronic interactions in
  {${\mathrm{La}}_{2}{\mathrm{CuO}}_{4}$}},\ }\href
  {https://doi.org/10.1103/PhysRevLett.86.5377} {\bibfield  {journal} {\bibinfo
   {journal} {Phys. Rev. Lett.}\ }\textbf {\bibinfo {volume} {86}},\ \bibinfo
  {pages} {5377} (\bibinfo {year} {2001})}\BibitemShut {NoStop}%
\bibitem [{\citenamefont {MacDonald}\ \emph {et~al.}(1990)\citenamefont
  {MacDonald}, \citenamefont {Girvin},\ and\ \citenamefont
  {Yoshioka}}]{macdonald}%
  \BibitemOpen
  \bibfield  {author} {\bibinfo {author} {\bibfnamefont {A.~H.}\ \bibnamefont
  {MacDonald}}, \bibinfo {author} {\bibfnamefont {S.~M.}\ \bibnamefont
  {Girvin}},\ and\ \bibinfo {author} {\bibfnamefont {D.}~\bibnamefont
  {Yoshioka}},\ }\bibfield  {title} {\bibinfo {title} {Reply to ``comment on
  `t/u expansion for the hubbard model'''},\ }\href
  {https://doi.org/10.1103/PhysRevB.41.2565} {\bibfield  {journal} {\bibinfo
  {journal} {Phys. Rev. B}\ }\textbf {\bibinfo {volume} {41}},\ \bibinfo
  {pages} {2565} (\bibinfo {year} {1990})}\BibitemShut {NoStop}%
\end{thebibliography}%

\section*{Acknowledgments}
PAM acknowledges a useful discussion with K.~Penc. RC acknowledges a useful comment from G.~Khaliullin. This research was partially supported by the European Research Council under the European Union's Horizon 2020 research and innovation programme Grant Agreement Number 788814 (EQFT). ME acknowledges support from a doctoral studentship funded by Lincoln College and the University of Oxford. RDJ acknowledges support from a Royal Society University Research Fellowship. RC acknowledges support from the National Science Foundation under Grant No. NSF PHY-1748958 and hospitality from KITP where part of this work was completed. The neutron scattering measurements at the ISIS Facility were supported by a beamtime allocation \cite{exp_doi} from the Science and Technology Facilities Council.

\section*{Author Contributions}
RC and PAM conceived research, DP synthesized the crystal and powder samples, ME prepared the multi-crystal mount, ME, RC, PAM and HCW performed the INS experiments, ME, RC and PAM performed the analysis, PAM developed relevant theoretical models, PAM, ME and RC performed theoretical calculations, PM and RDJ performed powder neutron diffraction measurements and RDJ analysed this data, PAM, RC, ME and RDJ wrote the paper and the supplementary information with input from all co-authors.
\newline
\section*{Competing Interests}
The authors declare no competing interests.
\newline

\vspace{0.5 cm}
\begin{center}
\large{\bf Supplementary Information}
\end{center}

Here we provide additional technical details on 1) the refinement of the crystal and magnetic structures from powder neutron diffraction, 2-3) calculation of the single-ion levels in the presence of spin-orbit coupling, trigonal crystal field, and exchange mean field to determine the spin and orbital contributions to the ordered moment in the ground state, 4) a tight binding model to capture the exciton dispersion, 5) the spin-wave calculations for the minimal effective $S=1/2$ XXZ model used to parametrize the magnon dispersions, 6) details of the INS experiments and quantitative fit of the observed dispersions, 7) symmetry-allowed bond-dependent anisotropic exchanges, 8) quantum order-by-disorder from those terms as the origin of the spectral gap, 9) topology of the nodal lines of Dirac magnons and Hamiltonian symmetries, 10) flavor-wave theory based on a model with spin-orbital exchange that captures the discrete symmetry-breaking, the magnons and their spectral gap, as well as the dispersive excitons in a single model.

\renewcommand{\thesection}{{Supplementary Note} \arabic{section}}
\renewcommand{\theequation}{\arabic{equation}} 
\renewcommand{\thetable}{\Roman{table}}
\renewcommand{\tablename}{Supplementary Table}
\renewcommand{\thefigure}{\arabic{figure}}
\renewcommand{\figurename}{Supplementary Figure}
\makeatother
\setcounter{figure}{0}
\setcounter{equation}{0}

\section{Refinement of crystal and magnetic structures}
\label{sec:refinement}
Here we present neutron powder diffraction (NPD) measurements to determine the magnitude of the ordered moment in the ground state, which is an important ingredient in the parametrization of spin and orbital character of the cobalt magnetic moments. The experiments were performed using the WISH time-of-flight diffractometer \cite{Chapon11} at ISIS, the UK Neutron and Muon Source. A high quality, single phase powder sample of \cto\ (mass 3.125~g) was loaded into a 6~mm diameter vanadium can and mounted within an Oxford Instruments $^4$He cryostat. High counting statistics data were collected at 1.5 and 150~K, representative of the magnetically ordered and paramagnetic phases, respectively (N.B. the paramagnetic data were collected well above $T_\mathrm{N}=38(3)$~K as magnetic diffuse scattering was found to persist above the transition). Additional lower counting statistics data were also collected on warming in 2.5~K steps between 1.5 and 50~K to obtain the order parameter. In the following analysis, Rietveld refinements of nuclear and magnetic structural models were performed using Fullprof \cite{rodriguezcarvaja93}, simultaneously against data measured in detector banks 2 and 9 (medium resolution, large $d$-spacing range) and banks 5 and 6 (high resolution, short $d$-spacing) of the WISH instrument. A small absorption correction was included in the refinements to account for moderate neutron absorption by cobalt.

\begin{figure*}[!htbp]
\includegraphics[width=\linewidth,keepaspectratio]{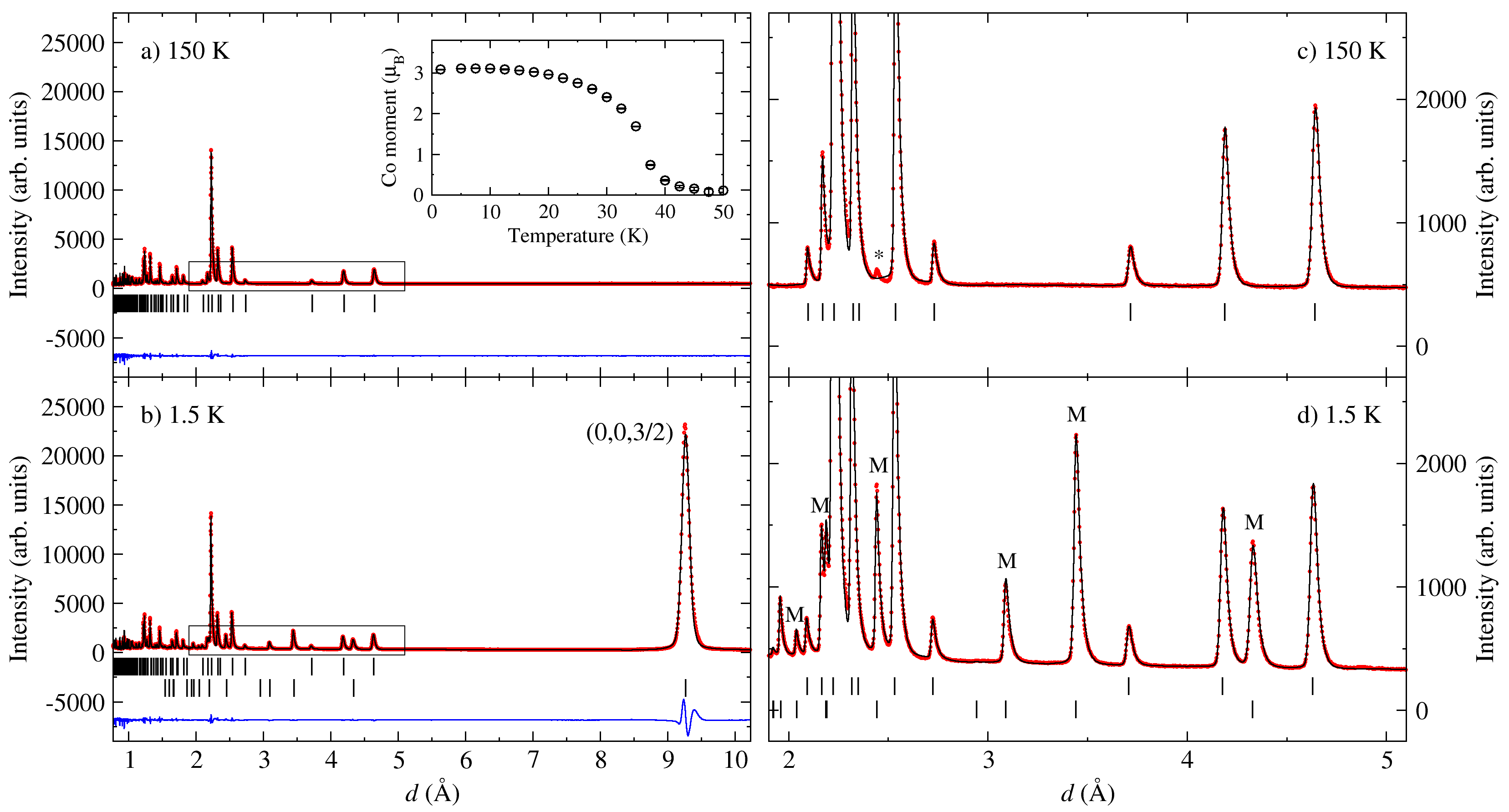}
\figcaption{Neutron powder diffraction data}{Data (red circles) was collected at (a),(c) 150~K and (b),(d) 1.5~K. Panels (c) and~(d) show an enlarged region of the diffraction pattern, as indicated by the black rectangle in (a) and (b), highlighting magnetic diffraction intensities labelled ``M''. The weak diffraction peak labelled by an asterisk (*) in (c) likely originates from a small CoTi$_2$O$_5$ impurity. Its presence did not affect the quantitative analysis of the diffraction pattern. Fits to the data are shown as solid black lines, and the difference $I_\mathrm{obs} - I_\mathrm{calc}$ is given as a blue line at the bottom of the panes. In (a) and (c) nuclear peak positions are denoted by black tick marks, and in (b) and (d) nuclear and magnetic peak positions are denoted by top and bottom black tick marks, respectively. The temperature dependence of the cobalt magnetic moment evaluated by fitting variable temperature neutron powder diffraction data is shown in the inset to panel a).\label{FIG::NPD}}
\end{figure*}

The published ilmenite crystal structure of \cto\ \cite{Newnham64} (space group $R\bar{3}$, herein defined using hexagonal axes in the obverse setting) was refined against the paramagnetic data (Supplementary Figures~\ref{FIG::NPD}a-c). Excellent agreement between model and data was achieved and the crystal structure parameters are summarised in Supplementary Table~\ref{TAB::crystal_structure}.

\begin{table}
\tabcaption{ Refined 150 K crystal structure parameters of \cto}{($R_\mathrm{p}=4.1\%$, $R_\mathrm{wp}=3.7\%$, $R_\mathrm{Bragg}=4.1\%$). Wyckoff positions are given in parenthesis.}\label{TAB::crystal_structure}
\begin{ruledtabular}
\begin{tabular}{c c c c c}
\multicolumn{5}{l}{\textbf{Cell parameters}} \\
\multicolumn{5}{l}{Space group: $R\bar{3}$ (\#148, hexagonal axes, obverse setting)} \\
$a,b,c$ ($\mathrm{\AA}$) & 5.06383(3) & 5.06383(3) & 13.9076(1) \\
\multicolumn{5}{l}{Volume ($\mathrm{\AA}^3$) 308.845(4)}\\
\\
\multicolumn{5}{l}{\textbf{Atomic fractional coordinates}} \\
Atom & $x$ & $y$ & $z$ & $U_\mathrm{iso}({\mathrm \AA}^{-2})$\\
\hline
Co ($6c$)  & 0         & 0         & 0.3562(3)  & 0.0132(9) \\
Ti ($6c$)  & 0         & 0         & 0.1454(2)  & 0.0066(6) \\
O  ($18f$) & 0.3161(2) & 0.0203(2) & 0.24605(5) & 0.0074(2) \\
\\
\end{tabular}
\end{ruledtabular}
\end{table}

Below $T_\mathrm{N}$, more than 10 new diffraction peaks appeared (labelled ``M'' in Supplementary Figure~\ref{FIG::NPD}d), which could be indexed using the T-point propagation vector $\mathbf{Q} = (0,0,3/2)$. Symmetry analysis performed using \textsc{isodistort} \cite{Campbell06,Stokes07}, showed that the full T-point magnetic representation for the cobalt Wyckoff positions decomposed into two 1D irreducible representations, T$_1^+$ and T$_1^-$, and two physically real, 2D reducible representations, $\mathrm{T}_2^+\oplus\mathrm{T}_3^+$ and $\mathrm{T}_2^-\oplus\mathrm{T}_3^-$. There exist four, symmetry distinct magnetic structures that transform by these four representations, respectively:
\begin{enumerate}[(a)]
\item Ferromagnetic (FM) honeycomb layers stacked via antiferromagnetic (AFM) bonds with magnetic moments parallel to the $c$-axis (magnetic space group $R_\mathrm{I}\bar{3}$),
\item AFM honeycomb layers (N\'{e}el-type) stacked via FM bonds with magnetic moments parallel to the $c$-axis (magnetic space group $R_\mathrm{I}\bar{3}$),
\item FM honeycomb layers stacked via AFM bonds with magnetic moments perpendicular to the $c$-axis (magnetic space group $P_\mathrm{S}\bar{1}$), and
\item AFM honeycomb layers (N\'{e}el-type) stacked via FM bonds with magnetic moments perpendicular to the $c$-axis (magnetic space group $P_\mathrm{S}\bar{1}$).
\end{enumerate}
We note that for structures (c) and (d) all in-plane moment directions are indistinguishable by symmetry. Furthermore, the $\mathrm{T}_2^+\oplus\mathrm{T}_3^+$($\mathrm{T}_2^-\oplus\mathrm{T}_3^-$) symmetry allows the T$_1^+$(T$_1^-$) mode to appear via a secondary order parameter, which describes a global rotation of all moments out of the $ab$ plane towards the hexagonal $c$ axis whilst maintaining a collinear magnetic structure.

The largest magnetic diffraction intensity occurs for the magnetic Bragg peak indexed by the propagation vector ${\mathbf{Q}}=(0,0,3/2)$. Given that the magnetic neutron diffraction intensity is proportional to the component of the magnetic moments perpendicular to the scattering vector, this observation alone conclusively rules out structures (a) and (b) that have moments strictly parallel to the $c$ axis. Furthermore, one can show that in the case of perfectly flat cobalt honeycomb planes ($z_{\rm Co}=1/3$ in Supplementary Table~\ref{TAB::crystal_structure}) the magnetic structure factor at $\mathbf{Q}$ is maximal for FM honeycomb planes and exactly zero for AFM honeycomb planes. The honeycomb planes of the true crystal structure are not perfectly flat, but the small buckling of these planes leads to only a few percent change in the predicted diffraction intensities. Hence, case (c) (illustrated in Supplementary Figure~\ref{fig:FNN}) is uniquely identified as the primary magnetic structure of \cto\ by the observation of the largest intensity at the propagation vector alone, in agreement with earlier neutron powder diffraction results \cite{Newnham64}.

A magnetic structure model based on (c) was refined against the neutron powder diffraction data collected at 1.5~K (Supplementary Figures~\ref{FIG::NPD}b and d). Excellent agreement between model and data was achieved ($R_\mathrm{p}=4.9\%$, $R_\mathrm{wp}=4.3\%$, $R_\mathrm{Mag}=3.1\%$). The in-plane direction of the magnetic moments cannot be determined from powder averaged diffraction data, and symmetry allowed out-of-plane tilting of the magnetic moments was found to be statistically insignificant. At 1.5~K the cobalt magnetic moment refined to 3.08(1)~$\mu_\mathrm{B}$. The temperature dependence of the magnetic moment was extracted from fits to data collected on warming and is shown in the inset to Supplementary Figure~\ref{FIG::NPD}a).

The above magnetic structure has lower symmetry ($P_\mathrm{S}\bar{1}$) than the paramagnetic crystal structure ($R\bar{3}$). In this case the crystal symmetry can be lowered via magnetostriction. However we found that a hexagonal unit cell metric could be used to achieve excellent fits to our data at all measured temperatures and no peak splitting or significant peak broadening could be observed within the experimental resolution upon cooling below $T_\mathrm{N}$. We therefore estimate that any symmetry lowering of the hexagonal metric by the magnetic ordering involves changes in the lattice parameters below a conservative threshold of 0.02\%.

\section{Single-Ion Physics}
\label{sec:cef}
Here we discuss the ground state and higher-energy excited states of the Co$^{2+}$ ($3d^7$) ions given their local, octahedrally-coordinated crystal field environment and spin-orbit interaction, fitted to inter-level transitions observed in INS data. Hund's rules - appropriate to the case where the Coulomb interaction is greater than the crystal field - give a bare $d^7$ shell orbital triplet $L=3$ and high spin $\fs{S}=3/2$. For an ideal octahedron, the crystal field acting on those levels has Hamiltonian ${\cal H}_{\rm CF}=-B\left( {\cal O}_4^0 + 5 {\cal O}_4^4 \right)$ with $B>0$, leading to a ground state triplet ($\Gamma_4$), and excited triplet and singlet levels, above energy gaps of $480~B$ and $1080~B$, respectively. Those level splittings are of the order of $1$~eV. Viewed another way,  the crystal field levels are populated in the high spin $t_{\rm 2g}^5e_{\rm g}^2$ configuration, which is the aforementioned $\fs{S}=3/2$ orbital triplet. Cobalt(II) ions in octahedral environments may also occur in a low spin configuration with spin-1/2 degree of freedom and a two-fold orbital degeneracy, however \cto\ is consistent with the high spin single-ion configuration because, as we show below, this offers a natural explanation for i) the observed transitions to higher single-ion levels and ii) the experimentally determined magnitude of the ordered moment in the ground state determined in~\ref{sec:refinement}.

Empirically, the exchange scale and spin-orbit coupling in \cto\ are both of order $10$~meV. Since the octahedral crystal field splitting is larger than any other relevant magnetic scales, we may focus on the ground state orbital triplet as an effective $l=1$ orbital angular momentum state with wavefunctions \cite{Abragam}
\begin{align*}
& \vert l_z= +1 \rangle =  \sqrt{\frac{5}{8}}\vert -3 \rangle +  \sqrt{\frac{3}{8}}\vert +1 \rangle \\
& \vert l_z= 0 \rangle =  - \vert 0 \rangle  \\
& \vert l_z= -1 \rangle =  \sqrt{\frac{5}{8}}\vert 3 \rangle +  \sqrt{\frac{3}{8}}\vert -1 \rangle
\end{align*}
in terms of the $\vert L_z\rangle$ states of the full $L=3$ Hilbert space. The full angular momentum operator when projected onto the restricted $l=1$ Hilbert space is expressed as $\mathbf{L} \equiv (-3/2) \mathbf{l}$.

The spin-orbit coupling ${\cal H}_{\rm SO}= (3/2)\lambda\mathbf{l}\cdot\bm{\mathsf{S}}$ with $\lambda>0$ acts on the $l=1$ and $\fs{S}=3/2$ states numbering $12$ in all. It is convenient to define an effective angular momentum $\mathbf{J}_{\rm eff}=\mathbf{l}+\bm{\mathsf{S}}$ as $J_{\rm eff}$ is a good quantum number for the eigenstates of ${\cal H}_{\rm SO}$. The spectrum is a spin-orbital ground state doublet with $J_{\rm eff}=1/2$ at energy $-15\lambda/4$, a quartet ($J_{\rm eff}=3/2$) at $-3\lambda/2$ and a 6-fold degenerate set ($J_{\rm eff}=5/2$) at $9\lambda/4$. The lowest doublet wavefunctions take the form
\begin{equation}
\frac{1}{\sqrt{2}}\vert \pm1 ,\mp 3/2 \rangle + \frac{1}{\sqrt{6}} \vert \mp 1,\pm 1/2\rangle - \frac{1}{\sqrt{3}} \vert 0,\mp 1/2\rangle
\end{equation}
in the $\vert l_z,\fs{S}_z \rangle$ basis.

\begin{figure}[htbp!]%
    \centering
    \subfloat[]{{\includegraphics[width=0.98\columnwidth]{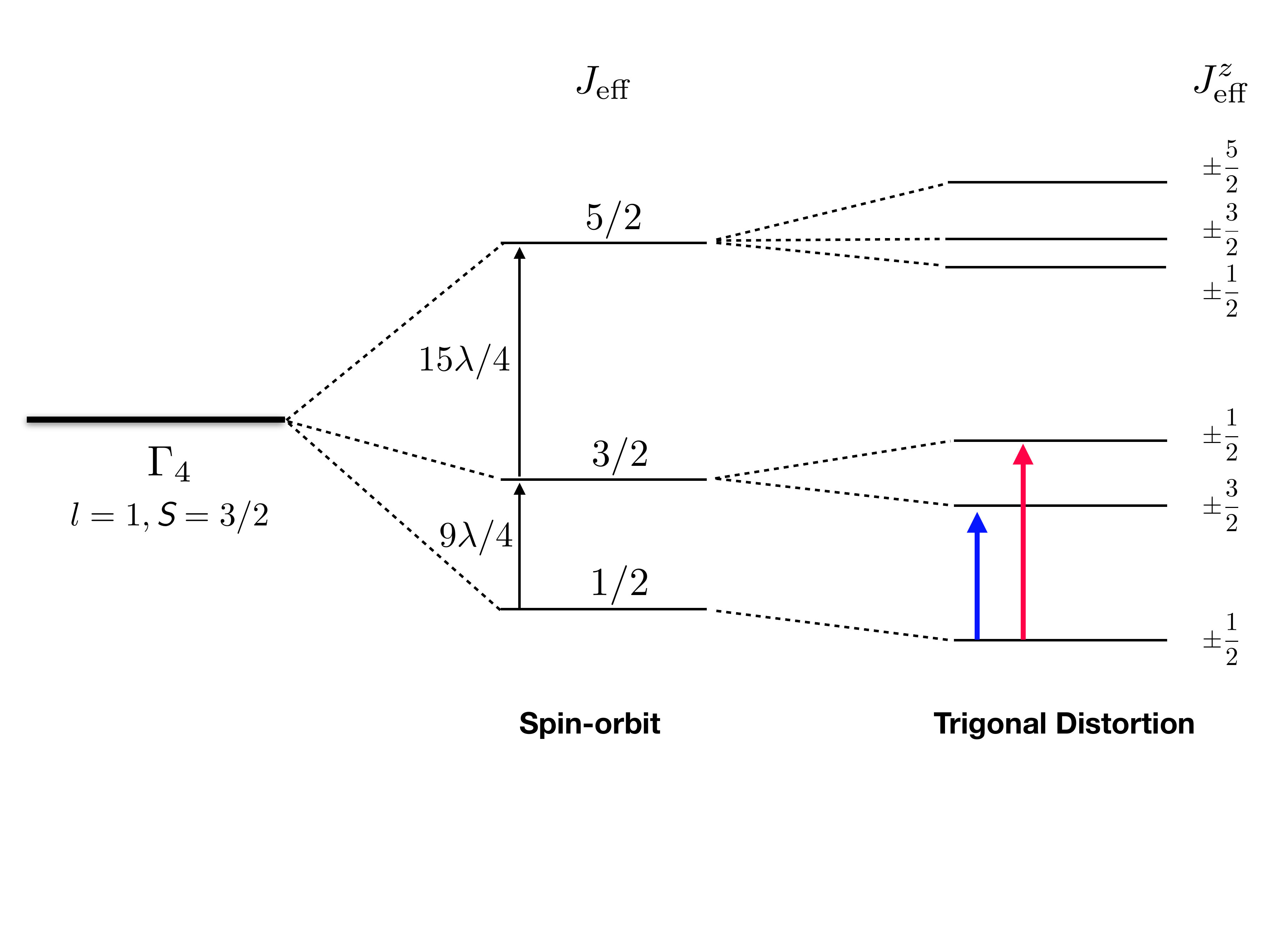} }}%
    \qquad
    \subfloat[]{{\includegraphics[width=0.95\columnwidth]{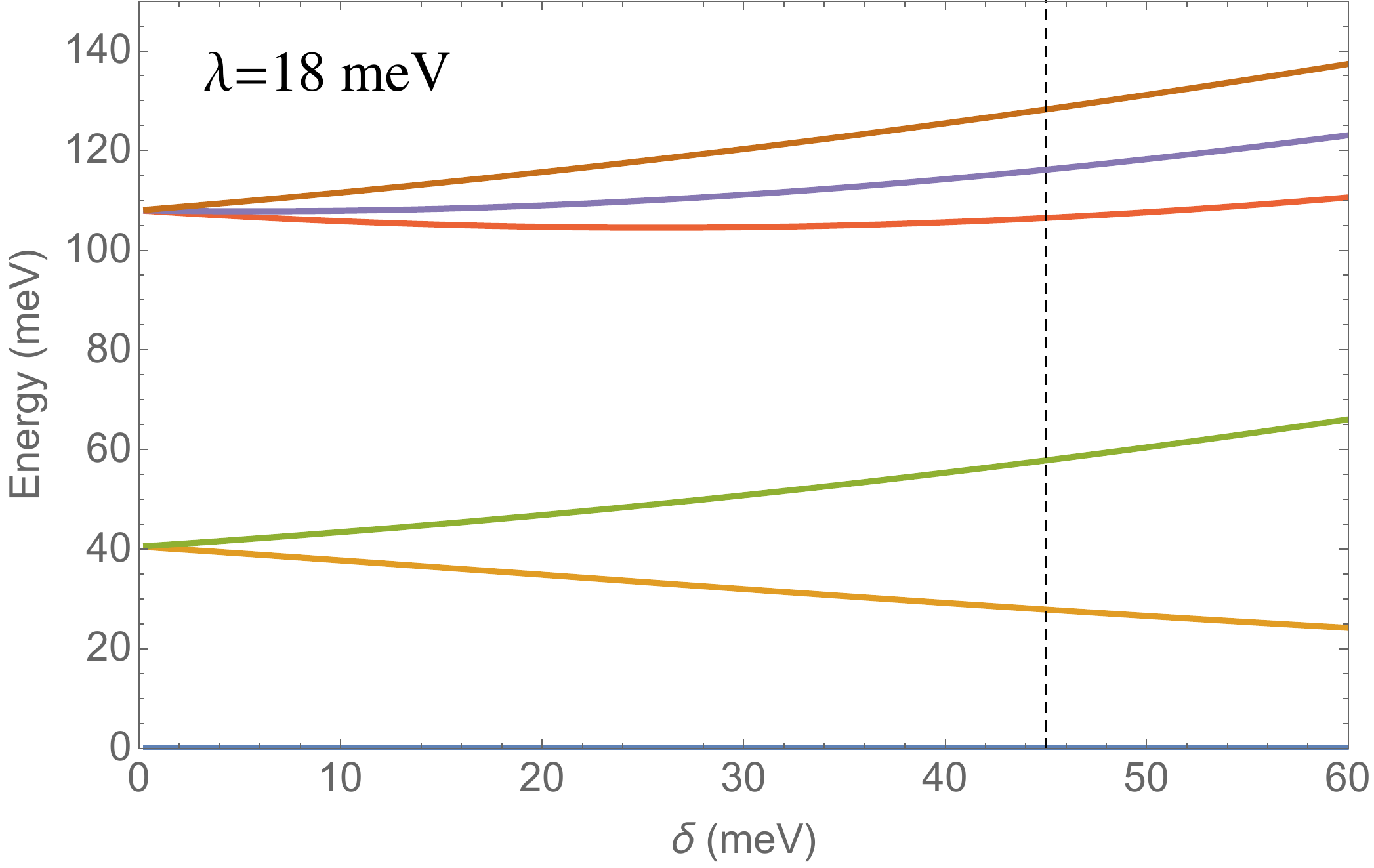} }}%
    \qquad
    \subfloat[]{{\includegraphics[width=0.93\columnwidth]{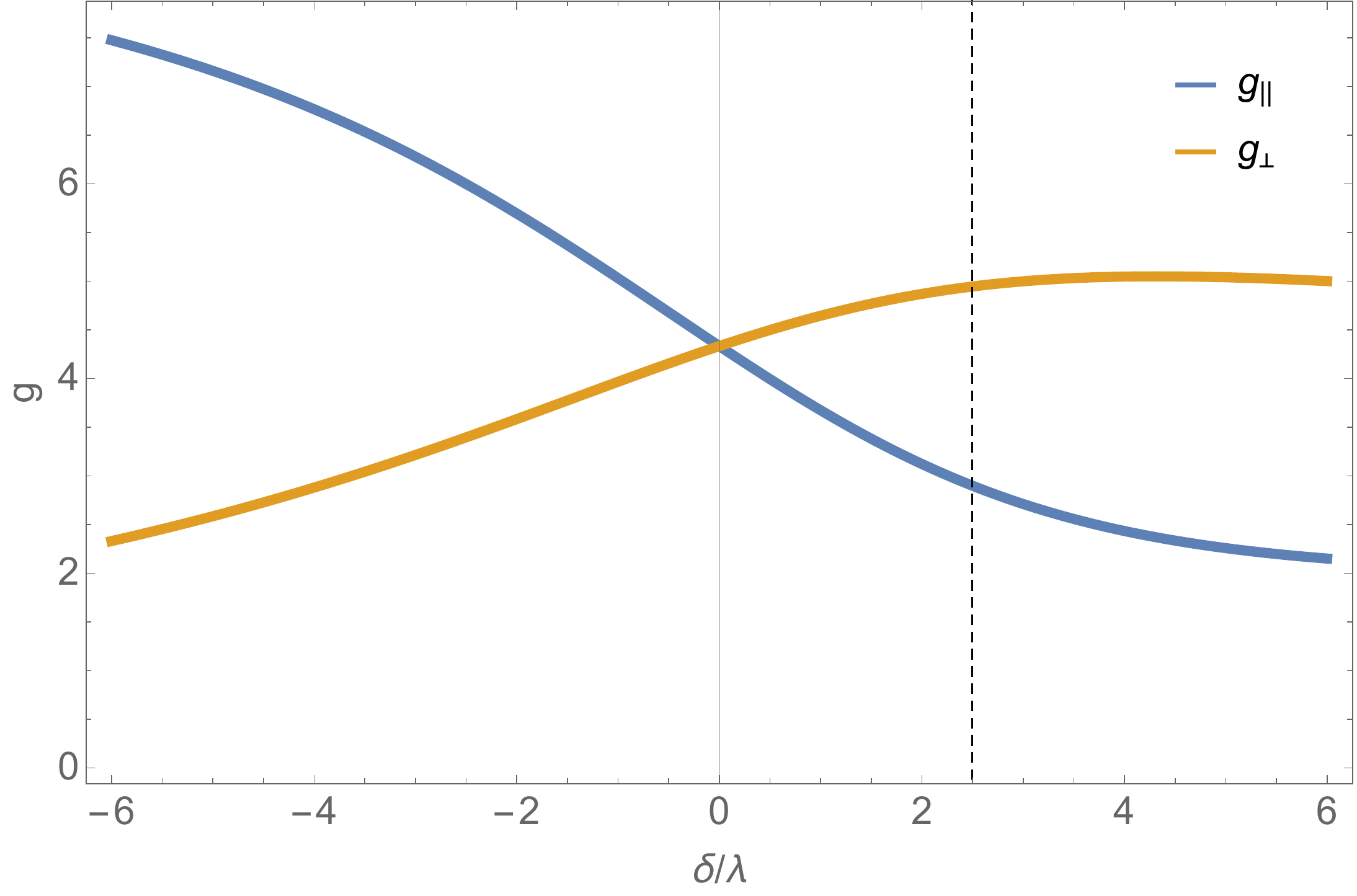} }}%
    \figcaption{Single-ion levels in the presence of spin-orbit coupling and trigonal distortion}{(a) Single-ion spectrum within the lowest $l=1$, $\bm{\mathsf{S}}=3/2$ sector. (b) Splitting of levels as a function of the trigonal distortion $\delta$ at fixed $\lambda=18$~meV. (c) Anisotropic $g$-factors along and perpendicular to the trigonal axis as a function of $\delta/\lambda$. In b) and c) the vertical dashed line corresponds to parameter values for the best fit to the experimentally observed transitions to excited single-ion levels (vertical blue arrows in (a) bottom right).}%
    \label{fig:CEF}%
\end{figure}

In the actual crystal structure the oxygen octahedra around the Co ions are slightly trigonally distorted (the local point group symmetry at the Co sites is $3$ instead of $\bar{4}3m$ for a cubic octahedron) and this distortion can be  parameterized by the term ${\cal H}_{\rm trig}=\delta [l_z^2 -(2/3)]$, where $z$ denotes the $c$-axis. The level scheme in the presence of spin-orbit and trigonal distortion is summarized in Supplementary Figure~\ref{fig:CEF}(a). The trigonal distortion splits the levels into six Kramers doublets, with $J_{\rm eff}^z$ remaining a good quantum number. We now use the available experimental data to constrain the single-ion parameters. Inelastic neutron scattering has revealed the existence of single-ion levels at $28$ and $58$~meV (see Fig.~4a). A best fit to those levels gives $\delta = 45$~meV and $\lambda = 18$~meV, consistent with earlier reports \cite{Yuan}. In the above analysis we have assumed $\delta>0$, as this gives larger magnetic moment in the $ab$-plane compared to along the $c$-axis, in agreement with single-crystal magnetic susceptibility measurements \cite{Balbashov}.

For the trigonally distorted case we examine the anisotropy of the magnetic moment in the ground state. This means that we compute matrix elements of the Zeeman coupled moment $g_l \mathbf{l} + g_{\fs{S}} \bm{\mathsf{S}}$ within the ground state Kramers doublet described by an effective spin $S=1/2$. Throughout we use Serif symbols $S$ and $\mathbf{S}$ to refer to the effective
spin-1/2 and SansSerif symbols $\fs{S}$ and $\bm{\mathsf{S}}$ to refer to the real spin-3/2. Here $g_l = -3/2$ and $g_{\fs{S}} \approx 2$. Supplementary Figure~\ref{fig:CEF}(c) shows how the $g$-factors along and perpendicular to the trigonal axis vary with the reduced parameter $\delta/\lambda$. The moments are isotropic for no trigonal distortion, but develop a strong easy-plane/axis anisotropy in the $g$-factor for $+/-$ve $\delta/\lambda$. For the fitted crystal field scheme, $g_{\parallel}=2.9$ and $g_{\perp}=4.95$ so the ratio $g_{\perp}/g_{\parallel}\approx1.7$, and the expected in-plane moment would be $g_{\perp}\mu_{\rm B}S=2.5~\mu_{\rm B}$, which falls short of the 3~$\mu_{\rm B}$ found experimentally.


\section{Magnitude of the Ordered Moment}
\label{sec:moment}

\begin{figure}[htbp!]%
    \subfloat[]{{\includegraphics[width=0.95\columnwidth]{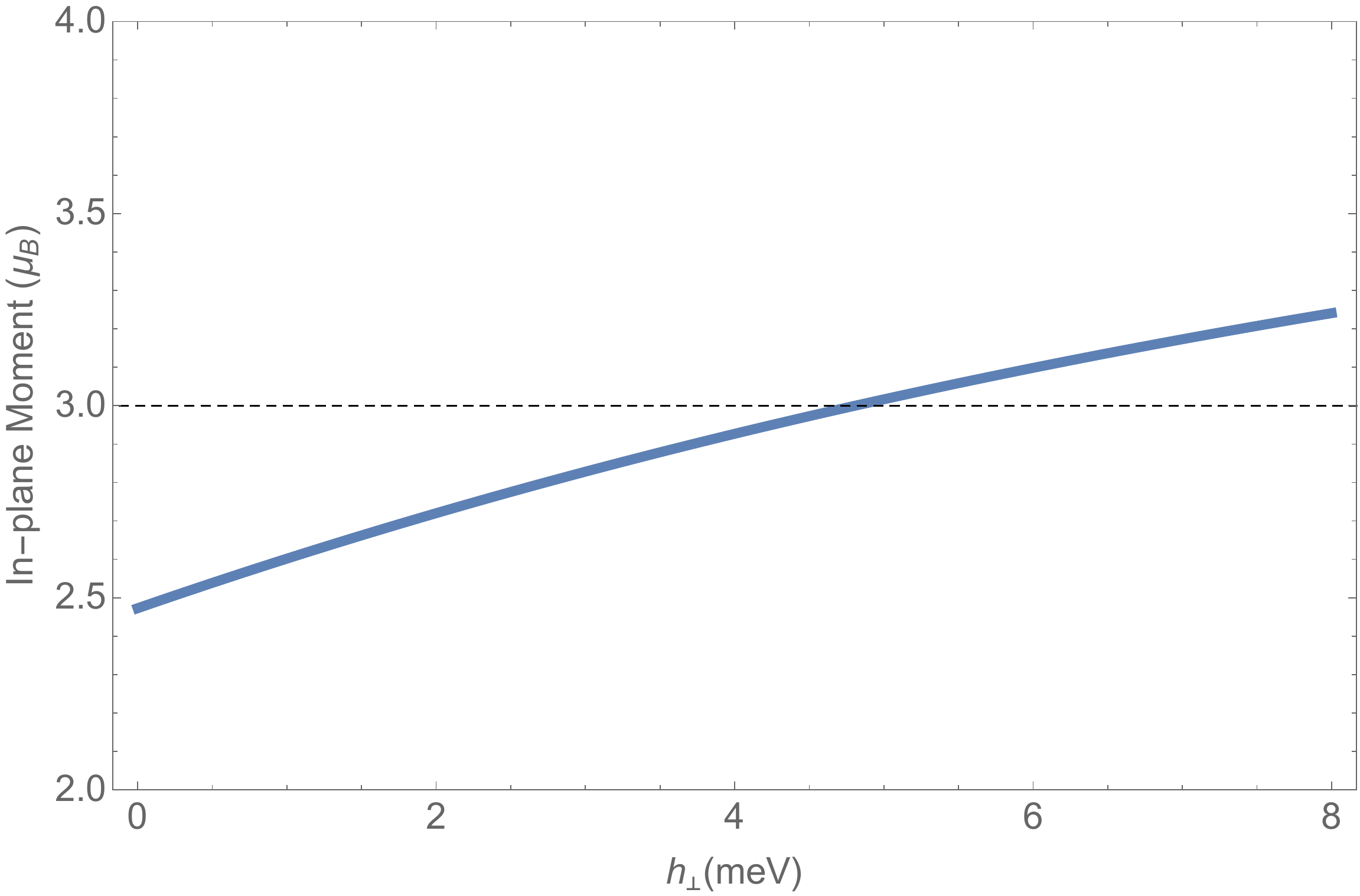} }}%
    \qquad
    \subfloat[]{{\includegraphics[width=0.95\columnwidth]{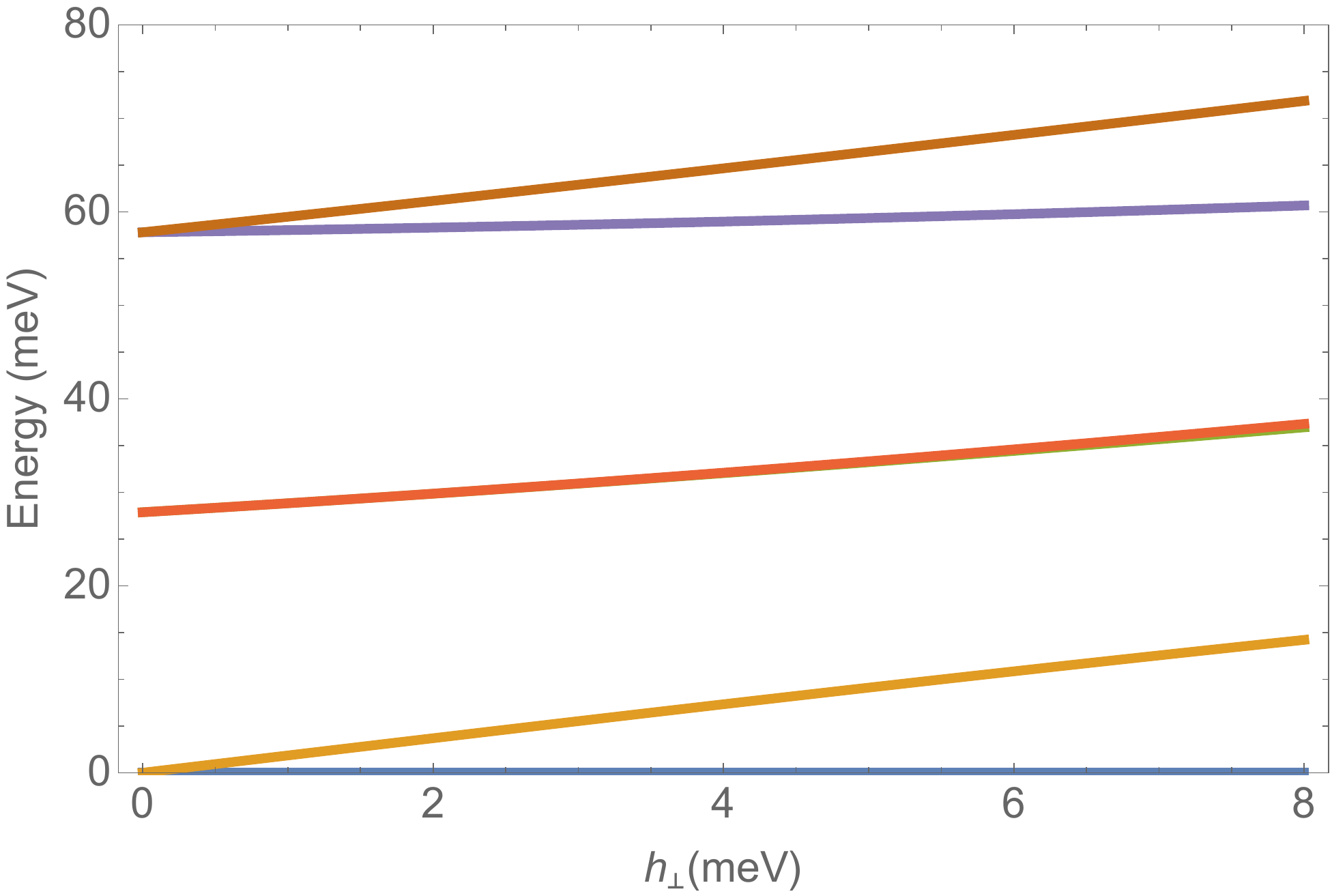} }}%
    \figcaption{Ground state moment and energy levels as a function of in-plane field}{(a) In-plane magnetic moment in the ground state as a function of in-plane mean field $\fs{h}_{\perp}$. The experimental value of 3~$\mu_{\rm B}$ is indicated by the horizontal dashed line. (b) Splitting of the single-ion levels as a function of ${h}_{\perp}$.}%
    \label{fig:meanfieldmoment}%
\end{figure}

Here we show that the shortfall between the calculated magnetic moment in the single-ion picture and the experimentally-determined ordered moment can be explained naturally if exchange mean-field effects are incorporated into the single-ion picture. These can be parameterized by the Zeeman Hamiltonian
${\cal H}_{\rm MF}=-\bm{\mathsf{h}}_{\perp}\cdot\bm{\mathsf{S}}_{\perp}$ where the ${\perp}$ subscript indicates that the mean field is oriented in-plane, along the ordered moment direction in the magnetic structure.

We treat $\fs{h}_{\perp}$ as a variable parameter and solve for the magnetic moment in the ground state with the result shown in Supplementary Figure~\ref{fig:meanfieldmoment}(a). We also compute the single-ion spectrum as a function of $\fs{h}_{\perp}$ in Supplementary Figure~\ref{fig:meanfieldmoment}(b) showing that the splitting of the first exciton around $28$~meV is very small, consistent with the experimental finding.

We can estimate the magnitude of the actual mean field experienced by the magnetic moments from the XXZ exchange parameters that provide a quantitative description of the observed magnon dispersions in~\ref{sec:fits}. Those exchange parameters refer to an effective $S=1/2$ spin model and, for this on-site moment, the magnitude of the mean field is $h_{\perp}=\vert S \sum^6_{n=1} (-)^{n+1} z_n  J^{\perp}_n \vert$ where $z_n$ is the number of $n$'th nearest neighbors and $J^{\perp}_n$ is the $n$'th neighbor in-plane exchange. The sign $(-)^{n+1}$ takes care of the antiferromagnetic arrangement between layers. For the exchange couplings given in Supplementary Equation~(\ref{eq:bestfit}), we find $h_{\perp}=9.1$~meV. We must then adjust for the bare moment by matching the Zeeman splittings for $E_{\rm MF}= -\mathbf{h}_{\perp}\cdot\mathbf{S}_{\perp}$ giving $\fs{h}_{\perp}=5.0$ meV from which we read off from Supplementary Figure~\ref{fig:meanfieldmoment}(a) a renormalized moment of about $3.0$~$\mu_{\rm B}$, as deduced experimentally. We also note that the spin-orbital mean field of~\ref{sec:MFTFW}, with parameters chosen principally to match the spin wave bandwidth, gives a ground state ordered moment of about $2.8$~$\mu_{\rm B}$, again predicting an enhancement of the ordered moment compared to that of isolated ions and towards the value seen experimentally. In summary, we conclude that the enhanced magnetic moment seen experimentally is due to mean-field exchange effects.

\section{Tight-binding model for the exciton dispersion}
\label{sec:exciton}

Here we outline the tight-binding model used to describe the observed dispersion [in Fig.~4c)] of the lowest crystal field excited level near 28~meV, attributed to hopping due to spin and orbital exchange. In a first approximation we neglect the effect of magnetic ordering on the crystal-field excitations and consider hopping of crystal-field excitations only between sites in the same honeycomb layer, so between sites of the A and B sublattices indicated in Supplementary Figure~\ref{fig:FNN}. Because the Kramers degeneracy of the crystal field modes is preserved and there are two sublattices in the paramagnetic regime, two dispersive bands are expected, analogous to the two bands of mobile electrons in graphene, which touch at the corners (K points) of the two-dimensional Brillouin zone \cite{graphene1}. A tight-binding description including 1st, 2nd and 3rd nearest neighbor in-plane (1st, 3rd and 5th in the full crystal structure) with hopping integrals $t_{1,3,5}$ on the same paths as $J_{1,3,5}$ in Supplementary Figure~\ref{fig:FNN} gives the dispersions relations
\begin{equation}
E_{\pm}(\mathbf{k})=E_0\pm \vert \Gamma_{\mathbf{k}} \vert +t_3 \gamma_{3\mathbf{k}}
\label{eq:tb_cf}
\end{equation}
where
\begin{equation}
\vert \Gamma_{\mathbf{k}} \vert e^{i\varphi_{\mathbf{k}}}=\left(t_1 \gamma_{1\mathbf{k}}+t_5 \gamma_{5\mathbf{k}}\right)e^{i\mathbf{k}\cdot(\mathbf{r}_2-\mathbf{r}_1)}
\label{eq:Gamma_k}
\end{equation}
and the cobalt positions $\mathbf{r}_{1,2}$ and geometric factors $\gamma_{n\mathbf{k}}$ are defined (later) in Supplementary Equation.~(\ref{eq:r12}) and (\ref{eq:gamma}). The above equations can capture well the observed dispersions of the exciton modes, see solid and dashed lines in Figure~4c). To find the model parameters experimental dispersion points were extracted from fitting Gaussian peaks to constant-energy and -wavevector scans through the high-energy INS data. From a best-fit to the experimental dispersion points we obtain
\begin{align}
E_0 & = 28.01(1) ~{\rm meV} \nonumber \\
t_1 & = -0.846(6) ~{\rm meV}  \nonumber\\
t_3 & = -0.027(4) ~{\rm meV} \nonumber\\
t_5 & = -0.084(6) ~{\rm meV}.
\label{eq:tb-bestfit}
\end{align}
The uncertainties in the fit parameters were obtained by adding Gaussian noise with a representative standard deviation $\sigma=0.3$~meV to the energies of the experimentally extracted exciton dispersion points and fitting the model parameters for many such data sets. This resulted in a distribution of values for each of the model parameters, the quoted uncertainties are the standard deviations of those distributions.
The hopping terms obtained above are of the same order of magnitude as the fitted exchange couplings presented in~\ref{sec:parameters}.

Fig.~4d) shows the intensity dependence assuming it is determined solely by interference scattering from the A and B sublattices, which takes the form \cite{graphene2} $I_{\pm}(\mathbf{k})\sim 1\pm\cos\varphi_{\mathbf{k}}$ with the upper/lower sign for the top/bottom band and the phase angle $\varphi_{\mathbf{k}}$ defined above. For wavevectors with in-plane projection in the vicinity of a K point, the phase $\varphi_{\mathbf{k}}$ is directly related to the polar (azimuthal) angle $\alpha_{\delta\mathbf{k}}$ of the in-plane wavevector displacement $\delta\mathbf{k}$ away from K. In the limit $\vert \delta \mathbf{k} \vert \rightarrow 0$ and $t_5/t_1 \rightarrow 0$, the following relations are obtained for representative K-points
\begin{eqnarray}
\!(2/3,5/3) \quad \quad & \varphi_{\mathbf{k}} &=\alpha_{\delta\mathbf{k}}-\pi+2\pi\epsilon L \label{eq:varphi_k1}\\
\!(2/3,2/3) \quad \quad & \varphi_{\mathbf{k}}&=\alpha_{\delta\mathbf{k}}+\pi/3+2\pi\epsilon L \label{eq:varphi_k2}\\
\!(4/3,1/3) \quad \quad & \varphi_{\mathbf{k}}&=-\alpha_{\delta\mathbf{k}}+2\pi\epsilon L
\label{eq:varphi_k3}
\end{eqnarray}
where the azimuthal angle $\alpha_{\delta\mathbf{k}}$ is measured with reference to the ($\frac{1}{2}\bar{1}0$) direction (horizontal axis in Fig.~2d). The relations near other K-points are obtained by $\bar{3}m$ symmetry. Here $\epsilon=2(z_{\rm{Co}}-1/3)$ characterises the buckling of the cobalt honeycomb layer ($\epsilon=0$ for flat planes). The above intensity form captures well the overall intensity distribution and explains why only one exciton mode carries significant weight for most wavevectors in Fig.~4c) except the last panel where both modes are visible. Note that by simultaneously changing the sign of both $t_1$ and $t_5$ leaves the dispersion relations in Supplementary Equation~(\ref{eq:tb_cf}) unchanged, however the intensities of the two modes then become completely the opposite way round to what is seen experimentally in Fig.~4c), so the parameter signs as listed in Supplementary Equation~(\ref{eq:tb-bestfit}) are uniquely determined by combining dispersions and intensities constraints.

The above intensity form $I(\mathbf{k})_{\pm}$ also explains the angular dependence of the intensity in the azimuthal scans in Fig.~4e) with maximum intensity in the top band occurring near $\alpha_{\delta_{\mathbf k}}=155(5)^{\circ}$, compared to the predicted value of $153(1)^{\circ}$ based on Supplementary Equation~(\ref{eq:varphi_k1}) [$\pi-2\pi\epsilon L$ averaged for the appropriate $L$-integration range of the scan]. The tight-binding model in this Section provides a good empirical fit to the observed exciton data. In \ref{sec:MFTFW} we treat the excitons and magnons in a unified way showing that the antiferromagnetic order should lead to further splitting of the exciton modes, however such splittings are expected to be small, beyond the resolution of the present measurements.

\section{Spin wave calculations for the minimal $S=1/2$ XXZ model}
\subsection{Structural and magnetic Brillouin zones}
\label{sec:BZ}
Here we describe the Brillouin zone relevant for the reciprocal space periodicity of the magnetic dispersion relations.  We introduce the hexagonal primitive vectors $\mathbf{a}$, $\mathbf{b}$ and $\mathbf{c}$ indicated in Supplementary Figure~\ref{fig:FNN} and the primitive rhombohedral unit cell with vectors
\begin{equation}
\left(\begin{array}{c}
\mathbf{A}_1 \\
\mathbf{A}_2 \\
\mathbf{A}_3
\end{array}\right)
= \frac{1}{3}
\left(\begin{array}{cccc}
-1 & -2 & 1 \\
2  & 1  & 1  \\
-1 & 1  & 1
\end{array}\right)
\left(\begin{array}{c}
\mathbf{a} \\
\mathbf{b} \\
\mathbf{c}
\end{array}\right),
\label{eq:Rh}
\end{equation}
such that $\mathbf{c} =\mathbf{A}_1+\mathbf{A}_2+\mathbf{A}_3$. The Brillouin zone corresponding to this primitive structural cell is illustrated in Supplementary Figure~\ref{fig:BZ}a) and belongs to the elongated ($c>\sqrt{3/2}a$) rhombohedral case \cite{Bradley}. It has top and bottom regular hexagonal faces with midpoints at $\pm\left(00\frac{3}{2}\right)$ and twelve side faces that alternate between rectangular and slightly-distorted hexagonal with midpoints at $\left(\frac{1}{2}0\bar{1}\right)$ and $\left(\frac{1}{2}0\frac{1}{2}\right)$, respectively, with other faces obtained by $\bar{3}$ symmetry.

\begin{figure}[!tbph]
    \includegraphics[width=\columnwidth]{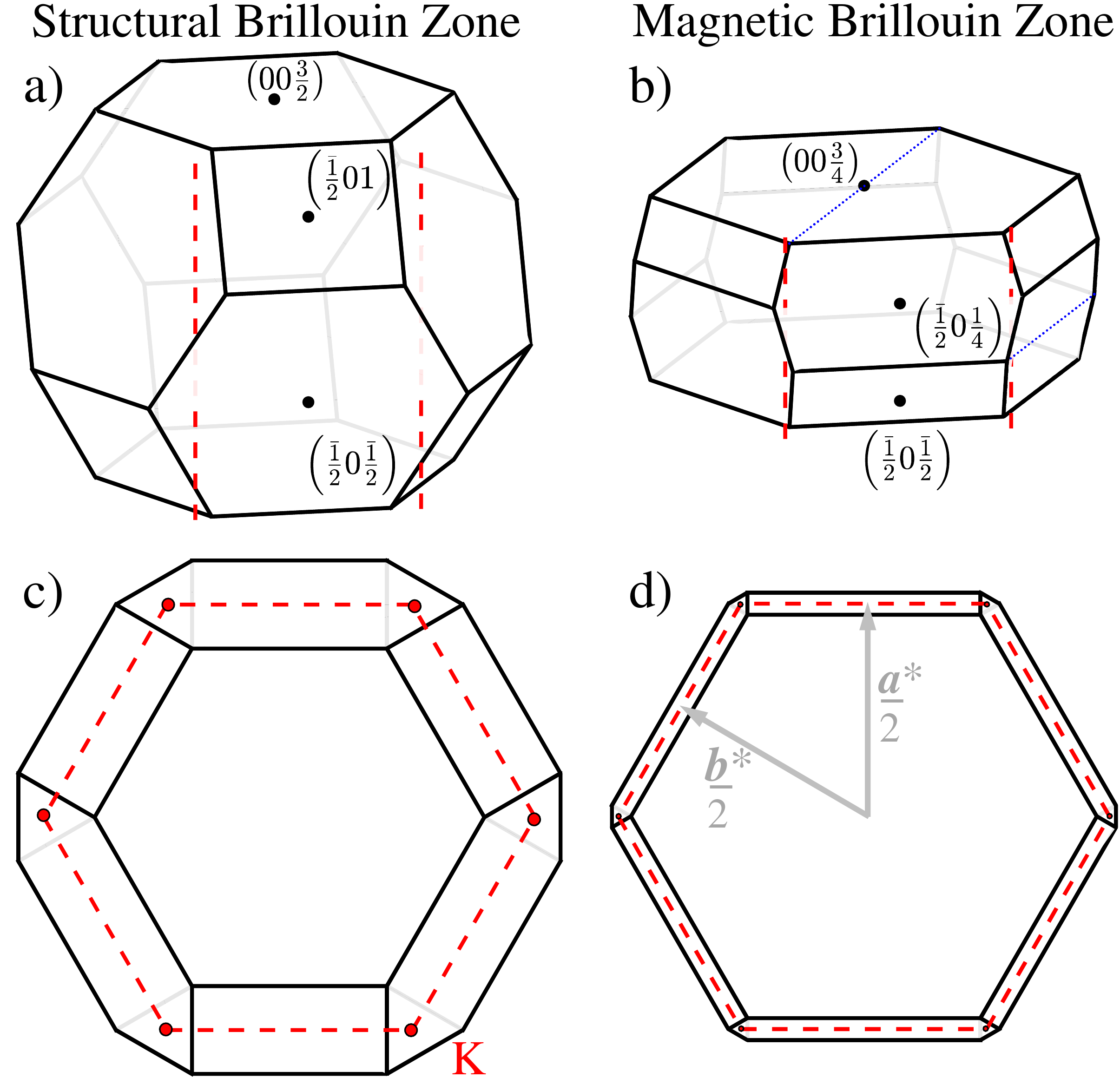}
    \figcaption{Brillouin zones}{a) Structural and b) magnetic Brillouin zones with selected mid-face points labelled and projected in c) and d), respectively, onto the $hk$ plane, where the dashed outline is the 2D hexagonal Brillouin zone of a single honeycomb layer. In a) and b) the side Brillouin zone edges wrap round vertical lines (dashed red) that project onto the corner K-points of the 2D hexagonal Brillouin zone. In b) the parallel dotted blue line segments on the top and a side face show a path equivalent to the scan direction in Supplementary Figure~\ref{fig:w12_splitting_mzb}.}
    \label{fig:BZ}
\end{figure}

The magnetic structure is illustrated in Supplementary Figure~\ref{fig:FNN} and has moments parallel in each layer and antiparallel between layers. This magnetic periodicity can be captured by a doubled-volume rhombohedral primitive cell shown by the dashed outline in Supplementary Figure~\ref{fig:FNN}, with basis vectors rotated by 60$^{\circ}$ and elongated a factor of 2 along $\mathbf{c}$ compared to the rhombohedral primitive structural cell in Supplementary Equation~(\ref{eq:Rh}), with the magnetic primitive unit cell vectors given by
\begin{equation}
\left(\begin{array}{c}
\mathbf{M}_1 \\
\mathbf{M}_2 \\
\mathbf{M}_3
\end{array}\right)
= \frac{1}{3}
\left(\begin{array}{cccc}
1 & 2 & 2 \\
-2  & -1 & 2 \\
1 & -1  & 2
\end{array}\right)
\left(\begin{array}{c}
\mathbf{a} \\
\mathbf{b} \\
\mathbf{c}
\end{array}\right).
\nonumber
\end{equation}
The Brillouin zone corresponding to this primitive magnetic unit cell is shown in Supplementary Figure~\ref{fig:BZ}c) and is half the volume of the structural Brillouin zone in Supplementary Figure~\ref{fig:BZ}a), with similar topology, but 60$^{\circ}$ rotated around (001). The top and bottom faces are hexagonal with midpoints at $\pm\left(00\frac{3}{4}\right)$ and the twelve side faces alternate between rectangular and strongly distorted hexagonal with midpoints at $\left(\frac{1}{2}0\frac{1}{2}\right)$ and $\left(\frac{1}{2}0\frac{\bar{1}}{4}\right)$, respectively, with other faces obtained by $\bar{3}$ symmetry.

The projection of the magnetic Brillouin zone in the $hk$ plane is illustrated in Supplementary Figure~\ref{fig:BZ}d), where the inner hexagon corresponds to the top face at $L=3/4$. In projection, this is located inside the 2D hexagonal Brillouin zone (red dashed outline) of a single honeycomb layer. Upon decreasing $L$, the corners of the magnetic Brillouin zone move initially outwards, along the set of small black segments, followed by two other small segments, such that in projection they describe small equilateral triangles centred at the nominal K-points of the 2D hexagonal Brillouin zone. Viewed in 3D, the magnetic Brillouin zone edges wrap around the straight lines that project onto K-points, as illustrated in Supplementary Figure~\ref{fig:BZ}b). This has the consequence that points along those straight lines have no special symmetry, they act like general points in the magnetic Brillouin zone. Therefore, there are no symmetry-imposed constraints for touching points in the magnetic dispersion bands to be pinned at those positions. Indeed, as illustrated in Fig.~3b) and detailed in~\ref{sec:eta} and \ref{sec:dirac}, we find that in the general case nodal lines wind along $L$ and precess in-plane around positions that can be displaced away from the nominal K-points.

\subsection{The XXZ Model and Further Neighbor Couplings}
\label{sec:XXZ}

Here we give details of the analytical spin-wave calculations for the minimal easy-plane exchange model that captures the principal features of the observed magnon dispersion relations. To describe the full structural arrangement of the Co ions we use the following primitive unit cell with vectors
\begin{equation}
\left(\begin{array}{c}
\mathbf{R}_1 \\
\mathbf{R}_2 \\
\mathbf{R}_3
\end{array}\right)
=\left(\begin{array}{cccc}
1 & 0 & 0 \\
0  & 1 & 0 \\
-\frac{1}{3} & -\frac{2}{3}  & \frac{1}{3}
\end{array}\right)
\left(\begin{array}{c}
\mathbf{a} \\
\mathbf{b} \\
\mathbf{c}
\end{array}\right)
\nonumber
\end{equation}
and vectors defining the positions of the two cobalt ions in this primitive cell
\begin{eqnarray}
& \mathbf{r}_1 & = -\frac{\epsilon}{2}\mathbf{c}  \nonumber \\
& \mathbf{r}_2 & = -\frac{1}{3}\mathbf{a} -\frac{2}{3}\mathbf{b}+\frac{\epsilon}{2}\mathbf{c},
\label{eq:r12}
\end{eqnarray}
where $\epsilon=2(z_{\rm Co}-1/3)$ characterises the buckling of the cobalt honeycombs, with the Co $z$-coordinate $z_{\rm Co}$ given in Supplementary Table~\ref{TAB::crystal_structure}. The above primitive cell was chosen to emphasize the $ab$ planes as a `natural' building block of the Co structural arrangement.

The full structure of cobalt ions is then generated by the set of positional vectors
\begin{equation}
\mathbf{R}_{n_i,m} =\sum_{a=1,2,3} n_{ia}\mathbf{R}_a  + \mathbf{r}_m \equiv \mathbf{R}_i +\mathbf{r}_m \,
\label{eq:latticevec}
\end{equation}
where the integers $n_{ia}$ select the primitive unit cell and $m=1,2$ is the cobalt sublattice index.

\begin{figure}[!tbph]
    \includegraphics[width=0.9\columnwidth]{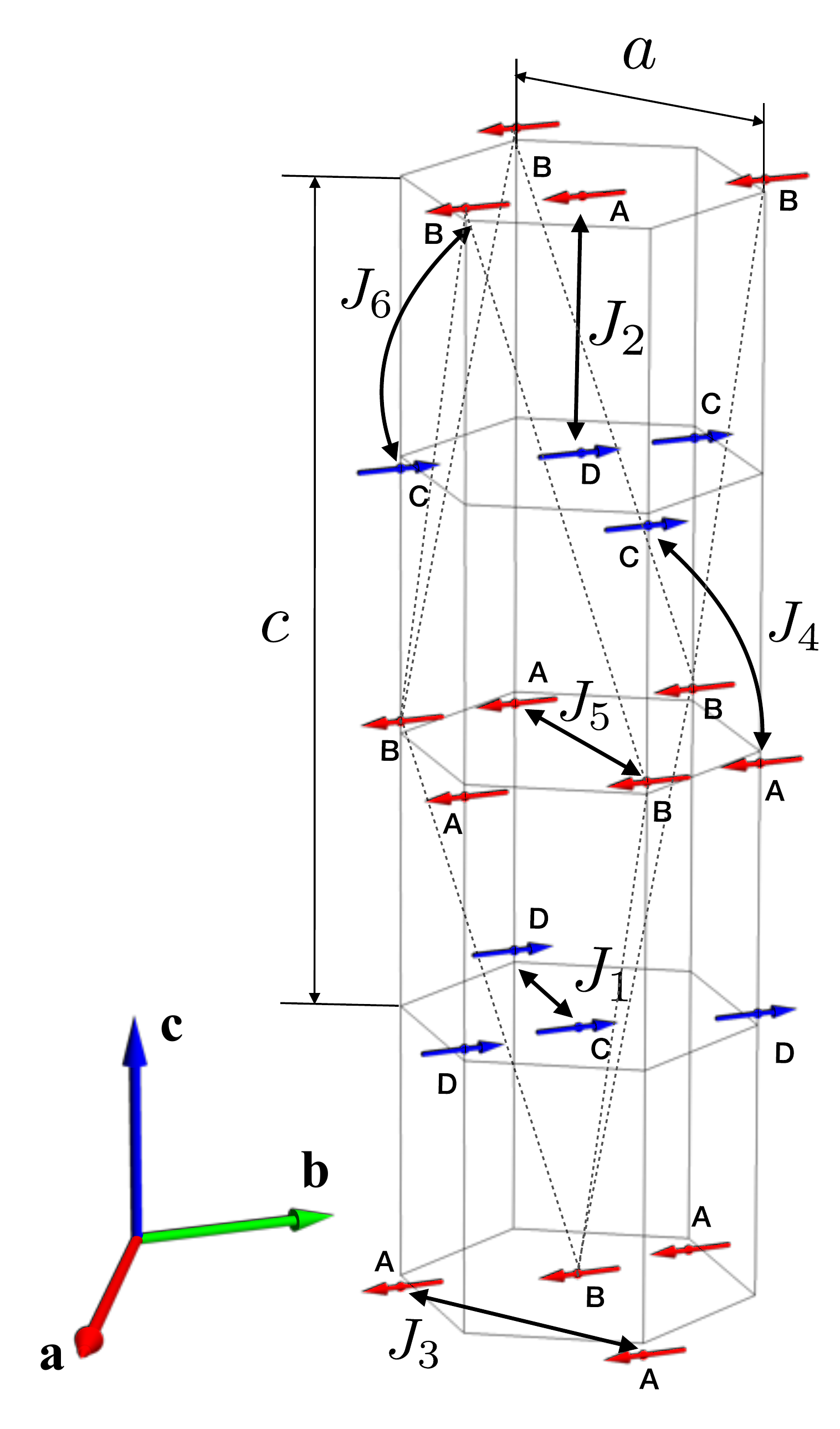}
    \figcaption{Magnetic structure and exchange paths}{Moments (thick red and blue arrows) are confined to the $ab$ plane, FM-aligned in the honeycomb layers with AFM stacking. Labels A,B,C,D indicate the four magnetic sublattices used in the spin-wave calculation and the thick double arrowed lines labeled $J_1$ to $J_6$ show the exchange paths up to 6th nearest neighbor. Dashed lines indicate the outline of the primitive rhombohedral magnetic unit cell corresponding to the magnetic propagation vector $\mathbf{Q}$.}
    \label{fig:FNN}
\end{figure}

The minimal model that we find describes all but the fine structure of the spin wave spectrum is an XXZ model on all these couplings:
\begin{equation}
{\cal H}_{\rm XXZ} = \sum_{\langle i,j \rangle_n} \frac{J^{\perp}_n}{2} \left( S_i^+ S_j^- + S_i^- S_j^+ \right) + J^z_n S^z_i S^z_j  ,
\label{eq:XXZ}
\end{equation}
where $J_n^z$ is the (Ising) coupling for spin components along $z\parallel c$ and $J_n^{\perp}$ is the coupling for spin components in the $ab$ plane, the summation is over all interacting pairs $\langle i,j \rangle_n$ of $n$'th nearest neighbors counted once, and we include $n$=1 to 6.

\subsection{Spin Wave Calculations}
\label{sec:sw}

The magnetic structure shown in Supplementary Figure~\ref{fig:FNN} has collinear moments that are ferromagnetic in each honeycomb plane and antiparallel between adjacent planes, with four magnetic sublattices (labelled A-D) per primitive magnetic cell. For the spin-wave calculation it is convenient to define a global Cartesian $xyz$ frame with $\mathbf{x} \parallel \mathbf{a}$, $\mathbf{z} \parallel \mathbf{c}$ and $\mathbf{y} = \mathbf{z} \times \mathbf{x}$, as illustrated in Supplementary Figure~\ref{fig:FW}d). We also define a local frame denoted $\tilde{x}\tilde{y}\tilde{z}$ where $\tilde{z}$ lies along the direction of the local ordered moment and $\tilde{x}$ is in-plane, where the two frames are related by
\begin{eqnarray}
\mathbf{x} & = & \cos(\phi+n\pi) \mathbf{\tilde{z}} - \sin(\phi+n\pi) \mathbf{\tilde{x}}, \nonumber\\
\mathbf{y} & = & \sin(\phi+n\pi) \mathbf{\tilde{z}} + \cos(\phi+n\pi) \mathbf{\tilde{x}},
\nonumber\\
\mathbf{z} & = & \mathbf{\tilde{y}},
\label{eq:local}
\end{eqnarray}
where $\phi$ is the in-plane angle of the ordered spins of sublattices $A$ and $B$, measured relative to $\mathbf{x}$ in the positive sense if rotating around $\mathbf{z}$ and $n=0$ for sublattices A and B, and $n=1$ for sublattices C and D.

In the local frame all moments are parallel and the magnetic primitive cell is reduced to the structural primitive cell, i.e. in this frame there are only two magnetic sublattices as opposed to four in the original frame. For the XXZ model in Supplementary Equation~(\ref{eq:XXZ}) the spin Hamiltonian expressed in the local frame has the same periodicity as that of the structural cell and in this case the problem is reduced to obtaining the dispersion relations for a Hamiltonian of the generic form
$${\cal H}=\frac{1}{2}\sum_{\mathbf{k}} \mathbf{\Upsilon}_{\mathbf{k}}^\dagger
{\cal D}(\mathbf{k})\mathbf{\Upsilon}_{\mathbf{k}}$$
where $\mathbf{\Upsilon}_{\mathbf{k}}^\dagger=(a_{\mathbf{k}}^\dagger, b_{\mathbf{k}}^\dagger, a_{-\mathbf{k}}, b_{-\mathbf{k}})$, with $a_{\mathbf{k}}^\dagger$ and $b_{\mathbf{k}}^\dagger$ being magnon creation operators for the two magnetic sublattices. Here the sum extends over all wavevectors $\mathbf{k}$ in the structural Brillouin zone in Supplementary Figure~\ref{fig:BZ}a).

Introducing $A$, $B$, $C$ and $D$ as implicit functions of $\mathbf{k}$, the dynamical matrix ${\cal D}(\mathbf{k})$ has the form
\begin{equation}
{\cal D}(\mathbf{k})= \left( \begin{array}{cccc}
A & B & C & D^\star  \\
B^\star & A & D & C  \\
C^\star & D^\star & A & B  \\
D & C^\star & B^\star & A
\end{array}  \right).
\label{eq:calD}
\end{equation}

Including couplings up to 6th nearest neighbor we find

\begin{widetext}
\begin{eqnarray}
A& = &[-3J^{\perp}_1 +J^{\perp}_2 - 6J^{\perp}_3 + \frac{1}{2}(J^{\perp}_3 + J_3^z) \gamma_{3\mathbf{k}} + 6J^{\perp}_4 - \frac{1}{2}(J^{\perp}_4 - J_4^z)  \gamma_{4\mathbf{k}} - 3J^{\perp}_5  + 3J^{\perp}_6]S \nonumber \\
B & = &[\frac{1}{2}(J^{\perp}_1 + J_1^z)  \gamma_{1\mathbf{k}} -  \frac{1}{2}(J^{\perp}_2 - J_2^z)  \gamma_{2\mathbf{k}} + \frac{1}{2}(J^{\perp}_5 + J_5^z)  \gamma_{5\mathbf{k}} - \frac{1}{2}(J^{\perp}_6 - J_6^z)  \gamma_{6\mathbf{k}}]S \nonumber \\
C & = &[\frac{1}{2}(J^{\perp}_3 - J_3^z)  \gamma_{3\mathbf{k}} - \frac{1}{2}(J^{\perp}_4 + J_4^z)  \gamma_{4\mathbf{k}}]S \nonumber \\
D & = &[ \frac{1}{2}(J^{\perp}_1 - J_1^z) \gamma^\star_{1\mathbf{k}}  - \frac{1}{2}(J^{\perp}_2 + J_2^z)  \gamma^\star_{2\mathbf{k}} + \frac{1}{2}(J^{\perp}_5 - J_5^z)  \gamma^\star_{5\mathbf{k}} - \frac{1}{2}(J^{\perp}_6 + J_6^z)  \gamma^\star_{6\mathbf{k}}]S.
\label{eq:ABCD}
\end{eqnarray}
\end{widetext}
Here
\begin{equation}
\gamma_{n\mathbf{k}} \equiv \sum_{ni \in {\rm NN}(n)} e^{i\mathbf{k}\cdot \mathbf{R}_{ni}}
\label{eq:gamma}
\end{equation}
where the sum runs over the set of $N(n)$ $n$'th nearest neighbors, with $N(1)=3$, $N(2)=1$, $N(3)=6$, $N(4)=6$, $N(5)=3$ and $N(6)=3$. The vectors $\mathbf{R}_{ni}$ define the relative displacement of the primitive cells where the $i$'th members in the set of $N(n)$ neighbors are located, and can be decomposed in terms of the primitive basis vectors as
\begin{equation}
\mathbf{R}_{ni}=p_{ni} \mathbf{R}_1+q_{ni} \mathbf{R}_2 +r_{ni} \mathbf{R}_3
\label{eq:gamR}
\end{equation}
with coefficients given in Supplementary Table~\ref{tab:gam} for all nearest neighbors up to $n=6$, with representative bonds illustrated in Supplementary  Figure~\ref{fig:FNN}.

\begin{table}[h!]
  \begin{center}
    \tabcaption{The integer coefficients $p_{ni}$, $q_{ni}$ and $r_{ni}$ defined in Supplementary Equation~(\ref{eq:gamR})}{}
    \label{tab:gam}
    \begin{tabular}{|c|c|c|c|c| | c|c|c|c|c|}
    \hline
      $n$ & $i$ & $p_{ni}$ & $q_{ni}$ & $r_{ni}$ & $n$ & $i$ & $p_{ni}$ & $q_{ni}$ & $r_{ni}$ \\ %
      \hline
      1 & 1 & 0 & 1 & 0 & 4 & 2 & 0 & 0 & 1 \\
      1 & 2 & 0 & 0 & 0 & 4 & 3 & 0 & -1 & -1 \\
      1 & 3 & 1 & 1 &  0 & 4 & 4 & 0 & 1 & 1 \\
      2 & 1 & 0 & 0 & -1 & 4 & 5 & 1 & 1 & 1 \\
      3 & 1 & 1 & 0 & 0 & 4 & 6 & -1 & -1 & -1 \\
      3 & 2 & -1 & 0 & 0 & 5 & 1 & 1 & 0 & 0 \\
      3 & 3 & 0 & 1 &  0 & 5 & 2 & -1 & 0 & 0 \\
      3 & 4 & 0 & -1 & 0 & 5 & 3 & 1 & 2 & 0 \\
      3 & 5 & 1 & 1 & 0 & 6 & 1 & 1 & 2 & 1 \\
      3 & 6 & -1 & -1 &  0 & 6 & 2 & 0 & 1 & 1 \\
      4 & 1 & 0 & 0 & -1 & 6 & 3 & 1 & 1 & 1 \\
      \hline
    \end{tabular}
  \end{center}
\end{table}

The dispersion relations are obtained by diagonalizing the matrix $\mathcal{G}{\cal D}(\mathbf{k})$, where $\mathcal{G}={\rm diag}(1,1,-1,-1)$, and are given by \begin{widetext}
\begin{equation}
\omega_{\pm}^2 = A^2 + \vert B \vert^2 - \vert C \vert^2 -\vert D \vert^2 \pm \sqrt{\left( 2AB-CD^\star-D^\star C^\star \right)\left(2AB^\star-CD-C^\star D\right) + \left( B D - B^\star D^\star \right)^2}.
\label{eq:omega}
\end{equation}
\end{widetext}
In order to compute the neutron scattering intensities, we require the right eigenvectors of $\mathcal{G}\mathcal{D}$. The components are
\begin{widetext}
\begin{align*}
W(\omega) & \equiv -(A+\omega) (A^2 + \vert B \vert^2 - \vert C \vert^2 - \vert D \vert^2 - \omega^2) + 2A\vert B\vert^2 - BDC^\star - B^\star D^\star C \\
X(\omega) & \equiv \left( A^2 C^\star + \vert B\vert^2 C^\star - \vert C\vert^2 C^\star + \vert D\vert^2 C - C^\star \omega^2 \right) - A\left( BD + B^\star D^\star \right) + \omega\left( BD - B^\star D^\star \right) \\
Y(\omega) & \equiv B^\star \left[ \left( A+\omega \right)^2 - \vert B \vert^2 + \vert C\vert^2 \right] -  \left( ACD + ADC^\star + CD\omega + C^\star D\omega  \right) + BD^2 \\
Z(\omega) & \equiv D\left( A^2 +  C^{\star 2} -\vert D \vert^2 - \omega^2 \right) + B^{\star 2}D^\star - 2AB^\star C^\star
\end{align*}
\end{widetext}
up to a normalization
\[
N(\omega) = \lvert -\vert W \vert^2 + \vert X \vert^2 - \vert Y \vert^2 + \vert Z \vert^2  \rvert.
\]
The eigenvectors are then the columns of
\[
\mathcal{U} = \left( \begin{array}{cccc} \bar{W}(\omega_-)& \bar{W}(\omega_+) & \bar{W}(-\omega_-) & \bar{W}(-\omega_+)  \\
 \bar{Y}(\omega_-)& \bar{Y}(\omega_+) & \bar{Y}(-\omega_-) & \bar{Y}(-\omega_+)  \\
 \bar{X}(\omega_-)& \bar{X}(\omega_+) & \bar{X}(-\omega_-) & \bar{X}(-\omega_+)  \\
 \bar{Z}(\omega_-)& \bar{Z}(\omega_+) & \bar{Z}(-\omega_-) & \bar{Z}(-\omega_+)
\end{array} \right)
\]
where the bar means $\bar{W}(\omega)\equiv W(\omega)/\sqrt{N(\omega)}$ and so on.

Let us define
\begin{align}
S_{\mathbf{q}}^{\tilde{x}\tilde{x}}(\mathbf{k},\omega) \equiv  & \vert W+X+(Y+Z)e^{-i(\mathbf{k}-\mathbf{q})\cdot(\mathbf{r}_1-\mathbf{r}_2)} \vert^2 \times \nonumber \\
& \frac{S}{4N} [\delta(\omega-\omega_+(\mathbf{k})) + \delta(\omega-\omega_-(\mathbf{k}))] \nonumber \\
S^{\tilde{y}\tilde{y}}(\mathbf{k},\omega) \equiv & \vert W-X+(Y-Z)e^{i\mathbf{k}\cdot(\mathbf{r}_1-\mathbf{r}_2)}\vert^2 \times \nonumber\\
& \frac{S}{4N} [\delta(\omega-\omega_+(\mathbf{k})) + \delta(\omega-\omega_-(\mathbf{k}))], \label{eq:Sxx_Syy_local}
\end{align} where implicitly the functions $W$, $X$, $Y$, $Z$, $N$ are evaluated at wavevector $\mathbf{k}$, and energy $\omega$ that satisfies the delta functions $\delta(\omega-\omega_{\pm}(\mathbf{k}))$. The in- and out-of-plane dynamical correlations in the local frame are obtained as $S_{\bm{0}}^{\tilde{x}\tilde{x}}(\mathbf{k},\omega)$ and $S^{\tilde{y}\tilde{y}}(\mathbf{k},\omega)$, respectively, so both magnon modes $\omega_{\pm}(\mathbf{k})$ occur in both polarizations with different intensities. Note that the magnon dispersion relations are independent of the buckling parameter $\epsilon$, which only has the effect of modulating the intensities through the exponential phase factor in Supplementary Equation~(\ref{eq:Sxx_Syy_local}).

Upon transformation to the global frame the out-of-plane correlations are unchanged. The in-plane polarized dispersions are momentum shifted by the propagation vector $\mathbf{Q}$ such that the total INS intensity is proportional to
\begin{equation} p_{\tilde{x}} g^2_{\perp}S_{\mathbf{Q}}^{\tilde{x}\tilde{x}}(\mathbf{k}+\mathbf{Q},\omega)+
p_{\tilde{y}}g^2_{\parallel}S^{\tilde{y}\tilde{y}}(\mathbf{k},\omega)
\label{eq:intensities}
\end{equation}
where we have included the neutron polarization factors $p_\zeta=1-(k_{\zeta}/k)^2$ and the anisotropic $g$-factor components for in- and out-of-plane moment directions, $g_{\perp}$ and $g_{\parallel}$, respectively. Here $k_{\zeta}=\mathbf{k}\cdot
\bm{\upzeta}$ is the wavevector transfer component along the $\bm{\upzeta}=\mathbf{\tilde{x}},\mathbf{\tilde{y}}$ direction. In the global frame at a given wavevector ${\mathbf{k}}$ there are four magnon branches, $\omega_{\pm}(\mathbf{k})$ (polarised out-of-plane along $z=\tilde{y}$) and  $\omega_{\pm}(\mathbf{k}+\mathbf{Q})$ (polarized in-plane), i.e. one recovers the expected number of modes for the four magnetic sublattices in the original problem. In the above we have used the fact that $2\mathbf{Q}$ is a vector of the reciprocal lattice of the structural cell, so $\omega(\mathbf{k}+\mathbf{Q})$ and $\omega(\mathbf{k}-\mathbf{Q})$ are identical by reciprocal space translational symmetry. The above analytical expressions for the dispersion relations and intensities have been checked against a numerical spin-wave code for the full four-sublattice spin-wave Hamiltonian, and also against SpinW \cite{SpinW}.

The spectrum is gapless at the origin and at the magnetic propagation vector $\mathbf{Q}$, as expected given the $U(1)$ symmetry of the XXZ Hamiltonian, i.e. there is no energy cost in rotating the spins in the $ab$ plane. The dispersion relations as well as the functional form of the dynamical structure factors are independent of the in-plane angle $\phi$, the only dependence of the INS intensity on $\phi$ comes through the neutron polarization factor for in-plane fluctuations, $p_{\tilde{x}}$. This is the case for a {single} magnetic domain, however assuming six equally-populated magnetic domains with moment directions at $\phi+n\pi/3$ ($n=0$ to $5$) as expected due to the $\bar{3}$ lattice point group symmetry of the crystal structure, the domain average $\langle p_{\tilde{x}} \rangle$ is independent of $\phi$ and the total INS intensity in this case is proportional to
\begin{equation}
\frac{1}{2}\left(1+\frac{k^2_z}{k^2}\right) g^2_{\perp}S_{\mathbf{Q}}^{\tilde{x}\tilde{x}}(\mathbf{k}+\mathbf{Q},\omega)+ \left(1-\frac{k^2_z}{k^2}\right)g^2_{\parallel}S^{\tilde{y}\tilde{y}}(\mathbf{k},\omega)
\label{eq:total_intensity}
\end{equation}
where $k_z=\mathbf{k}\cdot \mathbf{z}$ is the wavevector transfer component along the $c$-axis.

For the purpose of comparison with data it is helpful to discuss the overall reciprocal space periodicity and symmetry of the dispersions and dynamical correlations. The Brillouin zone folding that occurs upon going from the local to the global frame has the consequence that each of the dispersive modes in the global frame when labelled as $\omega_{m}(\mathbf{k})$ with $m=1$ to $4$ in order of increasing energy, has the translational periodicity of the magnetic Brillouin zone. The magnetic structure in Supplementary Figure~\ref{fig:FNN} breaks the 3-fold rotational symmetry of the crystal structure, however the XXZ Hamiltonian has a higher symmetry, $U(1)$, than required by the crystal structure, with the consequence that the magnon dispersions $\omega_m(\mathbf{k})$ are independent of the in-plane moment's angle $\phi$ and have the rotational symmetry of the cobalt structural arrangement, which is $\bar{3}m$. This rotational symmetry implies also that all magnetic domains will have identical dispersion relations, which justifies Supplementary Equation~(\ref{eq:total_intensity}) for a multi magnetic domain sample. Each of the two intensity terms in that equation, separately has the same rotational point group symmetry $\bar{3}m$. In the case of flat plane honeycombs ($\epsilon=0$), each of those two intensity terms also has the translational periodicity of the structural Brillouin zone, however the buckling of the layers breaks the translational periodicity of the intensity along the $L$-direction as it introduces an intensity modulation factor due to interference scattering from the two cobalt sites in the same honeycomb layer being offset along $z$, this intensity modulation term has a long period, $1/\epsilon$ along $L$.

Finally, we note that following the general arguments presented in \cite{shivam2017neutron}, the dynamical structure factor for small in-plane wavevector displacements $\delta\mathbf{k}$ away from the nodal points will have the azimuthal angular dependence $1\pm \cos \varphi_{\mathbf{k}}$, where the phase angle $\varphi_{\mathbf{k}}$ is related to the azimuthal angle $\alpha_{\delta\mathbf{k}}$ of $\delta \mathbf{k}$ via Supplementary Equations~(\ref{eq:varphi_k1})-(\ref{eq:varphi_k3}). This leads to a two-fold intensity modulation in azimuthal scans, in anti-phase between the two touching bands, as observed by the data in Fig.~2c). The relation between $\varphi_{\mathbf{k}}$ and $\alpha_{\delta\mathbf{k}}$ varies between neighboring K-points following a $\bar{3}m$ symmetry. This is illustrated in Fig.~2d) by the radial thick magenta arrows which show the directions $\hat{\mathbf{n}}$ {away} from the nearby K-points along which the intensity in the the top band is maximal in azimuthal scans at $L=0$. At finite $L$, due to scattering interference from the two cobalt sites offset along $z$, the $\hat{\mathbf{n}}$ vectors rotate in-plane by an angle $2\pi \epsilon L$, in opposite senses for adjacent K-points, following a $\bar{3}m$ symmetry. This $L$-dependence provides a natural explanation for the observed angular intensity dependence in the azimuthal scan in Fig.~2c) around the (2/3,2/3) Dirac node, with maximum intensity in the top band observed near $\alpha_{\delta_{\mathbf k}}=-80(3)^{\circ}$, compared to $-81(1)^{\circ}$ calculated based on Supplementary Equation~(\ref{eq:varphi_k2}) averaged for the appropriate $L$-integration range of the scan. The $L$-dependence of the azimuthal scans is illustrated in Supplementary Figure~\ref{fig:azimuthal_scans}.

\begin{figure}[tbph!]
    \includegraphics[width=\columnwidth]{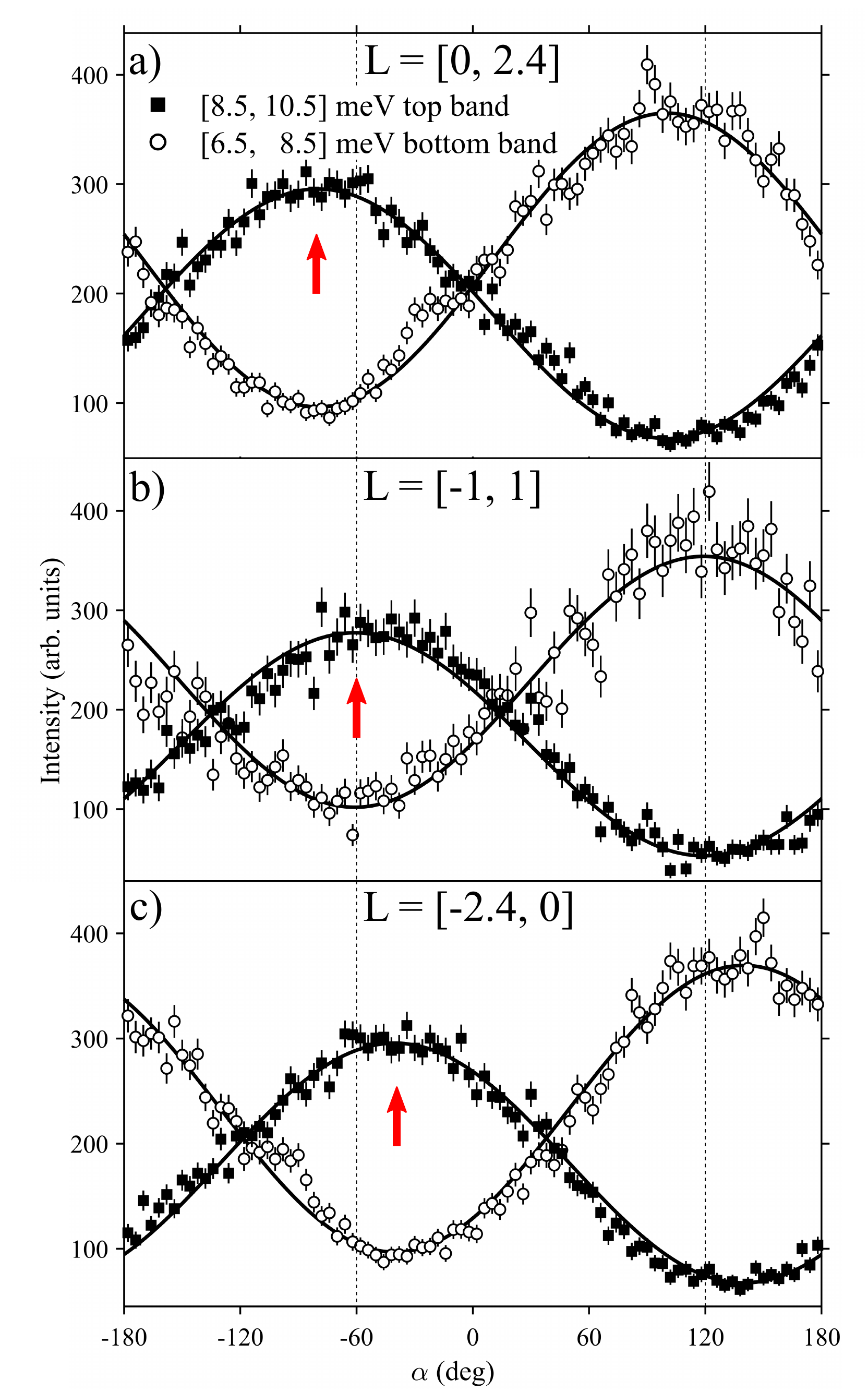}
    \figcaption{Azimuthal intensity dependence for different $L$ values}{Azimuthal intensity scans as per Fig.~2c), for increasing $L$-values from panel c) to a), fitted to cosinusoidal forms (solid lines), with error bars representing one standard deviation. Thin dashed vertical lines at $\alpha=-60$ and $120^\circ$ show the expected intensity extremal positions for flat honeycombs ($\epsilon=0$). Red thick arrows show where the top band (filled symbols) would be expected to be maximal for the case of buckled honeycombs [$-\pi/3-2\pi\epsilon L$ as per Supplementary Equation~(\ref{eq:varphi_k2}) appropriately averaged for the $L$-range of the scans].}
    \label{fig:azimuthal_scans}
\end{figure}

\section{Inelastic neutron scattering experiments and fitting of magnon dispersions to an XXZ$\Delta$ model}
\label{sec:fits}

\subsection{Experimental Details}
Here we provide details of the INS experiments \cite{exp_doi} to probe the spin dynamics, performed using the MERLIN direct-geometry time-of-flight spectrometer\cite{MERLIN} at ISIS. The sample consisted of two co-aligned single crystals of CoTiO$_3$ (total mass 5.8~g) grown via the floating zone method, mounted with the ($hk0$) scattering plane horizontal. Full Horace maps of the inelastic scattering were collected at a base temperature of 8~K (cooling was provided by a closed-cycle refrigerator) by rotating the sample around the vertical axis in steps of $0.5^\circ$ over an angular range of 120$^\circ$, with each step counted for 9~mins at an average proton current of 170~$\mu$A. The temperature-dependence of the inelastic scattering up to 300~K was measured for one representative sample orientation. The spectrometer was operated in repetition rate multiplication (RRM) mode to collect the inelastic scattering simultaneously for monochromatic incident neutrons with energies $E_\mathrm{i}$=9.6, 18 and 45~meV, with energy resolutions on the elastic line of 0.36(2), 0.72(2) and 2.7(1) (full width at half maximum, FWHM), respectively. Additional measurements were collected with $E_\mathrm{i}$=83~meV to probe transitions to higher crystal field levels. The elastic line in all runs was centred on zero energy transfer to better than 0.25\% of $E_{\rm i}$. The time-of-flight neutron data were processed using the \textsc{mantid}\cite{Arnold2014} and \textsc{horace}\cite{Ewings2016} data analysis packages.
\begin{figure}[!tbph]
    \includegraphics[width=\columnwidth]{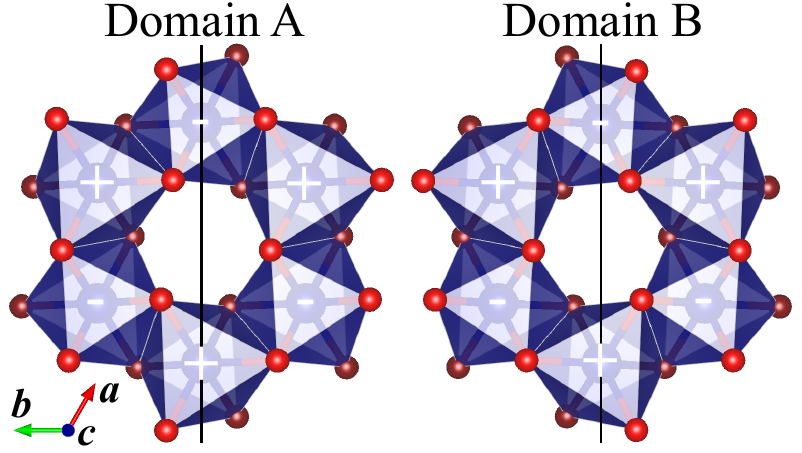}
    \figcaption{Structural domains}{Projection of a single honeycomb layer onto the $ab$ plane for the reference structural domain A (left, Supplementary Table~\ref{TAB::crystal_structure}) and domain B (right) related by a (101) mirror plane (solid vertical line). $\pm$ signs for the cobalt ions (large blue balls) and red/brown color for the oxygens (small balls) indicate positions above/below the plane. The structural and magnetic Brillouin zones in Supplementary Figure~\ref{fig:BZ} apply to both domain A and B.}
    \label{fig:twins}
\end{figure}

\subsection{Structural domains and data symmetrization}
Careful examination of the observed diffraction signal (integrated elastic line) showed that the sample contained two almost equal-weight structural twins, related by a 2-fold rotation around the (110) axis, or equivalently mirrored with respect to the ($101$) plane. Under this transformation the Co and Ti positions are unchanged, only the oxygen positions are affected as illustrated in Supplementary Figure~\ref{fig:twins}. With reference to the crystal structure in Supplementary Table~\ref{TAB::crystal_structure} called structural domain A with oxygens at ($x_{\rm O},y_{\rm O},z_{\rm O}$), the mirrored domain B has oxygens located at positions equivalent to ($y_{\rm O},x_{\rm O},-z_{\rm O}$). Both structural twins scatter into the same Bragg positions at $\bm{\uptau}=(h,k,l)$ with $-h+k+l=3n$, $n$ integer, and are most easily distinguished by analysing the diffraction signal at ($1\bar{3}1$) (and equivalent positions by $\bar{3}$ symmetry) where interference scattering from Co, Ti and O leads to a strong intensity for domain A, but near cancellation for domain B, and viceversa for reflection ($\bar{2}31$). The observed diffraction pattern showed almost equal intensities for those two reference reflections, so we conclude that the sample contained equal amounts of the A and B domains, which would imply a $\bar{3}m$ point group symmetry for the (diffraction and inelastic) signal. Indeed the inelastic intensity showed to a very good degree $\bar{3}m$ symmetry with mirrors at ($h0l$) and to enhance the counting statistics the wavevector transfers $\mathbf{k}$ of the pixels in the four-dimensional Horace scans were remapped using symmetry operations of the above point group to a minimal 60$^{\circ}$ sector in the $hk$ plane and $l>0$.

We note that magnetic ordering with moments in plane breaks the $3$-fold rotation, so the dispersion relations and dynamical structure factor for a {single} magnetic domain would in principle have point group $\bar{1}$ (a minimal model that exhibits this lower symmetry of its magnetic spectrum is the XXZ Hamiltonian augmented by finite diagonal exchange anisotropy $\eta\neq 0$ discussed in~\ref{sec:eta}. However, averaging over three equal-weight magnetic domains with moments rotated by $\pm120^{\circ}$ as expected in a macroscopic sample, would restore the 3-fold symmetry for the intensity pattern. This combined with the A and B structural domains then would restore the higher point group symmetry $\bar{3}m$ for the intensity pattern, justifying the pixel averaging used.

\begin{figure}[!btph]
    \includegraphics[width=\columnwidth,keepaspectratio]{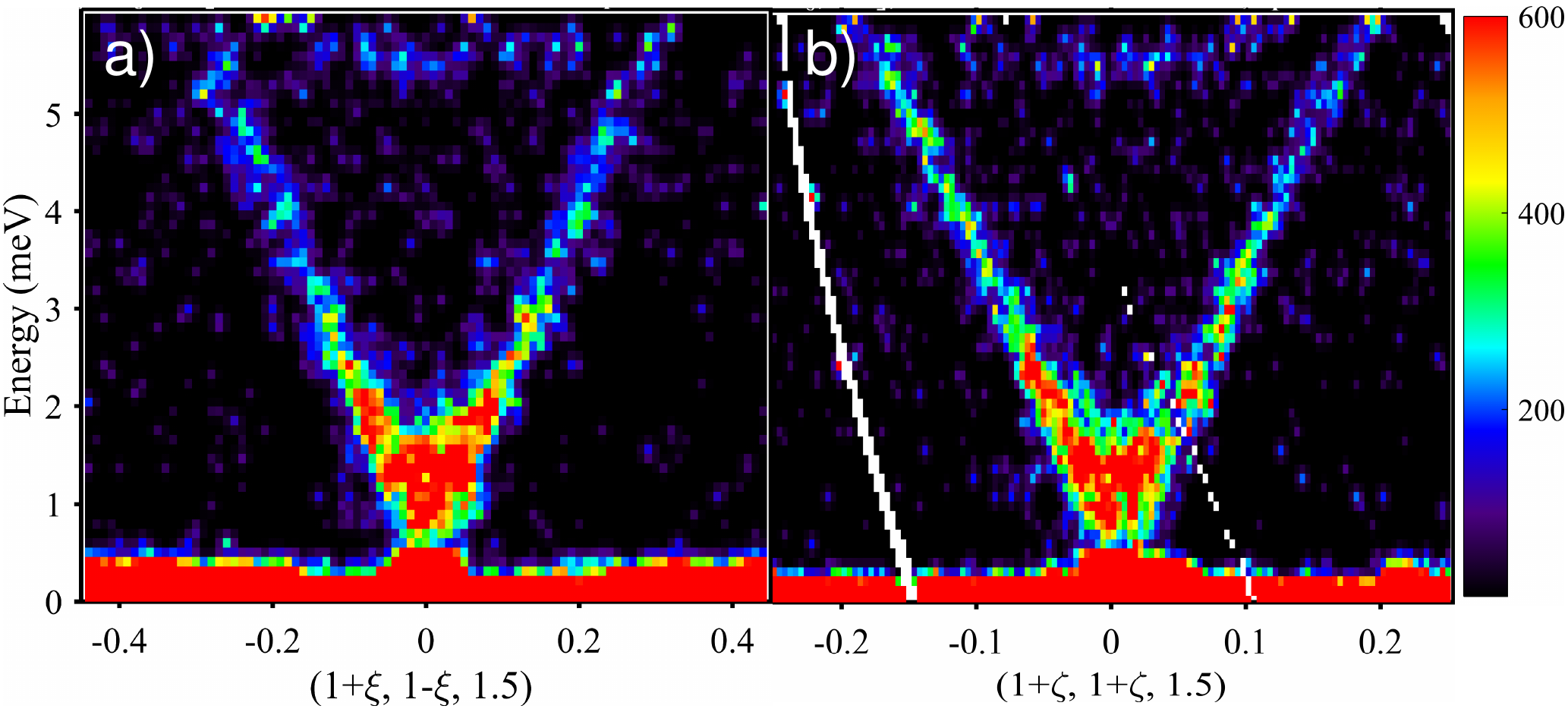}
    \figcaption{Spectral gap and low-energy in-plane dispersions}{a,b) INS intensity along orthogonal in-plane wavevector directions through the ($1,1,\frac{3}{2}$) magnetic Bragg peak, showing a strong V-shaped magnetic signal above a gap, with a clear separation from the elastic line. Intensities are averaged for $L=[1.375,1.625]$ and a transverse in-plane wavevector range of $\pm0.062$ and $\pm0.072$~\AA$^{-1}$, respectively. The measurement configuration was as in Fig.~1c). The colour bar indicates the scattering intensity in arbitrary units on a linear scale.\label{fig:gap_1_1_1.5}}
\end{figure}

\subsection{Parameterization of magnon dispersions by an XXZ$\Delta$ model}
\label{sec:parameters}

\begin{figure}[!tbph]
    \includegraphics[width=\columnwidth]{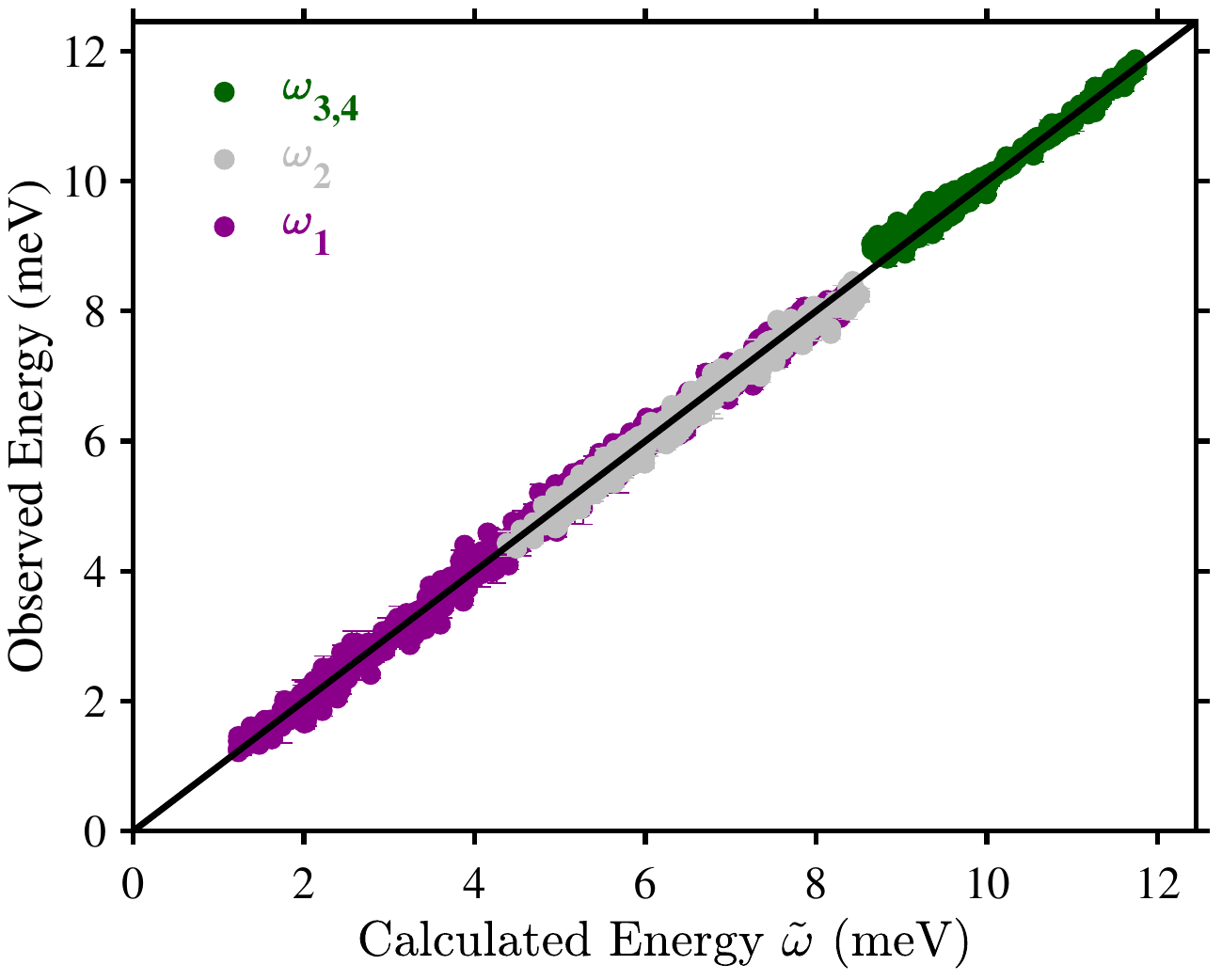}
    \figcaption{Agreement between observed and calculated magnon energies}{Magenta, grey, and green symbols correspond to magnon bands indexed in order of increasing energy. A single symbol is used for the un-resolved $\tilde{\omega}_{3,4}$ modes (corresponding to their weighted average by the dynamical structure factor including the neutron polarization factor), the error bars representing one standard deviation. The solid line shows the 1:1 agreement.}
    \label{fig:omega_obs_vs_calc}
\end{figure}

The XXZ Hamiltonian discussed in the previous section has a $U(1)$ symmetry, however the crystal structure has only discrete rotational symmetry and moreover the observed magnon spectrum is clearly gapped, as shown in Fig.~1c) and Supplementary Figure~\ref{fig:gap_1_1_1.5}, indicating not a continuous, but a discrete set of allowed $\phi$ values. To account for the presence of a spectral gap at this stage we introduce a phenomenological gap parameter $\Delta$ and assume that the effect of the symmetry-breaking interactions that generate this gap can be accounted for, in a first approximation, by simply adding this gap in quadrature to the analytical XXZ dispersions, i.e. the experimental dispersion points are compared with $\tilde{\omega}_m(\mathbf{k})=\sqrt{\Delta^2+\omega^2_m(\mathbf{k})}$, where $m=1$ to 4 labels the four magnon modes at a given wavevector ${\mathbf{k}}$ in order of increasing energy. Empirical $(h,k,l,E,m)$ dispersion points (where $E$ is energy) were extracted from fitting Gaussian peaks to constant-wavevector and/or -energy scans through the four-dimensional inelastic neutron scattering data. Supplementary Figure~\ref{fig:omega_obs_vs_calc} illustrates the level of agreement that can be obtained when comparing nearly $2,000$ empirical dispersion points with dispersions of the XXZ$\Delta$ model for a representative set of exchange parameters below, all in meV,
\begin{align}
J^{\perp}_1 & = -6.36 \hspace{-0.3cm} & J^{z}_1 & = 1.97 \nonumber \\
J^{\perp}_2 & = -0.33  & J^{z}_2 & = 0.30 \nonumber \\
J^{\perp}_3 & = 0.78 & J^{z}_3 & = 0.15 \nonumber  \\
J^{\perp}_4 & = 0.11  & J^{z}_4 & = 0.32 \nonumber \\
J^{\perp}_5 & = -0.39  & J^{z}_5 & = 0.20 \nonumber \\
J^{\perp}_6 & = 0.79  & J^{z}_6 & = 0.68 \label{eq:bestfit}
\end{align}
and $\tilde{\Delta} =1.23(7)$~meV, where $-/+$ve signs for the exchanges mean FM/AFM coupling.

In the above we have used the symbol $\tilde{\Delta}$ to indicate that the gap is overestimated through this analysis. There is a net shift of the scattering weight towards higher energies originating from the finite wavevector integration around the lowest energy mode because the integration range captures intensity from the mode away from the minimum, thus shifting the average upwards. We account for this effect in a first approximation by assuming all exchange parameters fixed as per Supplementary Equation~(\ref{eq:bestfit}) and calculating the expected scattering for the slice in Fig.~1c), which is most sensitive to the gap, allowing for a variable $\Delta$ in the fit. We include the full wavevector averaging in the transverse (highly-dispersive, in-plane) direction as in the data slice, and optimise $\Delta$ to get the best agreement between scans through the data and simulation, as shown in Supplementary Figure~\ref{fig:scan_en_gap}. This gave $\Delta=1.0(1)$~meV, renormalized down from $\tilde{\Delta}$. Fig.~1c) illustrates the effect of the wavevector averaging in the transverse direction: near the bottom of the dispersion there is a systematic upwards energy shift between the position of the dominant scattering weight and the dispersion energy (solid line) at the nominal wavevector positions. The position of the scattering weight in the data and simulations agree once wavevector averaging is included, compare Fig.~1c) and d).

\begin{figure}[!tbph]
    \includegraphics[width=\columnwidth]{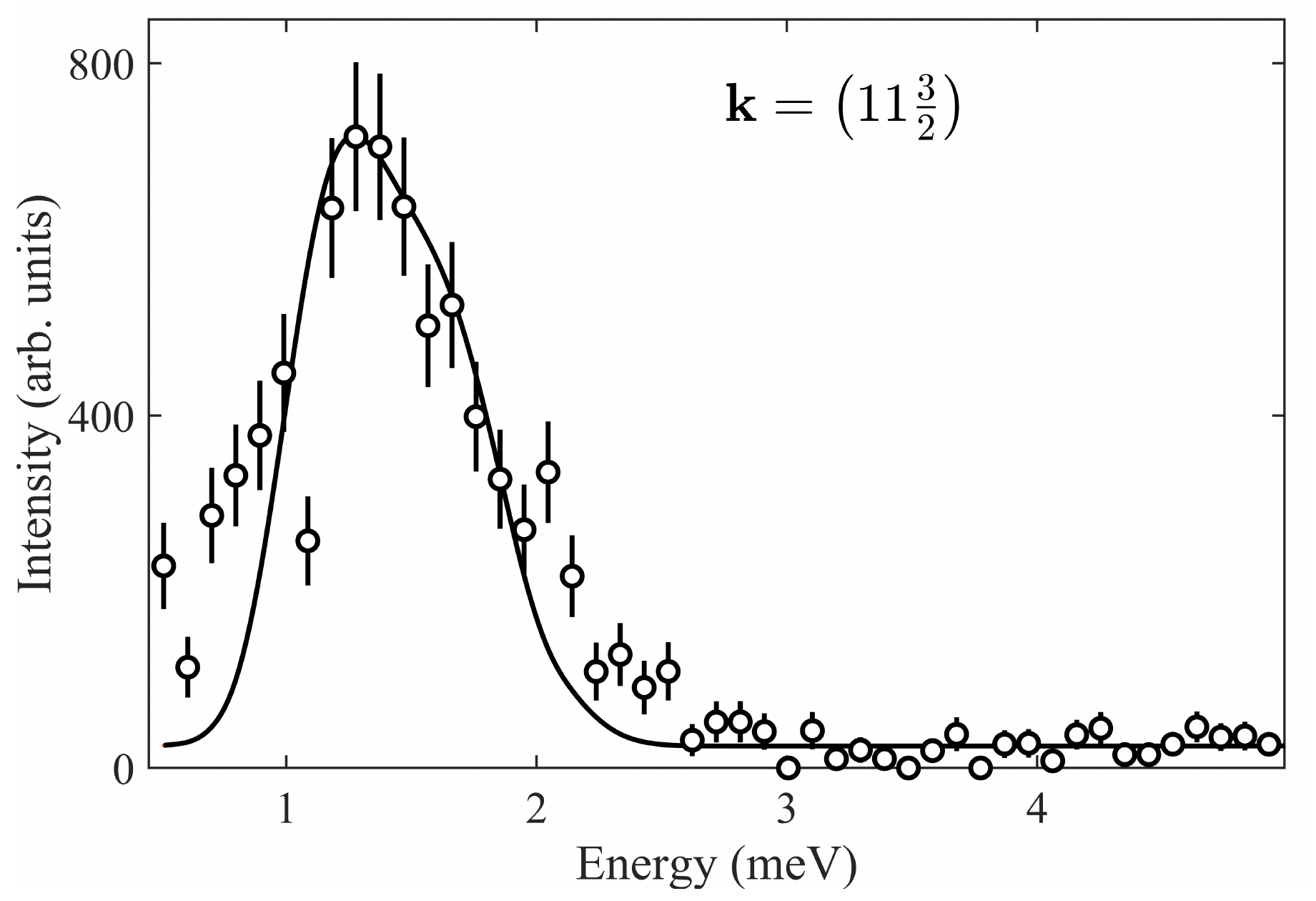}
    \figcaption{Energy scan above the $\left(11\frac{3}{2}\right)$ magnetic Bragg peak}{ The white circles denote the inelastic neutron scattering intensity, averaged for $L=[1.4,1.6]$ and transverse in-plane wavevector ranges of $\pm0.1$~\AA$^{-1}$ in the $(1,0,0)$ and $(1/2,-1,0)$ directions, with error bars representing one standard deviation. The solid line is the fit described in the text to determine the gap value.}
    \label{fig:scan_en_gap}
\end{figure}

The parameter set in Supplementary Equation~(\ref{eq:bestfit}) provides a quantitative account of the observed magnon dispersions (up to the fine structure around the K points to be discussed later) and qualitatively of the intensities as well, compare Fig.~1a) and b). However, the number of Hamiltonian parameters considered is large and we have found that some parameters are strongly correlated in their effects on the dispersions. Therefore, the dispersions alone are not constraining enough to uniquely determine the values of all the individual exchange parameters. For example, moving in parameter space away from the set in Supplementary Equation~(\ref{eq:bestfit}) by varying the value of $J_1^z$, fixing $J_1^{\perp}=-J_1^z-4.4$~meV and optimising all the remaining exchange parameters results in a very small relative change in $\chi^2=\sum_i\vert\omega_{\rm obs}(i)-\omega_{\rm calc}(i)\vert^2$, of only a few percent when $J_1^z$ is reduced all the way to $0$, such a small variation in $\chi^2$ is at the level that the change in the agreement with the data is almost indistinguishable. Here $\omega_{\rm obs}(i)/\omega_{\rm calc}(i)$ is the observed/calculated energy for the $i$th dispersion point. With the exception of $J_1^{\perp}$, any one of the other $11$ exchange parameters can be set equal to $0$ and optimising the rest of the parameters gives a comparable agreement to that in Supplementary Figure~\ref{fig:omega_obs_vs_calc}. Therefore, more constraints are needed to uniquely identify the values of the individual exchange parameters and we therefore regard the set in Supplementary Equation~(\ref{eq:bestfit}) as representative of the best agreement that can be obtained with the measured dispersions, and in the following we focus on the key features of the measured dispersions.

\begin{figure}[!tbph]
    \includegraphics[width=\columnwidth]{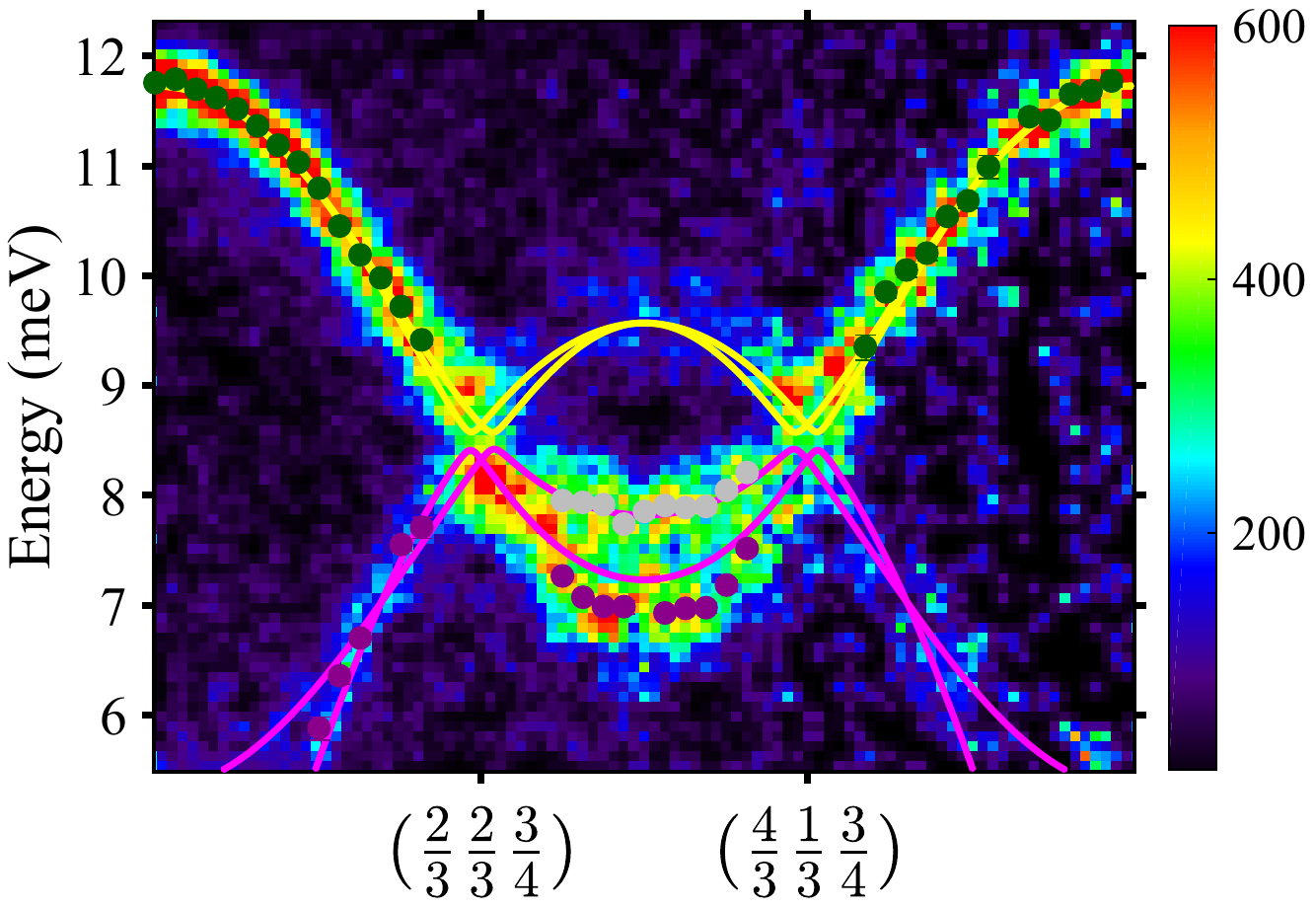}
    \figcaption{Magnon spectrum along a direction passing through two adjacent K points at fixed $L=3/4$}{Along this path a clear splitting of the two lower modes (purple and gray solid dots) is observed. Overplotted lines are the dispersions for the model in Supplementary Equation~(\ref{eq:bestfit}) with anisotropic XXZ inter-layer couplings. Heisenberg inter-layer couplings as in \cite{Yuan} would give almost degenerate lower modes in this region (not shown). Solid dots are empirical dispersion points (colour indicates the mode index in order of increasing energy as per the legend in Supplementary Figure~\ref{fig:omega_obs_vs_calc}. The upper (yellow) and lower (magenta) pair of modes do not touch along this wavevector path. The nodal lines are present for these parameters, but are displaced away from the two K points along directions that make a finite angle with the plotted wavevector path. The colour bar indicates scattering intensity in arbitrary units on a linear scale.}
    \label{fig:w12_splitting}
\end{figure}

Whilst the overall dispersion trends are in general in agreement with the minimal model parametrization proposed in \cite{Yuan}, our higher-resolution INS data reveal additional dispersion modulations and fine structure (splitting of modes) that require additional couplings and anisotropies. For example Supplementary Figure~\ref{fig:w12_splitting} shows a clear splitting between the two lower modes (gray and magenta solid dots) in the region in-between the two labelled K-points, those two lower modes would be almost degenerate in this region in the parametrization used in \cite{Yuan}. The model in Supplementary Equation~(\ref{eq:bestfit}) (lines) predicts a substantial splitting between those modes, although it still underestimates the magnitude of the splitting seen experimentally. The agreement with the data can be improved by adding a bond-dependent anisotropic exchange $\eta$, which, we argue, is physically responsible also for generating the finite spectral gap above the magnetic Bragg peaks, to be discussed in the following two sections.

For completeness, we note that the dipolar couplings are negligible compared to the scale of the above exchanges, i.e. the dipolar energy scale is $\mu_0 \mu^2/4\pi a_0^3 = 0.018$~meV, where $\mu$ is the ordered magnetic moment per site in the ground state (3~$\mu_{\rm B}$) and $a_0\approx 3$~\AA\ is the nearest-neighbor Co-Co distance.

For the above $J_1-J_6$ XXZ model the reduction of the ordered moment due to zero point fluctuations $\Delta S$ within linear spin wave theory is merely $0.021$  compared to about $0.1$ for the nearest-neighbor XXZ model (with only $J^{\perp}_1$ and $J^z_1$ nonzero).

\section{Symmetry and Anisotropic Exchange}
\label{sec:exchange}

In~\ref{sec:fits} we showed that the spin waves computed from an XXZ model capture most of the features of the experimental neutron scattering data very well. However, it was necessary to include a phenomenological gap, which is absent in the XXZ model. We now begin to address the microscopic origin of the gap and we do this in two parts. The first is to recognize that spin-orbit coupling can lead to anisotropic exchanges beyond the XXZ model which arises from a projection of a Heisenberg model down to the spin-orbital ground state Kramers doublet. In this section we discuss the possible anisotropic exchange from a phenomenological point of view. These additional anisotropies in the effective spin one-half description generally lead to quantum order-by-disorder as outlined in~\ref{sec:gap} in which we also compute the order-by-disorder gap exactly to leading order in $1/S$ for particular anisotropic exchange couplings in order to estimate the required magnitude of the anisotropies. In the second part,~\ref{sec:MFTFW}, we consider the microscopic origin of the anisotropic exchange and directly compute the spectral gap through a flavor wave mean field theory.

We start with the nearest neighbor bonds. The inversion symmetry at the midpoint of the bond forces this exchange to be symmetric. In other words, six independent coupling terms are allowed, which, of course, includes the XXZ form. Once the anisotropic exchange is defined for one of the bonds chosen as ``reference'', the exchange on all other bonds of the same type in the full crystal lattice is obtained using crystal lattice symmetry operations such as the 3-fold rotation at the cobalt sites or primitive lattice translations. It is insightful to consider two (bond-dependent) reference frames to define the anisotropic exchange. The global $xyz$ frame used so far is a natural frame for the vertical diamond-shaded bond in Supplementary Figure~\ref{fig:FW}d) as the $y$ axis is along the bond direction and $z$ is along $c$. In this frame, the exchange matrix is
\begin{equation}
{\cal{J}}_1=\left(\begin{array}{ccc} J^{xx} & J^{xy} & J^{xz} \\ J^{xy} & J^{yy} & J^{yz} \\ J^{xz} & J^{yz} & J^{zz}  \end{array}  \right).
\label{eq:exchange}
\end{equation}

It is  also convenient to write the exchange in a Cartesian $111$ frame (also illustrated in Supplementary Figure~\ref{fig:FW}d) and denoted by SansSerif symbols $\fs{x}\fs{y}\fs{z}$), where the axes have the property that the hexagonal $c$-axis is along the symmetric combination $(\hat{\bfs{x}}+\hat{\bfs{y}}+\hat{\bfs{z}})/\sqrt{3}$. This frame transforms simply under the lattice point group symmetry $3$ and is natural from the point of view of the underlying superexchange mechanism. In this frame the exchange matrix for the same bond has the form
\begin{equation}
\left( \begin{array}{ccc} J - J_T  & \Gamma'_1  & \Gamma  \\ \Gamma'_1 & J + K  & \Gamma'_2  \\ \Gamma  & \Gamma'_2  & J + J_T   \end{array}  \right)
\label{eq:111exchange}
\end{equation}
where three of the six independent terms have a natural interpretation, i.e. $J$ is the isotropic (Heisenberg) exchange, $K$ is a bond-Ising, or Kitaev-like, coupling, and $\Gamma$ is an off-diagonal symmetric exchange for components in the plane orthogonal to the Kitaev axis. It is possible to justify microscopically the origin of those three couplings using a minimal superexchange model for the pair of $90^{\circ}$ Co-O-Co bonds (\ref{sec:MFTFW}).

The transformation that converts between the exchange terms between the two coordinate frames is
\begin{align}
& \left( \begin{array}{cccccc} J^{xx} & J^{yy} & J^{zz} & J^{xy} & J^{xz} & J^{yz}   \end{array} \right)^{\rm T} = \nonumber \\
& \left( \begin{array}{cccccc}
1 & \frac{2}{3} & \frac{1}{3} & 0 & -\frac{2}{3} & -\frac{2}{3} \\
1 & 0 & -1 & 0 & 0 & 0 \\
1 & \frac{1}{3} & \frac{2}{3} & 0 & \frac{2}{3} & \frac{2}{3} \\
0 & 0 & 0 & -\frac{1}{\sqrt{3}} & -\frac{1}{\sqrt{3}} & \frac{1}{\sqrt{3}} \\
0 & -\frac{\sqrt{2}}{3} & \frac{\sqrt{2}}{3} & 0 & -\frac{1}{3\sqrt{2}} & -\frac{1}{3\sqrt{2}} \\
0 & 0 & 0 & -\frac{2}{\sqrt{6}} & \frac{1}{\sqrt{6}} & -\frac{1}{\sqrt{6}}
\end{array}  \right)
\left( \begin{array}{c} J \\ K \\ \Gamma \\ J_{T} \\ \Gamma'_1 \\ \Gamma'_2   \end{array}\right)
\end{align}

In the following we shall consider anisotropic exchange beyond nearest neighbor in particular to address the effect of these couplings on the fine structure of the spin wave spectrum at the Dirac nodes. For 2nd nearest neighbors, symmetry is highly constraining and restricts the exchange to the XXZ form expressed in the $xyz$ frame as
\begin{equation}
\left( \begin{array}{ccc} J^{\perp}_{2} & 0 & 0 \\ 0 & J^{\perp}_{2} & 0 \\ 0 & 0 & J_{2}^z   \end{array} \right).
\end{equation}
The 3rd nearest neighbor bonds are all constrained by lattice symmetries once a single such bond is fixed. However, a single bond has no symmetry on its own and nine independent exchange couplings are allowed including three diagonal couplings, three symmetric off-diagonal couplings and three Dzyaloshinskii-Moriya couplings. The same is true for $4$th nearest neighbor couplings, while $5$th and $6$th nearest neighbor exchange couplings are constrained not to have antisymmetric exchange (as is the case for nearest neighbors) so they have six couplings each.\\

\subsection{Bond-dependent Anisotropic Exchange $\eta=J^{yy}-J^{xx}$}
\label{sec:eta}
In the next section, we shall find it useful to consider the anisotropic exchange component on the nearest neighbor bond $\eta\equiv J^{yy}-J^{xx}$. One can compute the spin wave spectrum in the presence of this coupling within the local frame of~\ref{sec:sw} with the $B$ and $D$ parameters in Supplementary Equation~(\ref{eq:ABCD}) acquiring the additive terms $B'$ and $D'$, respectively, with
\begin{widetext}
\begin{eqnarray}
B'& = & -\eta \,\frac{S}{8} \left[\left(\cos 2\phi + \sqrt{3} \sin 2\phi  \right) e^{i\beta_1} -2\cos 2\phi \: e^{i\beta_2} + \left(\cos 2\phi - \sqrt{3} \sin 2\phi  \right)e^{i\beta_3}\right]\nonumber \\
D' & = & B'^\star
\label{eq:BD_eta}
\end{eqnarray}
\end{widetext}
where $\beta_i = \mathbf{k}\cdot \mathbf{R}_{1i}$ with the vectors $\mathbf{R}_{1i}$, $i=1-3$ given in Supplementary Table~\ref{tab:gam}. Note that for finite $\eta$ the spectrum depends on the spin orientation angle $\phi$ in the easy plane and we will show in the subsequent section that this feature is responsible for selecting discrete orientations via quantum order-by-disorder, $\phi=n\pi/3$ for $\eta>0$, and $\phi=\pi/2+n\pi/3$ for $\eta<0$, $n$ integer.

\begin{figure}[!tbph]
    \includegraphics[width=\columnwidth]{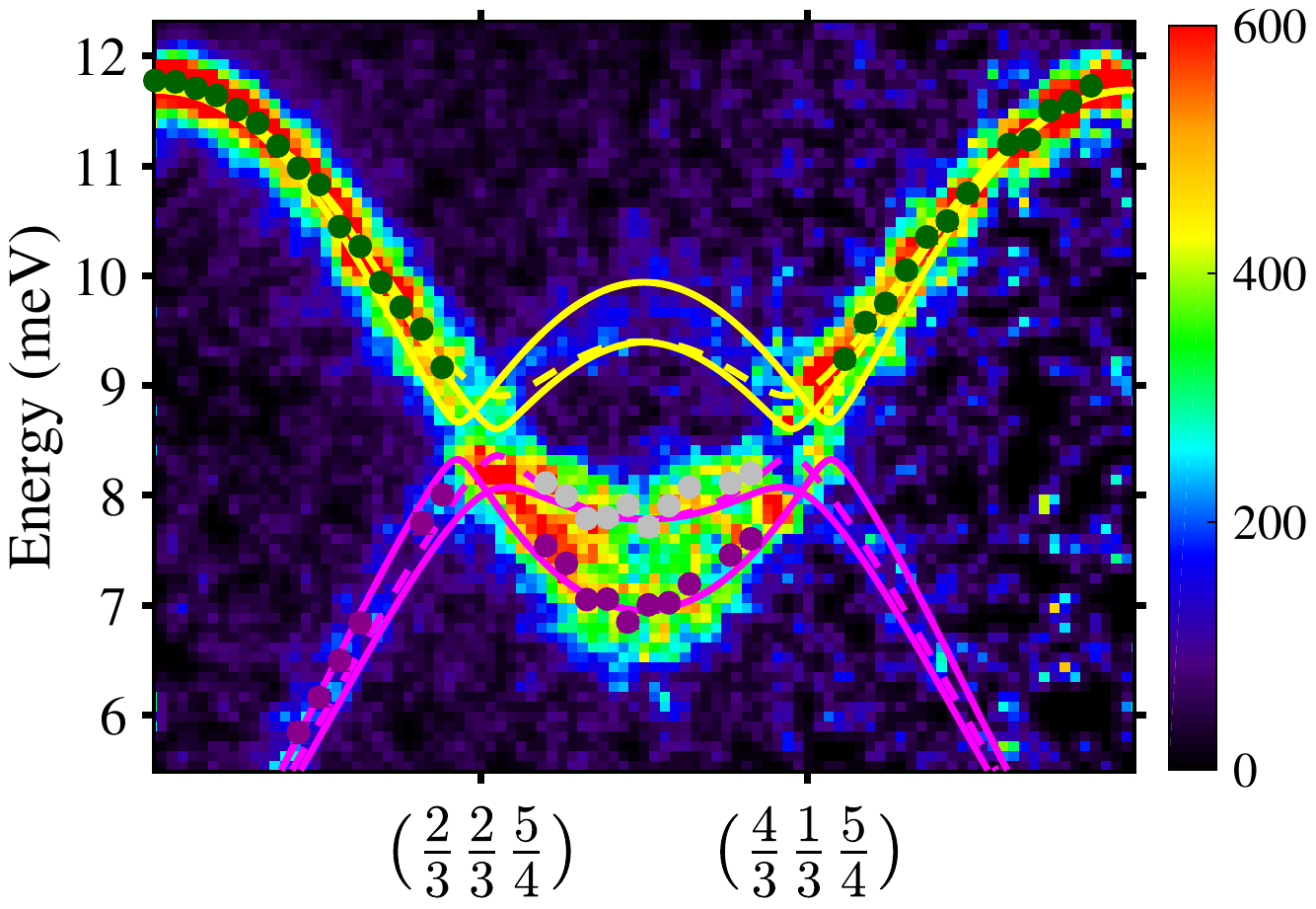}
    \figcaption{Magnon spectrum along a direction confined to the magnetic Brillouin zone boundary passing through two K points}{The slice direction is equivalent to the dotted blue line segments in Supplementary Figure~\ref{fig:BZ}b). Note the clear splitting of the two lower modes (purple and gray symbols) between the two labelled K-points. The overplotted lines are dispersions of the XXZ$\eta$ model with parameters in Supplementary Equation~(\ref{eq:bestfit}) and $\eta=-1.7$~meV, solid/dashed for magnetic domains with $\phi=\pi/2$ and $\pi/2+2\pi/3$, respectively (the other domains' dispersions overlap with the ones shown). Within an XXZ model the two lower modes are degenerate. The colour bar indicates scattering intensity in arbitrary units on a linear scale.}
    \label{fig:w12_splitting_mzb}
\end{figure}

For small $\eta$, significant changes to the spectrum occur only near the magnetic Brillouin zone boundary where the dispersion surfaces are distorted/shifted along a direction that correlates with the moments' orientation in the ground state, leading to an overall $\bar{1}$ point group symmetry, the same as the magnetic structure. Magnetic domains obtained by $\pm 120^{\circ}$ rotation therefore can have distinct dispersion relations and this can provide a physical mechanism to account for several features in the data, in particular the splitting between the two lower modes in Supplementary Figure~\ref{fig:w12_splitting_mzb}, where solid/dashed lines correspond to magnetic domains rotated by $120^{\circ}$. This path was chosen because the observed splitting cannot be captured within an XXZ model, where the combined rotational ($\bar{3}m$) and translational symmetry of the spectrum requires the two lower modes to be degenerate. This can be seen as follows: when mapped to the magnetic Brillouin zone in Supplementary Figure~\ref{fig:BZ}b), this path is equivalent to one of the main diagonals (blue dotted line) of the hexagonal top face (the parallel blue dotted segment on one of the side faces is simply an extension of the top segment into the next zone, mapped back into the first magnetic Brillouin zone). The top diagonal is mapped by rotational $\bar{3}m$ symmetry operations, i.e. $2 \parallel (\frac{\bar{1}}{2}10)$ followed by $3\parallel (001)$, onto the same diagonal on the bottom face, translated by $-\mathbf{Q}$. This means that for any wavevector ${\mathbf{k}}$ on the top diagonal $\omega_{-}(\mathbf{k})=\omega_{-}(\mathbf{k}-\mathbf{Q})$, implying that in the global frame $\omega_1(\mathbf{k})=\omega_2(\mathbf{k})$, i.e. the lower two modes are degenerate along this path. Furthermore, the same rotational symmetry implies that all magnetic domains have the same dispersion relations, so the observed splitting cannot be explained within an XXZ model. $\eta\neq0$ breaks this symmetry requirement allowing magnetic domains with spins rotated by 120$^{\circ}$ to have non-overlapping dispersions, thus providing a natural mechanism to explain the observed splitting.

\begin{figure}[!tbph]
    \includegraphics[width=\columnwidth]{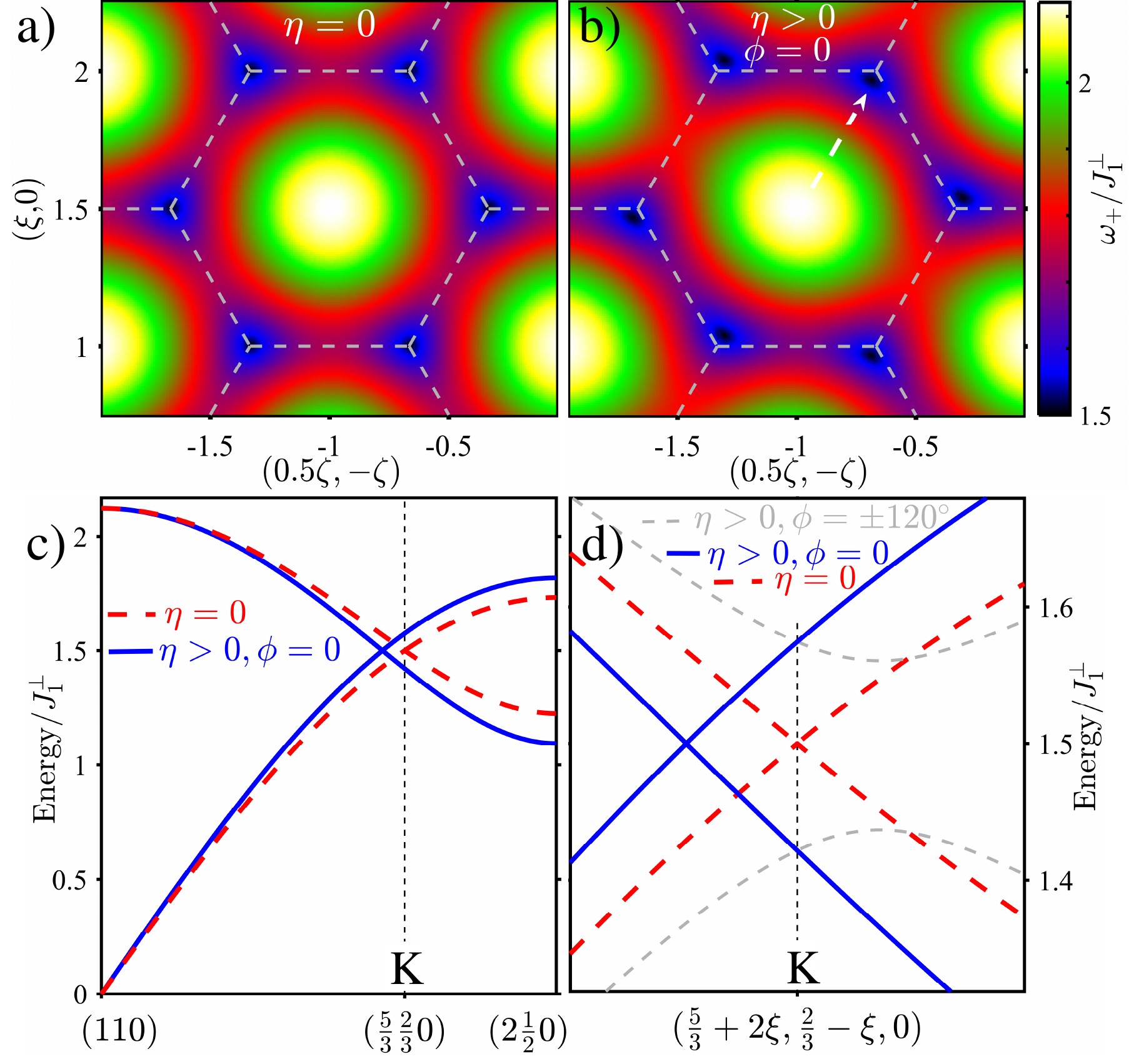}
    \figcaption{Shift of the Dirac node in momentum space for finite $\eta$}{a-b) Contour plot of the top magnon band $\omega_+$ for a single honeycomb with only the dominant exchange $J_1^{\perp}<0$ and in a) $\eta=0$, and in b) $\eta>0$ and moments' orientation in the ground state $\phi=0$ (thick dashed arrow in b). c) Dispersion relation along the dashed arrow direction in b) for the same parameters as in a) (dashed red lines) and b) (solid blue lines), d) shows a zoomed-in version near the K-point to emphasize the displacement of the Dirac node away from K for finite $\eta$. Dashed gray lines show the magnon dispersions of magnetic domains rotated by $\pm 120^\circ$. An energy scan at K would display a double-peak structure with a separation of $2\hbar v \, \delta \kappa $, where $v$ is the Dirac velocity and $\delta \kappa$ is the in-plane displacement of the Dirac node away from K.}
    \label{fig:XYeta}
\end{figure}

The $\eta$ interaction can also account for the fine structure in the energy scan centred at K-points in Fig.~3d), which cannot be explained by the XXZ model (dashed red line). To illustrate this, it is helpful to consider first the effects of adding a finite $\eta$ to a single, isolated honeycomb with only the dominant nearest-neighbor exchange $J_1^{\perp}<0$ with all other exchanges set to zero. A contour plot of the upper magnon band in this case is shown in Supplementary Figure~\ref{fig:XYeta}a), the Dirac cones are centred at the K-point corners of the 2D hexagonal Brillouin zone with the magnon band displaying a $\bar{3}m$ point group symmetry around the zone centre. Supplementary Figure~\ref{fig:XYeta}b) shows the corresponding plot for small $\eta>0$ and the moment's orientation in the ground state along the $\phi=0$ direction, indicated by the dashed white arrow. The magnon dispersion surface has now $\bar{1}$ symmetry, being distorted along the ordered in-plane spin direction in the ground state, with the Dirac nodes moving along this direction (two nodes move in and four nodes move out of the Brillouin zone hexagon). The corresponding dispersion plots for the above two cases are shown in Supplementary Figure~\ref{fig:XYeta}c), dashed red lines for case a), and solid blue lines for case b), with a zoomed-in version near the K point shown in Supplementary Figure~\ref{fig:XYeta}d), note the Dirac nodal points have moved away from K-points, so a scan at K would show a double-peak structure, as in Fig.~3d).

For the full XXZ model the nodal points occur in the form of double helix nodal lines that precess around K-points, adding a finite $\eta$ has the effect of shifting in-plane the centre of precession away from K-points along a direction parallel to the moments' direction in the ground state if $\eta>0$ and $\phi=n\pi/3$, and transverse to the moments' direction if $\eta<0$ and $\phi=\pi/2+n\pi/3$. In both cases, the precession centres of the double-helix nodal lines in Figure~3b) are displaced away from the K-points by an in-plane wavevector of magnitude $\delta \kappa$, with the consequence that at K-points there is an energy gap $\simeq 2\hbar v \, \delta \kappa $ between the mean of the top two bands and the mean of the bottom two bands, where $v$ is the Dirac velocity, similar to the simplified 2D case illustrated in Supplementary Figure~\ref{fig:XYeta}d). The scan in Fig.~3d)  includes wavevectors in a narrow cylindrical region centred at K and extending out to just touch the double helix nodal lines, so for all of those wavevectors there will be a finite energy gap between the top and bottom sets of bands, so a two-peak structure would be expected in the energy scan, as indeed seen experimentally. For a quantitative comparison with the data, we use the cross-section model in Supplementary Equation~(\ref{eq:intensities}) averaged over all magnetic domains. In the present case, it is sufficient to consider only three magnetic domains (of the A structural domain), so for $\eta>0$ we take $\phi=0,\pm2\pi/3$, as for those $\phi$ values each B magnetic domain has the same response as the A magnetic domain with spins along the same direction, and domains obtained by reversing the spins on each site also have identical signatures. We consider two scenarios related by the transformation $\eta\rightarrow -\eta$ and $\phi\rightarrow\phi+\pi/2$, as this leaves $B'$ and $D'$ in Supplementary Equation~(\ref{eq:BD_eta}) invariant, and subsequently the dispersions and dynamical structure factor in Supplementary Equations.~(\ref{eq:omega},\ref{eq:Sxx_Syy_local}) are invariant as well. The above two scenarios indeed produce a similar, but not identical intensity profile in Supplementary Figure~\ref{fig:scan_en_K_eta} (dashed gray and solid black lines), the small differences are due to the  polarization factor $p_{\tilde{x}}$ in Supplementary Equation~(\ref{eq:intensities}), which changes upon rotating the spins by 90$^{\circ}$ between the two scenarios. The first peak in the energy scan is identified with crossing the two lower bands (almost overlapping $\omega_{1,2}$ modes) and the higher peak with crossing the top two bands (almost overlapping $\omega_{3,4}$ modes), with the observed energy separation well accounted for in Supplementary Figure~\ref{fig:scan_en_K_eta} using $\vert \eta \vert =1.7$~meV. The calculated lineshape for the case $\eta<0$ (black solid line, lower peak more intense) is in better agreement with the data than for $\eta>0$ (gray dashed line), from which we conclude that the former is the more likely of the two scenarios. For the parameters with best agreement we show in Supplementary Figure~\ref{fig:8_9_meV}c-d) the calculated momentum intensity maps, the agreement with the data in panels a-b) is excellent. In particular, the observed dramatic change when moving from energies below (bottom panels) to energies above the nodal energy (top panels) - the intensity shift from inside to outside of the central hexagonal Brillouin zone (dashed hexagonal outline) - is well captured, even the subtle local rotations of the intensity patterns around the zone corners - most visible in the lower-right corners - are well reproduced (those intensity modulations arise from the finite buckling of the cobalt honeycombs).

\begin{figure}[!tbph]
    \includegraphics[width=\columnwidth]{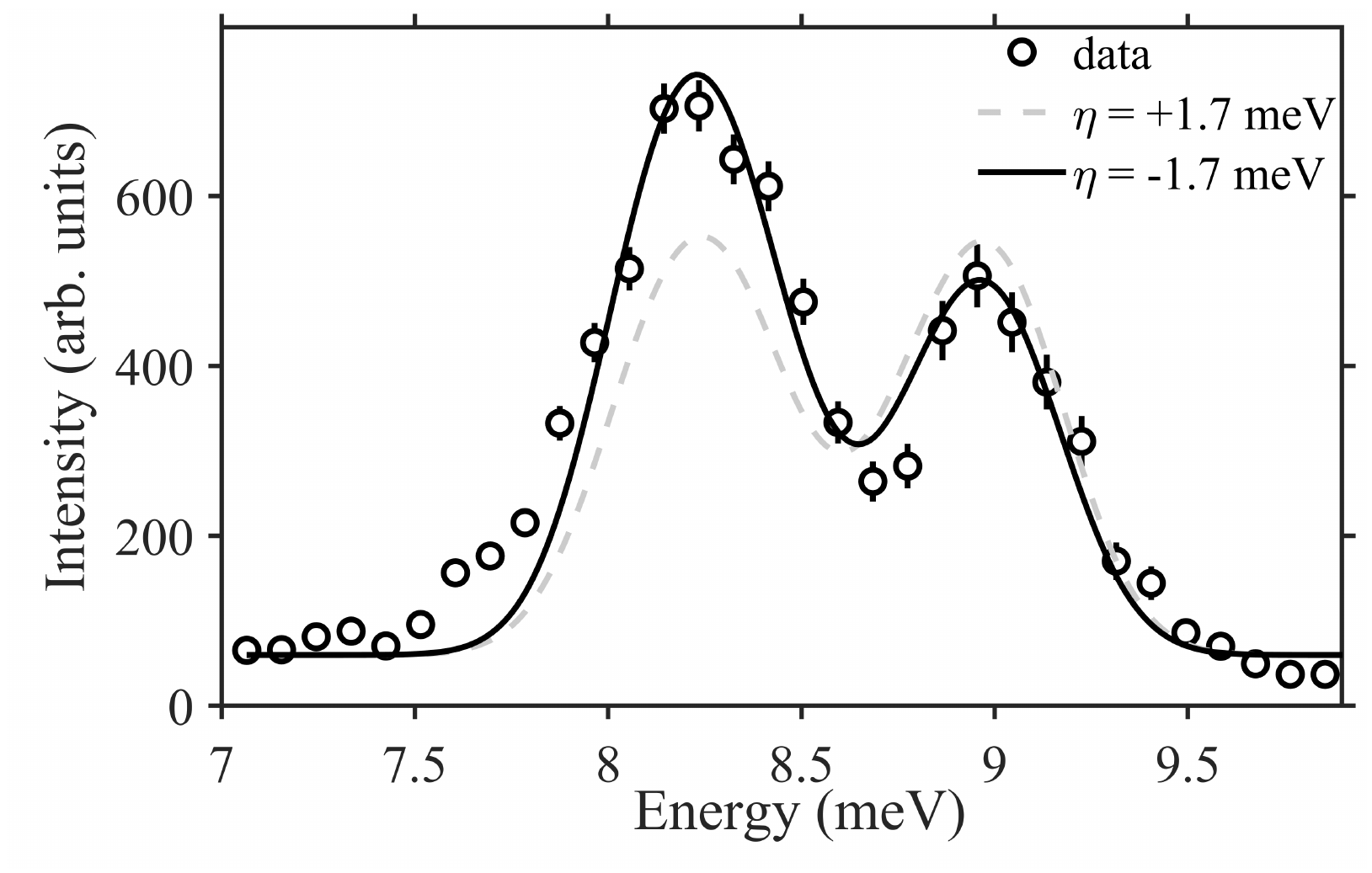}
    \figcaption{Fine structure of the energy scan at K compared with the model with finite $\eta$}{White circles denote the inelastic neutron scattering intensity in an energy scan, as in Fig.~3d), with error bars representing one standard deviation, compared to the appropriate domain average for $\eta<0$ (black solid) and $\eta>0$ (dashed gray) as described in the text. Both calculations have been shifted along the horizontal axis by $\Delta E=+0.11$~meV to provide a better agreement with the data if exchange parameters are fixed as per Supplementary Equation~(\ref{eq:bestfit}).}
    \label{fig:scan_en_K_eta}
\end{figure}

\begin{figure}[!htbp]
    \includegraphics[width=\linewidth,keepaspectratio]{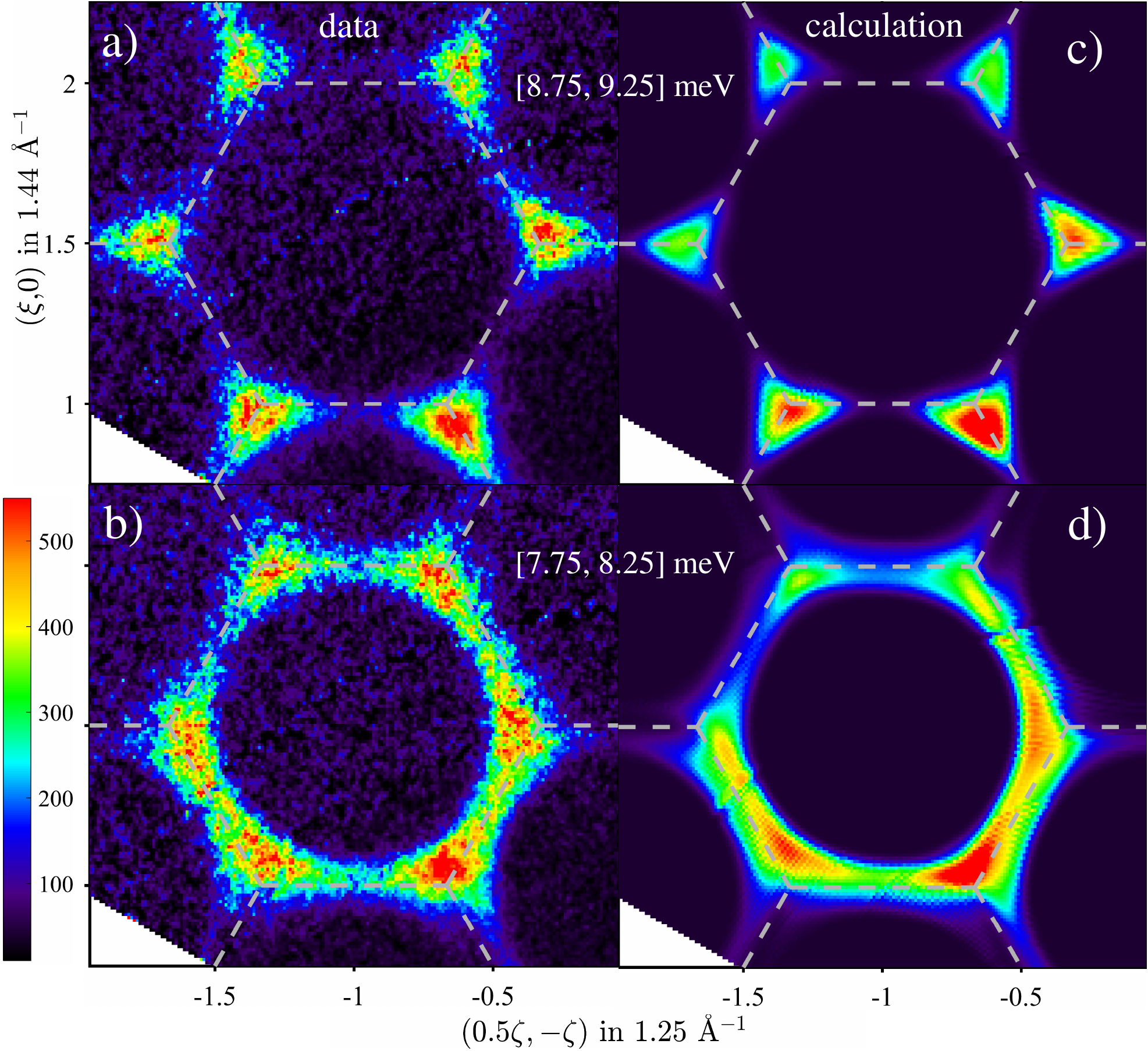}
    \figcaption{\bf Momentum intensity maps above/below the Dirac node energy for the model with finite $\eta$}{Left/right panels are data/calculation, top/bottom panels are above/below the Dirac node energy. The data panels are as in Fig.~2d-e) and the calculation parameters are as in Supplementary Figure~\ref{fig:scan_en_K_eta} (solid line). The colour bar applies to all sub-figures and indicates scattering intensity in arbitrary units on a linear scale.\label{fig:8_9_meV}}
\end{figure}

The coupling $\eta$ is one of four anisotropic nearest neighbor couplings beyond the XXZ model. Establishing in detail the magnitudes and signs of all such anisotropic exchanges is beyond the scope of the present work. However, we have demonstrated in this Section that a finite $\eta$ provides a natural explanation for the double peak structure in Supplementary Figure~\ref{fig:scan_en_K_eta} and the splittings observed on the magnetic zone boundary in Supplementary Figure~\ref{fig:w12_splitting_mzb}, whilst leaving the Dirac nodal lines intact. As we describe in the next Supplementary Note, the $\eta$ coupling can also provide a mechanism to account for the appearance of the spectral gap.

\section{Quantum Order-By-Disorder and the Spectral Gap}
\label{sec:gap}
\subsection{Mean-field ground-state degeneracy}
Upon including exchange anisotropy terms as discussed in the previous Section, the spin Hamiltonian symmetry is reduced down to discrete rotations, however at the mean-field level the ground state energy remains independent of the in-plane moment angle $\phi$ and consequently the linear spin-wave spectrum remains gapless. To see this, we parameterize the easy-plane spin configuration by $\mathbf{m}_i = m_0  {\rm Re}\left[ e^{i\phi}\left( \hat{\mathbf{\tilde{x}}}_i+i\hat{\mathbf{\tilde{z}}}_i \right) \right]$ where the $\hat{\mathbf{\tilde{x}}}$ and $\hat{\mathbf{\tilde{z}}}$ axes define the local frame, which rotates 180$^\circ$ around the $\tilde{x}$-axis between adjacent layers. Here, $m_0$ is the size of the ordered moment. We then note that the mean field free energy for quadratic spin interactions can be written as $F_{\rm MFT}[\Psi]=\alpha \vert \Psi \vert^2 + \beta \Psi^2 + \beta^\star (\Psi^*)^2$ where $\Psi = m_0 e^{i\phi}$ and invariance under 3-fold symmetry forces $\beta=0$ as $\Psi \rightarrow e^{2\pi i/3} \Psi$ under this symmetry operation. This means that the mean field free energy cannot depend on $\phi$, even when symmetry-allowed two-spin exchange anisotropy terms are included. A similar argument forbids symmetry breaking from four-body couplings that may arise from spin-lattice coupling. Indeed, the lowest order terms that lead to discrete symmetry breaking are six-body terms that may be attributed to fluctuations. In short, we expect order-by-disorder to arise in CoTiO$_3$ purely on the basis of the spectral gap.

\subsection{Quantum order-by-disorder in the presence of bond-dependent anisotropic exchange $\eta$}
A more direct way to see this is that anisotropic exchange couplings break the $U(1)$ Hamiltonian symmetry down to the lattice symmetries so one would expect the classical ground-state  degeneracy as a function of $\phi$ to be lifted by zero-point quantum fluctuations. To leading order the zero point contribution to the ground state energy (per magnetic unit cell) is
\begin{equation}
E_0 \equiv N_s S\epsilon_{\rm qu} = \frac{1}{N_c}\sum_{\mathbf{k},m} \frac{\omega_m({\mathbf{k}})}{2}
\label{eq:zpe}
\end{equation}
where $N_c$ is the number of magnetic unit cells, each containing $N_s=4$ magnetic sublattices. The sum runs over all wavevectors ${\mathbf{k}}$ in the magnetic Brillouin zone and all magnon branches indexed by $m=1$ to $N_s$.

\begin{figure}[!htbp]%
    \subfloat[]{{\includegraphics[width=\columnwidth]{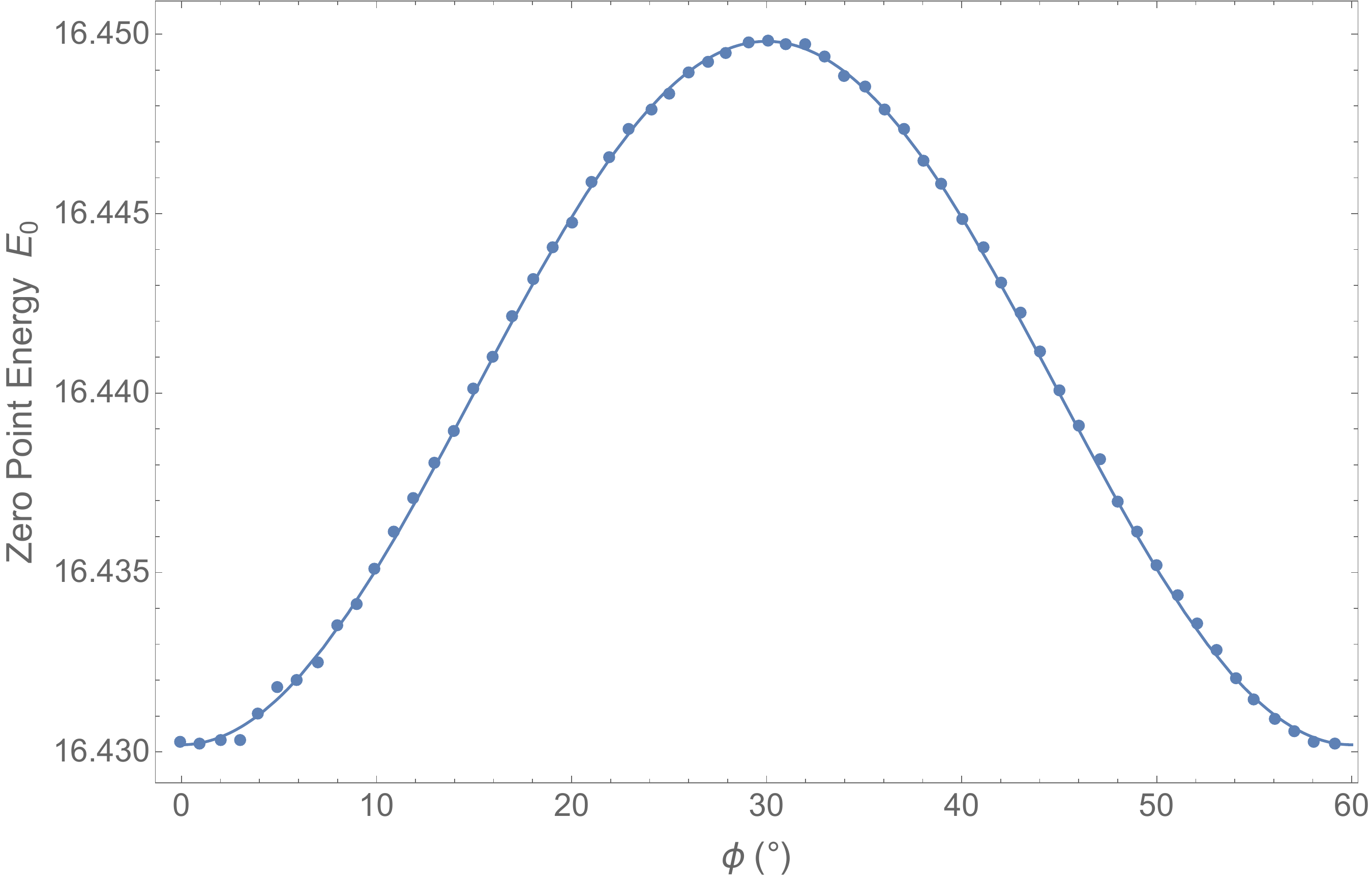} }}%
    \qquad
    \subfloat[]{{\includegraphics[width=\columnwidth]{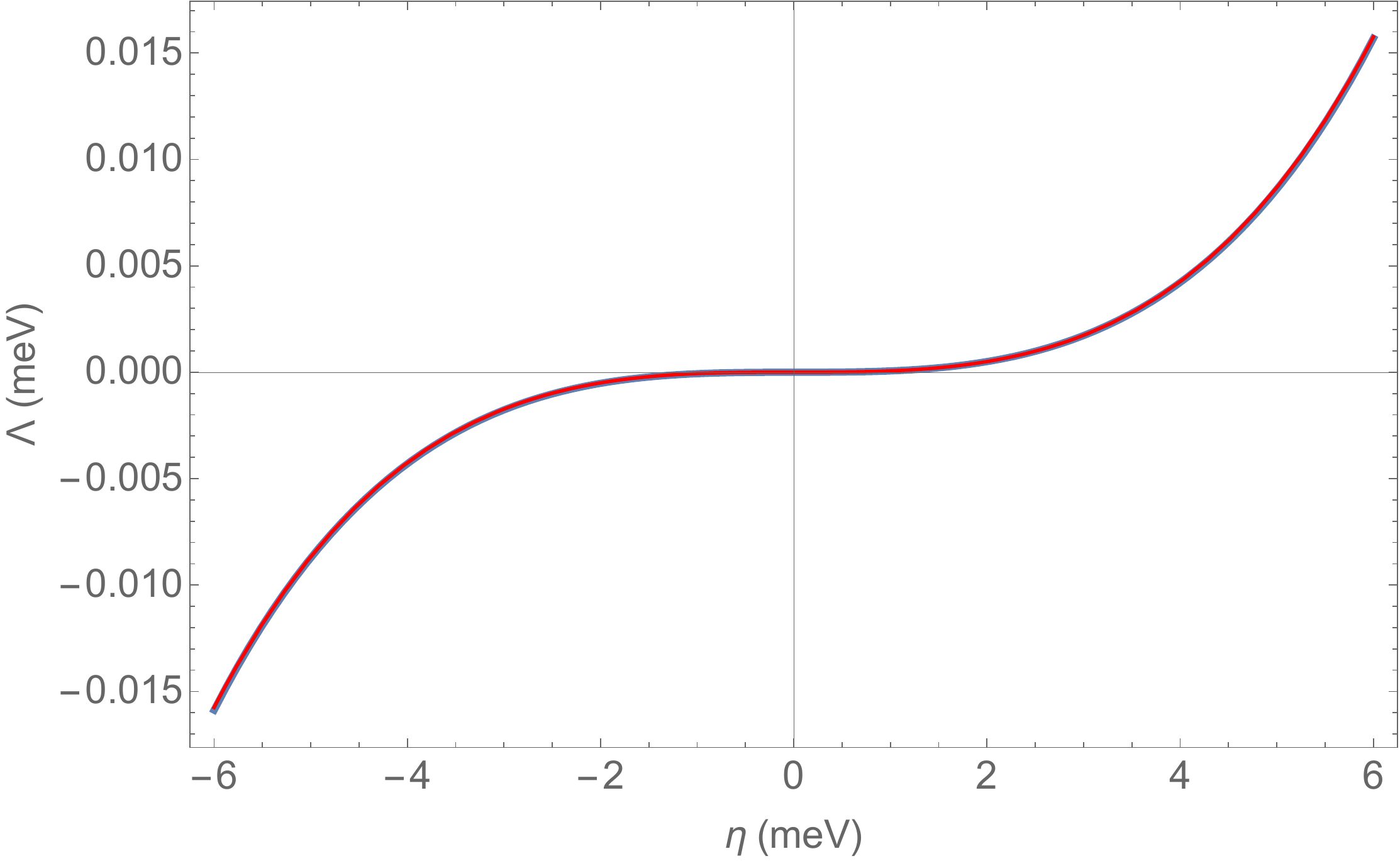}}}%
    \figcaption{Parameterisation of quantum zero-point energy}{(a)
     Leading zero-point contribution to the ground state energy, $E_0$ in Supplementary Equation~(\ref{eq:zpe}), as a function of in-plane angle $\phi$, for exchange anisotropy $\eta=6.36$~meV. The solid line is a fit to the sinusoidal form in Supplementary Equation~(\ref{eq:E_0}) with parameters $\bar{E}_0= 16.44$~meV and $\Lambda=0.0196$~meV at fixed $\phi_0=0$. (b) Zero-point energy oscillation amplitude $\Lambda$ (blue line) as a function of exchange anisotropy $\eta$ for fixed $\phi_0=0$, fitted to an odd polynomial form $\fs{a}\,\eta^3 + \fs{b}\,\eta^5$ (red line) with $\fs{a}=6.13(7)\times 10^{-5}$~meV$^{-2}$ and $\fs{b}=(3.23 \pm 1) \times 10^{-7}$~meV$^{-4}$ (the relatively large uncertainty in $\fs{b}$ reflects the range of values depending on whether higher order odd terms $\eta^7$ and $\eta^9$ are included or not in the polynomial fit).}%
    \label{fig:ObD}%
\end{figure}

To illustrate the phenomenon we add to the XXZ model in~\ref{sec:fits} a symmetry-allowed nearest-neighbor anisotropic diagonal coupling $\eta$ defined in the $xyz$ frame by $J^{xx}= J^{\perp}_{1}-\eta/2$ and $J^{yy}= J^{\perp}_{1}+\eta/2$, i.e. $+/-$ve $\eta$ favors spins pointing orthogonal/parallel to the bond direction (as $J^{\perp}_{1}<0$). At finite $\eta$ the (numerically calculated) spin-wave dispersions do depend on the in-plane moment angle $\phi$ and the leading zero-point correction to the ground state energy is six-fold modulated in $\phi$ with period $\pi/3$ as shown in Supplementary Figure~\ref{fig:ObD}a) and can be fitted (solid line) to a cosinusoidal form
\begin{equation}
E_0 = \bar{E}_0 - \frac{\Lambda}{2} \cos [6(\phi-\phi_0)]
\label{eq:E_0}
\end{equation}
where $\Lambda$ is an odd polynomial in the anisotropic exchange $\eta$ [see Supplementary Figure~\ref{fig:ObD}b)]. In other words, there is a quantum order-by-disorder mechanism that lifts the $U(1)$ classical ground state degeneracy leading to a six-fold symmetric set of ground states, i.e. $+/-$ve $\eta$ select the family of orientations $\phi=0$ or $\pi/6$ (modulo $\pi/3$), respectively.

\subsection{Order-by-disorder spectral gap}
To determine the anisotropy value that gives a gap comparable to what is seen experimentally we use the expression for the pseudo-Goldstone gap to leading order in $1/S$ given by \cite{PhysRevLett.121.237201}
\begin{equation}
\Delta = S^{1/2} \sqrt{ \left(\frac{\partial^2 \epsilon_{\rm cl}}{\partial \theta^2} \right)_0 \left(  \frac{\partial^2 \epsilon_{\rm qu}}{\partial \phi^2} \right)_0}
\label{eq:ObDGap}
\end{equation}
where the energy densities are defined through the total ground state energy \begin{equation}
E_{\rm GS} = {\cal N}S(S+1)\epsilon_{\rm cl} + {\cal N}S\epsilon_{\rm qu} + \ldots
\end{equation}
where ${\cal N}=N_c N_s$ is the total number of spins, $\theta$ is the uniform tilt of the moments out of the $ab$ plane towards the $c$-axis. The derivatives in Supplementary Equation~(\ref{eq:ObDGap}) are evaluated about the quantum selected ground state configuration.

We determine
 \begin{align}
  \epsilon_{\rm cl} & = \frac{3}{2}\left[ J^{\perp}_{1}\cos^2\theta + J^z_{1}\sin^2\theta \right] \nonumber \\ & - \frac{1}{2} \left[ J^{\perp}_{2}\cos^2\theta + J^z_{2}\sin^2\theta  \right] \nonumber \\ & + 3 \left[ J^{\perp}_{3}\cos^2\theta + J^z_{3}\sin^2\theta  \right] - 3 \left[  J^{\perp}_{4}\cos^2\theta + J^z_{4}\sin^2\theta  \right] \nonumber \\ & + \frac{3}{2} \left[ J^{\perp}_{5}\cos^2\theta + J^z_{5}\sin^2\theta  \right] - \frac{3}{2} \left[ J^{\perp}_{6}\cos^2\theta + J^z_{6}\sin^2\theta  \right]
 \end{align}
 and so
 \begin{align}
\left( \frac{ \partial^2 \epsilon_{\rm cl}}{\partial \theta^2} \right)_0 & = -3 \left[ J^{\perp}_{1} - J^z_{1}  \right] + \left[ J^{\perp}_{2} - J^z_{2}  \right] \nonumber \\ & + 6 \left[ -J^{\perp}_{3} + J^z_{3}  \right] + 6 \left[ J^{\perp}_{4} - J^z_{4}  \right] \nonumber \\ & + 3 \left[ -J^{\perp}_{5} + J^z_{5}  \right] + 3 \left[ J^{\perp}_{6} - J^z_{6}  \right].
\label{eq:ecl}
 \end{align}
Note that the classical energy $\epsilon_{\rm cl}$ is independent of the anisotropy parameter $\eta$. For completeness we note that in the 111 frame, more closely tied to the underlying microscopic superexchange mechanism, the nearest neighbor part of $\left( \frac{ \partial^2 \epsilon_{\rm cl}}{\partial \theta^2} \right)_0$ is $3[\Gamma+ \Gamma'_1 + \Gamma'_2]$.

For $\eta=J_1^{\perp}$ we find $\left(  \frac{\partial^2 \epsilon_{\rm qu}}{\partial \phi^2} \right)_0 = 36\times 0.00245/(1/2) = 0.1764$~meV having divided out $S=1/2$. Supplementary Equation~(\ref{eq:ecl}) gives $\left( \frac{ \partial^2 \epsilon_{\rm cl}}{\partial \theta^2} \right)_0 = 21.4$~meV. Then the gap is $\Delta = 1.908 S^{1/2} = 1.35$~meV, on the order of magnitude of the experimental value 1.0(1)~meV. The effect on the spin length of switching on $\eta$ is to increase the role of fluctuations $-$ for the pure XXZ model in Supplementary Equation~(\ref{eq:bestfit}) we found $\Delta S= 0.021$, this increases to $\Delta S=0.0237$ for $\eta=J_1^{\perp}$.

Such a large value of $\eta$ comparable to the largest exchange $J_1^{\perp}$, is however not realistic, as in this case the dispersion relations would not be compatible with the experimental data, based on the analysis of the previous section a value of $\vert \eta \vert \simeq 1.7$~meV would be more in line with the observed dispersions. This suggests that either (i) the order-by-disorder gap formula Supplementary Equation~(\ref{eq:ObDGap}) underestimates the gap, being derived in the large-$S$ limit and applied here for $S=1/2$, or ii) there are other effects present in the actual material that also contribute to the gap, and we consider such a possible gap generation mechanism via spin-orbital exchange later in~\ref{sec:MFTFW}.

\begin{figure}[!htbp]%
    \includegraphics[width=\columnwidth]{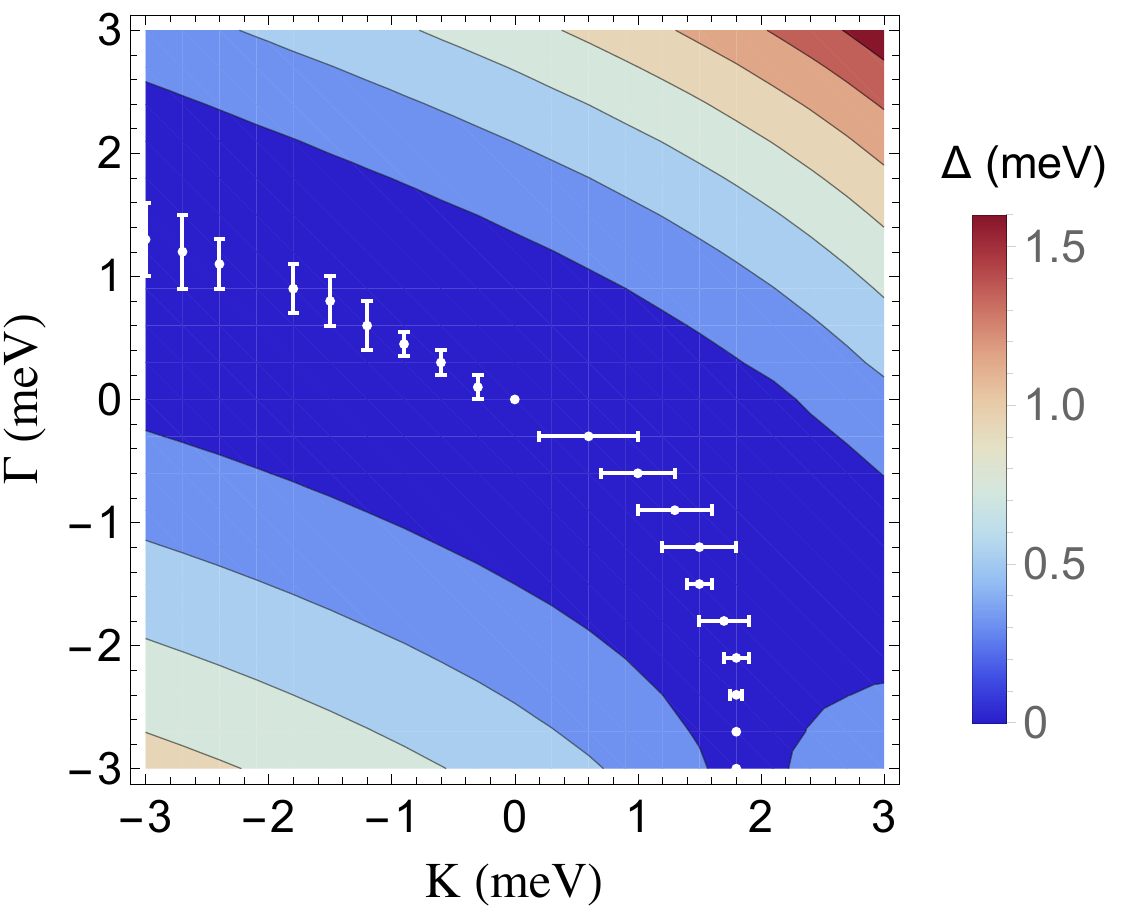} %
    \figcaption{Phase diagram of the XXZ model with additional Kitaev $K$ and $\Gamma$ interactions}{There are two phases selected via the quantum order-by-disorder mechanism. The phase in the lower left part of the phase diagram is parameterized by moment angles $\phi=0$ and the other phase has $\phi=\pi/6$. The contours and color scheme show the order-by-disorder pseudo-Goldstone gap computed from Supplementary Equation~(\ref{eq:ObDGap}). The phase boundary is the locus of points (solid white dots) where the gap $\Delta$ closes, vertical/horizontal error bars indicate the estimated uncertainty of the gap closing location in vertical/horizontal scans.}%
    \label{fig:K_Gamma}%
\end{figure}

To complete this section, in the following we discuss the effects of the other anisotropic exchange terms in Supplementary Equation~(\ref{eq:exchange}). The mixed in-plane-out-of-plane terms $J^{xz}$ and $J^{yz}$ on their own make no contribution to the classical and leading order quantum ground state energy. The effect of an in-plane mixed term $J^{xy}$ can most easily be described by working in a reference frame  $x'y'z'$ rotated with respect to the $xyz$ frame around $z$-axis by an angle $\phi_0$ such that we rotate into the principal axes of
\begin{equation}
\left( \begin{array}{cc} J^{\perp}_{1}-\eta/2 & J^{xy} \\ J^{xy} & J^{\perp}_{1}  + \eta/2 \end{array} \right)
\end{equation}
and select the axis that corresponds to the minimal eigenvalue. One finds that the angle $\phi_0$ is half the polar angle of the vector $(\eta,-2J^{xy})$. For example, this establishes that for $\eta<0$ and $J^{xy}=0$, $\phi_0=\pi/6$. Since the model has a 3 point group symmetry and is time reversal invariant the solution is determined by the above equation up to integer multiples of $\pi/3$. The zero-point quantum energy contribution has the same form as in Supplementary Equation~(\ref{eq:E_0}), but with an origin angle $\phi_0$ that depends on the value of $J^{xy}$ as above. In other words, the minimum energy angle is not symmetry-restricted to take only the discrete values 0 or $\pi/6$ (up to integer multiplets of $\pi/3$), but can take any real value in-between those reference ones for finite $J^{xy}$. We note that this exchange anisotropy term is symmetry-allowed by the absence at the midpoint of the bond of a mirror plane normal to the nearest-neighbor bond  due to (i) the layer stacking and (ii) the buckling of the cobalt honeycombs and (iii) distortions of the CoO$_6$ octahedra away from regular octahedra.

\subsection{Order-by-disorder in a generalized Kitaev-$\Gamma$ model}
For completeness, we also consider the effects of exchange anisotropy terms described in the 111 frame. Starting from the minimal $J_1-J_6$ XXZ model and adding Kitaev $K$ and $\Gamma$ terms in Supplementary Equation~(\ref{eq:111exchange}) we find that both terms can select the $\phi=0$ or $\pi/6$ (modulo $\pi/3$) family of ground states via order-by-disorder according to the phase diagram plotted in Supplementary Figure~\ref{fig:K_Gamma}, where the color represents the zero-point gap $\Delta$ and the white dots indicate the boundary between the two families of ground states.

\subsection{Single-ion anisotropy}
Before moving on to discuss the nodal lines, we briefly dwell on the question of whether discrete symmetry breaking can arise directly from single ion anisotropy in CoTiO$_3$. In other words, can the observed excitation gap, which implies discrete moment orientations in the $ab$ plane, occur in the absence of bond-dependent anisotropic exchange? The short answer is no because any high order single ion anisotropies have to be filtered through the octahedral crystal field splitting, the spin-orbit coupling and trigonal distortion, with mixing of states provided by exchange interactions, by which point it is more fruitful to view them as effective bond-dependent exchange between moments in the lowest Kramers doublet. The longer answer is as follows. The local site symmetry alone (point group ${3}$) constrains the possible single ion crystal-field Hamiltonian to have the form \cite{Bauer2010}
\begin{eqnarray}
{\cal H}_{\rm Single-Ion} & =&  B_{2}^{0}{\cal O}_2^0 + B_{4}^{0}{\cal O}_4^0 +
B_{4}^{3}{\cal O}_4^3 \nonumber \\
& + & B_{6}^{0}{\cal O}_6^0 + B_{6}^{3}{\cal O}_6^3 + B_{6}^{6} {\cal O}_6^6,
\end{eqnarray}
where ${\cal O}_{l}^{m}$ are Stevens operators acting on the orbital sector, expressed in a reference Cartesian $XYZ$ frame with $Z$ along $c$ and $X$ in the $ab$ plane along some reference direction, unconstrained by symmetry for point group $3$ (previously in Supplementary Note 2 we expressed the crystal field Hamiltonian ${\cal H}_{\rm CF}$ for an ideal octahedron in a reference frame with respect to the four-fold octahedron axes). In terms of the angular momentum operators ${\cal O}_2^0=3L_z^2 - L(L+1)$, which arises from the trigonal distortion of the ideal CoO$_6$ octahedron. The only operators that have a nontrivial in-plane anisotropy are ${\cal O}_{4}^{3}$, ${\cal O}_{6}^{3}$ and ${\cal O}_{6}^{6}$. Semiclassically, for $\vartheta$ the polar angle from the $Z$ axis and $\phi$ the azimuthal angle from the $X$-axis in the $XY$ plane, ${\cal O}_{4}^{3}\sim \sin^3\vartheta \cos\vartheta \cos 3\phi$ which vanishes for in-plane moments ($\vartheta=\pi/2$) as does ${\cal O}_{6}^{3}$.

The operator ${\cal O}_{6}^{6}\sim L_{-}^6+L_{+}^6$ or, semiclassically, $\cos 6\phi$, has a 6-fold periodic angular dependence in the $ab$ plane and can in principle lead to in-plane discrete symmetry breaking. However, within the low energy effective $l=1$ $\Gamma_4$ orbital sector (which is the orbital ground state for an ideal octahedron), this term is inoperative. To obtain an effect from this operator we are forced to extend the model to the full free-ion $L=3$ Hilbert space. We may estimate the size of the spectral gap realizing that it can only originate through perturbative mixing of excited orbital levels (of energy $G \sim 1$~eV above the $l=1$ $\Gamma_4$ orbital ground state) via the exchange $J$ (with energy scale $\sim 10$~meV). In other words, the mechanism, as previously argued, is order-by-disorder through virtual crystal field fluctuations \cite{McClarty_2009,PhysRevB.93.184408}. A strong coupling calculation \cite{PhysRevB.93.184408} reveals that the spectral gap will scale as $J^3/G^2 \sim 10^{-3}$~meV. In order to account for the magnitude of the observed spectral gap ($\Delta=1$~meV), multiplicative factors including those coming from  perturbation theory combinatorics and matrix elements would have to boost this by three orders of magnitude. At the level of the full $d^7$ Hilbert space, both the single ion anisotropy and the spin-orbital exchange will contribute to the observed gap. To the extent that these mechanisms can be disentangled (since the ${\cal O}_{6}^{6}$ anisotropy will also affect the exchange) the latter mechanism will be the primary one as this appears straightforwardly from mixing of an order $1$~meV spin-orbital exchange across a $10$ meV crystal field gap. This mechanism can also be viewed within the effective $l=1$, $\fs{S}=3/2$ subspace or within the pseudo-spin one-half picture where the virtual crystal field excitations would manifest themselves as bond-dependent multiple spin exchange anisotropies.

\section{Nodal Lines Topology and Symmetry}
\label{sec:dirac}

We now consider the implications of various discrete symmetries of the spin wave Hamiltonian for the magnon band structure and its topological properties. To consider the full generality of the problem, in the following we work with the (full) four-sublattice magnetic unit cell, for which the spin wave Hamiltonian has the form
\[
\mathcal{H}_{8\times 8}=\sum_{\mathbf{k}} \bm{\Phi}_{\mathbf{k}}^\dagger \mathcal{D}_{8\times 8}(\mathbf{k})\bm{\Phi}_{\mathbf{k}}
\]
where $\bm{\Phi}_{\mathbf{k}}^\dagger=(a_{\mathbf{k}}^\dagger, b_{\mathbf{k}}^\dagger,c_{\mathbf{k}}^\dagger, d_{\mathbf{k}}^\dagger, a_{-\mathbf{k}}, b_{-\mathbf{k}},c_{-\mathbf{k}}, d_{-\mathbf{k}})$
where the $a^\dagger_{\mathbf{k}}$, $b^{\dagger}_{\mathbf{k}}$, $c^{\dagger}_{\mathbf{k}}$ and $d^{\dagger}_{\mathbf{k}}$ operators create magnons on the A, B, C and D magnetic sublattices, respectively, and the dynamical matrix takes the block form
\[
\mathcal{D}_{8\times 8}(\mathbf{k}) = \left( \begin{array}{cc}
\bm{A}(\mathbf{k}) & \bm{B}(\mathbf{k})  \\
\bm{B}^\dagger(\mathbf{k}) & \bm{A}^{*}(-\mathbf{k})
\end{array}  \right).
\]
Without giving the explicit form of the $\bm{A}$ and $\bm{B}$ matrices for the $J_1$ to $J_6$ XXZ Hamiltonian, we note that the dynamical matrix $\mathcal{D}_{8\times 8}(\mathbf{k})$ has time reversal symmetry defined through    $\mathcal{T}^{-1}\mathcal{D}_{8\times 8}(\mathbf{k})\mathcal{T}=\mathcal{D}_{8\times 8}(-\mathbf{k})$, where $\mathcal{T}$ is antiunitary. The time reversal operator is, explicitly, the complex conjugation operator times the unit operator acting on the sublattice indices. The same Hamiltonian also has spatial inversion symmetry $-$ a unitary transformation: $\mathcal{P}^{-1}\mathcal{D}_{8\times 8}(\mathbf{k})\mathcal{P}=\mathcal{D}_{8\times 8}(-\mathbf{k})$ with $\mathcal{P}=\Gamma_1$, the anti-diagonal matrix with ones along the anti-diagonal.

It follows that the model has $\mathcal{A}\equiv\mathcal{PT}$ symmetry:
\[
\mathcal{A} \mathcal{D}_{8\times 8}(\mathbf{k})\mathcal{A}^{-1} = \mathcal{D}_{8\times 8}(\mathbf{k})
\]
where $\mathcal{A}$ is anti-unitary. It is this symmetry that is responsible for protecting the Dirac nodal lines. This is because the $\mathcal{PT}$ symmetry imposes a reality condition on the Hamiltonian. Consider now a closed loop in the 3D Brillouin zone. The Hamiltonian defined along this loop is real and along this loop there is a $\mathbb{Z}_2$ topological classification meaning that there is a winding number that assumes values $0$ (topological trivial) and $\pi$ (topologically nontrivial). If the winding number on the loop is $\pi$, it must enclose a singular point - a nodal point - and since the winding number cannot change continuously, say by deforming the loop in 3D, there must be a nodal line in 3D such that the winding number is $\pi$ on any loop winding around the nodal line.

We now examine the robustness of the $\mathcal{PT}$ symmetry under perturbations. From the foregoing explicit form of the operators, it is straightforward to see that
\begin{equation}
\bm{A}(\mathbf{k}) = \left( \begin{array}{cccc}
A_{11}(\mathbf{k}) & A_{12}(\mathbf{k}) & A_{13}(\mathbf{k}) & A_{14}(\mathbf{k})  \\
A_{12}^\star(\mathbf{k}) & A_{22}(\mathbf{k}) & A_{23}(\mathbf{k}) & A^{\star}_{13}(-\mathbf{k})  \\
A_{13}^\star(\mathbf{k}) & A_{23}^\star(\mathbf{k}) & A_{22}(-\mathbf{k}) & A^{\star}_{12}(-\mathbf{k})  \\
A_{14}^{\star}(\mathbf{k}) & A_{13}(-\mathbf{k}) & A_{12}(-\mathbf{k}) & A_{11}(-\mathbf{k})
\end{array}  \right)
\label{eq:Ablock}
\end{equation}
\begin{equation}
\bm{B}(\mathbf{k}) = \left( \begin{array}{cccc}
B_{11}(\mathbf{k}) & B_{12}(\mathbf{k}) & B_{13}(\mathbf{k}) & B_{14}(\mathbf{k})  \\
B_{21}(\mathbf{k}) & B_{22}(\mathbf{k}) & B_{23}(\mathbf{k}) & B_{13}(\mathbf{k})  \\
B_{31}(\mathbf{k}) & B_{32}(\mathbf{k}) & B_{22}(\mathbf{k}) & B_{12}(\mathbf{k})  \\
B_{41}(\mathbf{k}) & B_{31}(\mathbf{k}) & B_{21}(\mathbf{k}) & B_{11}(\mathbf{k})
\end{array}  \right)
\label{eq:Bblock}
\end{equation}
is the most general spin wave Hamiltonian that preserves $\mathcal{PT}$. In other words, in order to break $\mathcal{PT}$, we must find an exchange coupling that breaks these weak constraints on the Hamiltonian.

The $J_1$ to $J_6$ XXZ model has both time reversal symmetry and parity symmetry and is therefore compatible with Supplementary Equations~(\ref{eq:Ablock}) and (\ref{eq:Bblock}) with a much more restricted form - for example the diagonal elements are identical. If we go beyond the XXZ model to the full set of symmetry-allowed exchange couplings, we have checked that the resulting spin wave Hamiltonian preserves the $\mathcal{PT}$ symmetry out to and including the sixth nearest neighbor couplings which includes $38$ independent exchange terms. We have additionally verified explicitly (by numerical solution of the corresponding $8\times8$ spin-wave Hamiltonian) that none of these couplings gap out the nodal lines at least when the magnetic ground state is preserved.

We now consider the structure of the nodal lines. When the model is fine-tuned to have $J_2$ of Heisenberg form ($J_2^{\perp}=J_2^z$) and the third neighbor exchanges are set to zero ($J_3^{\perp}=J_3^z=0$), one finds that nodal lines are degenerate, occur at the K points, and are straight lines along $L$. For anisotropic $J_2$ ($J_2^{\perp}\neq J_2^z$), the nodal lines around each K-point split into a pair of helical nodal lines that each wind around the K-points as illustrated by the red and blue lines in  Fig.~3b) with periodicity $L=3$. The chirality of the double helix nodal lines, i.e. the sense in which they wind along the $L$ direction, is opposite at neighboring corner K-points due to the $\bar{3}$ symmetry of the lattice. The minimal model that produces these ``double helix'' nodal lines is the XXZ $J_1$-$J_2$ model. In this model, the anisotropy of the $J_2$ coupling is responsible for the splitting of the nodal lines and the winding along $L$, i.e. $-$ the radius of the helix in momentum space in planes perpendicular to the $L$ direction is
\begin{equation}
\vert \delta \mathbf{k} \vert=\frac{2}{\sqrt{3}a} \left| \frac{J^{\perp}_2 -J^{z}_2}{J^{\perp}_1+J^z_1}\right|.
\end{equation}
Note that $\delta \mathbf{k}=0$ (the pair of nodal lines are fused and straight) for isotropic $J_2$ couplings. In this case adding anisotropic $J_3$ couplings can also produce helical nodal lines, but the effect comes in at higher order with $\vert \delta \mathbf{k} \vert \simeq 2/(\sqrt{3}a)\vert J_2(J^{\perp}_3 -J^{z}_3)/[J^{\perp}_1(J^{\perp}_1+J^z_1)]\vert$. For completeness, we note that the magnitude of $\delta \mathbf{k}$ is not affected by the buckling parameter $\epsilon$, or by adding further neighbor XXZ couplings on the $J_4$, $J_5$ or $J_6$ bonds.

We also note that the two nodal lines in a pair that wind around each K-point are translated versions of each other along $L$ by the propagation vector $\mathbf{Q}$ (translation by $2\mathbf{Q}$ leaves each nodal line invariant). This is most easily seen by first working in the local frame, where a single nodal line appears near each K point (the red helical lines in Fig.~3b) corresponding to the nodal points $\bm{\upkappa}$ where $\omega_{+}(\bm{\upkappa})=\omega_{-}(\bm{\upkappa})$. In the global frame, each of the modes $\omega_{\pm}$ acquires a `pair' shifted in momentum by $\mathbf{Q}$ (as per Supplementary Equation~(\ref{eq:intensities})), so the original nodal points $\bm{\upkappa}$ are translated into a new set of nodal points $\bm{\upkappa}+\mathbf{Q}$, giving rise to the blue helical lines in Fig.~3b).

One can understand the origin of the double helix nodal lines in terms of the underlying exchange couplings by first observing that the $\gamma_{n\mathbf{k}}$ functions for $n=1,4,5,6$ connect groups of three sites in the same layer and those always appear in the spin-wave Hamiltonian multiplying a function of the form $\sum_j \exp(i \mathbf{k}\cdot \bm{\delta}_{j})$ where $\bm{\delta}_{j}$ vectors connect the central site to the three neighbors related by three-fold symmetry. These functions vanish at the hexagonal zone corner (independent of $L$) implying that all such terms must preserve a nodal line running along the $L$ direction. In our $J_1-J_6$ model, the relevant couplings for the existence of a helical nodal line are $J_2$ and $J_3$. If the third neighbor couplings vanish, the helical line appears only if $J_2$ is anisotropic. Examining the $4\times 4$ spin wave Hamiltonian, we observe that the only term that depends on the anisotropic part of $J_2$ is in the element $B$. We also note that such a term is identical to the one that appears in the related rhombohedral $J_1-J_2$ Heisenberg ferromagnet ($J_{1,2}<0$), already studied in \cite{pershoguba2018}. In the present cobalt lattice arrangement, the spin-wave Hamiltonian for this ferromagnetic model is
\begin{equation}
{\cal H}_{\rm FM}(\mathbf{k}) = S\left( \begin{array}{cc} -3J_{1}-J_2 & J_1\gamma'_{1\mathbf{k}} + J_2\gamma'_{2\mathbf{k}} \\ J_1\gamma'^{*}_{1\mathbf{k}} + J_2\gamma'^{*}_{2\mathbf{k}} & -3J_{1}-J_2 \end{array} \right)\label{eq:HToy}
\end{equation}
where $\gamma'_{j\mathbf{k}}=\gamma_{j\mathbf{k}}e^{i\mathbf{k}\cdot(\mathbf{r}_2-\mathbf{r}_1)}$ with $j=1,2$ and $\gamma_{j\mathbf{k}}$ defined in Supplementary Equation~(\ref{eq:gamma}).
The presence of nodal points is determined by the condition $\vert J_1 \gamma'_{1\mathbf{k}} + J_2 \gamma'_{2\mathbf{k}} \vert =0$. For $J_2=0$ this gives straight Dirac nodal lines at K-points. For finite $J_2$ the nodal lines become helical and precess around the K-points upon varying $L$. In detail, considering for concreteness the nodal line near the K-point (5/3,2/3), the above condition gives the in-plane wavevector offset $\delta\mathbf{k}$ from K, in Cartesian coordinates as
$$(\delta k_x,\delta k_y)=\frac{2}{\sqrt{3}a}\frac{J_2}{J_1}
\left( \cos \frac{2\pi L}{3},-\sin \frac{2\pi L}{3}\right)
$$
to first order in $J_2/J_1$. Here we have used the Cartesian $xyz$ frame defined in Supplementary Figure~\ref{fig:FW}d), where $k_x$ is along ($1\frac{\bar{1}}{2}0$) and $k_y$ along ($010$). The above equations describe a helical nodal line winding clockwise in the $+L$ direction with period $3$ along $L$ and with radius proportional to $J_2/J_1$.

\section{Spin-Orbital Flavor Wave Theory}
\label{sec:MFTFW}

We have presented detailed, quantitative parametrization of the experimental data within the context of an effective XXZ spin one-half model for the cobalt moment. The validity of this model can be argued on the basis of single-ion spectrum in the presence of spin-orbit coupling and trigonal distortion of the oxygen octahedra that leave a doublet on each magnetic site separated from the first excited state by about $28$~meV. The quality of the agreement with data provide an {ex post facto} justification for this effective model. However, the exchange scale is some significant fraction of the low-lying single-ion splitting so one is naturally led to consider a more microscopic route to describing \cto\ that incorporates the spin $\fs{S}=3/2$ and effective orbital $l=1$ degrees of freedom in full. One important motivation for building on the effective model is to provide a direct calculation of the clearly resolved spectral gap of about $1$~meV above the magnetic Bragg peaks. We have shown that the effective spin one-half model including exchange anisotropies cannot open up a gap at the mean field level, while the leading fluctuation contribution to the energy does select a discrete set of ground states. One then infers that a gap will arise in the spin wave spectrum and we have calculated this gap to leading order in $1/S$. However, by enlarging the local Hilbert space and considering spin-orbital exchange there is hope that mixing of states does lift this $U(1)$ degeneracy. In this section we develop a flavor wave or multi-boson expansion \cite{papanicolaou1988unusual,PhysRevB.86.174428} for the excitations in \cto\ that also allows us to include arbitrary couplings in the Hamiltonian and indeed find that the $U(1)$ degeneracy is lifted by spin-orbital exchange terms, which select discrete in-plane angular orientations for the magnetic moments in the ground state and open a gap in the magnon spectrum.

A further motivation for investigating a more microscopic model is to deepen our understanding of the order-by-disorder mechanism. We have investigated in detail the effect of bilinear anisotropic exchange terms acting within the pseudospin one-half subspace. From the point of view of this effective model, any microscopic exchange of higher order acting on two sites projects down to an such a pseudospin one-half exchange coupling to leading order in an expansion in the inverse crystal field gap. However, virtual crystal field fluctuations will generate multi-site effective exchange terms that will break the accidental degeneracy of the spin bilinear model down to a discrete set of ground states. These effects can be more efficiently captured by enlarging the Hilbert space to consider the low-lying spin and orbital coupled crystal field as we describe below.

In  this enlarged Hilbert space we shall consider the leading order spin-orbital exchange that produces order-by-disorder through virtual crystal field fluctuations. Higher order terms - such as four spin terms or two-site multipolar exchange - are possible in principle but they are suppressed by powers of the large charge gap.  In cases where higher order exchange couplings are significant, such as in cuprates, the reason for the size of these couplings is the combination of proximity to a Mott transition and the largeness of the exchange \cite{cyclicexchange,macdonald}. CoTiO$_3$, in contrast, is deep in the Mott insulating regime with an exchange about $5\%$ of that in the cuprates leading to an estimate of the ratio of the biquadratic to the Heisenberg exchange of $O(10^{-3})$. As we discuss next, there are spin-orbital exchange terms that are both much larger than this {and} that lead to discrete symmetry breaking.

\subsection{Flavor Wave Expansion}

The computation of the excitation spectrum proceeds as follows. We consider a general one- and two-body Hamiltonian
\[
{\cal{H}} = \sum_{ia,\alpha} h_{ia}^\alpha \hat{O}_{ia}^\alpha + \sum_{ia,jb}\sum_{\alpha\beta} J_{ia,jb}^{\alpha\beta} \hat{O}_{ia}^{\alpha} \hat{O}_{jb}^{\beta}
\]
acting on the spin $\fs{S}=3/2$, orbital $l=1$ subspace where it is understood that the one-body terms include the spin-orbit coupling ${\cal{H}_{\rm SO}}$ and the trigonal distortion ${\cal H}_{\rm trig}$. A local mean field theory yields a spectrum $\vert i,a, p\rangle = \sum_{l^z, \fs{S}^z} c^{(p)}_{l^z, S^z} \vert l,l^z, \fs{S}, \fs{S}^z \rangle$ on each distinct magnetic sublattice $a$ where $p=0,\ldots,d-1$ and $d=12$ is the local Hilbert space dimension. The ground state can then be written as $\vert \Psi_{\rm MF}\rangle = \prod_i \prod_a \vert i, a, 0 \rangle$.

Each operator can be written as
\[
\hat{O}_{ia}^{\alpha} = \sum_{n,m} [c_{a}^{\alpha}]_{pq} \vert i,a,p \rangle \langle i,a,q \vert \equiv \sum_{p,q} [c_{a}^{\alpha}]_{pq} A^{\dagger}_{i,a,p} A_{i,a,q}
\]
where $[c_{a}^{\alpha}]_{pq}\equiv \langle i,a,p \vert \hat{O}_{a}^{\alpha}\vert i,a,q \rangle$ and the bosonic operators we have introduced must satisfy the constraint
\[
\sum_{p=0}^{d-1} A^{\dagger}_{i,a,p} A_{i,a,p} = M
\]
and $M=1$ in the system of interest. Formally, we wish to have an expansion in powers of $1/M$ where the single-ion Hamiltonian appears to leading quadratic order so we rescale the single-ion Hamiltonian by a factor $M$. To compute experimentally relevant quantities, however, we set $M=1$.

The expansion of operator $\hat{O}_{ia}^{\alpha}$ in terms of $A$ bosons is
\begin{widetext}
\begin{align}
\hat{O}_{ia}^{\alpha} & = M[c_{a}^{\alpha}]_{00} - [c_{a}^{\alpha}]_{00}\sum_{p=1}^{d-1} A^{\dagger}_{ia,p} A_{ia,p}  + \sum_{p=1}^{d-1}\sum_{q=1}^{d-1} [c_{a}^{\alpha}]_{pq} A^{\dagger}_{ia,p} A_{ia,q} \nonumber \\
 & + \sum_{p=1}^{d-1}  [c_{a}^{\alpha}]_{0p} \sqrt{M - \sum_{q=1}^{d-1} A^{\dagger}_{i,a,q} A_{ia,q} } A_{ia,p}
 +  \sum_{p=1}^{d-1}  [c_{a}^{\alpha}]_{p0} A^{\dagger}_{ia,p} \sqrt{M - \sum_{q=1}^{d-1} A^{\dagger}_{ia,q} A_{ia,q} }
\end{align}
\end{widetext}

The leading order term is simply the mean field energy, the first order terms in $A$ vanish when the mean field energy is minimal and the quadratic contribution from the single-ion physics is:
\[
M\sum_{ia,\alpha} h_{ia}^\alpha \left( [c_{ia}^{\alpha}]_{pq} - \delta_{pq}[c_{ia}^{\alpha}]_{00} \right) A^{\dagger}_{ia,p}A_{ia,q}.
\]
The quadratic terms coming from the interactions are
\begin{widetext}
\begin{align}
H_{\rm FW} = M\sum_{ia,jb}\sum_{pq=1}^{d-1}\sum_{\alpha\beta} J_{ia,jb}^{\alpha\beta} \left[ [c^{\alpha}_{a}]_{00}\left( [c_b^\beta]_{pq} -\delta_{pq}[c_b^\beta]_{00} \right) A^\dagger_{jb,q} A_{jb,p} + [c^{\beta}_{b}]_{00}\left( [c_a^\alpha]_{pq} -\delta_{pq}[c_a^\alpha]_{00} \right) A^\dagger_{ia,q} A_{ia,p}  \right. \nonumber \\ + \left. [c_a^\alpha]_{0q}[c_b^\beta]_{0p} A_{ia,q}A_{jb,p} + [c_a^\alpha]_{0q}[c_b^\beta]_{p0} A_{ia,p}A^\dagger_{jb,q} + [c_a^\alpha]_{p0}[c_b^\beta]_{0p} A^\dagger_{ia,p}A_{jb,q} + [c_a^\alpha]_{p0}[c_b^\beta]_{p0} A^\dagger_{ia,p}A^\dagger_{jb,q}   \right].
\end{align}
\end{widetext}

In Fourier space this can be written as $H_{\rm FW} = \sum_{\mathbf{k}} \bm{\Upsilon}_{\mathbf{k}} \bm{H}_{\rm FW}(\mathbf{k})\bm{\Upsilon}_{\mathbf{k}}$ where $\bm{\Upsilon}_{\mathbf{k}}=\left( A^\dagger_{\mathbf{k}a,p} A_{-\mathbf{k}a,p} \right)$ and
\begin{equation}
\bm{H}_{\rm FW}(\mathbf{k}) = \left( \begin{array}{cc}  \mathbb{A}(\mathbf{k}) & \mathbb{B}(\mathbf{k}) \\ \mathbb{B}^\dagger(\mathbf{k}) & \mathbb{A}'(\mathbf{k}) \end{array} \right)
\end{equation}
is a $2(d-1)N_s \times 2(d-1)N_s$ matrix ($N_s=4$ being the number of magnetic sublattices).
The component matrices are
\begin{widetext}
\begin{align}
 \mathbb{A}^{pq}_{ab}(\mathbf{k}) & = \sum_{\alpha,\beta} J_{ab}^{\alpha\beta}(\mathbf{k}) [c_{a}^{\alpha}]_{p0} [c_{b}^{\beta}]_{0q} + \sum_{\alpha,\beta,c}  J_{ab}^{\alpha\beta}(\mathbf{k}=0)\left( [c_{a}^{\alpha}]_{pq} -\delta_{pq} [c_{a}^{\alpha}]_{00} \right)  [c_{c}^{\beta}]_{00} \\
 \mathbb{A}'^{pq}_{ab}(\mathbf{k}) & = \sum_{\alpha,\beta} J_{ab}^{\alpha\beta}(\mathbf{k}) [c_{a}^{\alpha}]_{0p} [c_{b}^{\beta}]_{q0} + \sum_{\alpha,\beta,c}  J_{ab}^{\alpha\beta}(\mathbf{k}=0)\left( [c_{a}^{\alpha}]_{pq} -\delta_{pq} [c_{a}^{\alpha}]_{00} \right)  [c_{c}^{\beta}]_{00} \\
  \mathbb{B}^{pq}_{ab}(\mathbf{k}) & = \sum_{\alpha,\beta} J_{ab}^{\alpha\beta}(\mathbf{k}) [c_{a}^{\alpha}]_{p0} [c_{b}^{\beta}]_{q0}.
\end{align}
\end{widetext}
The spectrum is determined by carrying out a Bogoliubov transformation which amounts to diagonalizing the matrix $\bm{\sigma}_3 \bm{H}_{\rm FW}(\mathbf{k})$ where $\bm{\sigma}_3 = {\rm diag}(1,\ldots,1,-1\ldots,-1)$ - the matrix with $(d-1)N_s$ ones and $(d-1)N_s$ minus ones along the diagonal. The transformation $\bm{T}_{\mathbf{k}}$ that brings the Hamiltonian to diagonal form $(\bm{T}_{\mathbf{k}}^\dagger)^{-1} \bm{H}_{\rm FW}(\mathbf{k})\bm{T}_{\mathbf{k}}^{-1}=\bm{\Lambda}_{\mathbf{k}}$  is generally non-unitary satisfying instead the constraint $\bm{T}_{\mathbf{k}}\bm{\sigma}_3 \bm{T}^\dagger_{\mathbf{k}}=\bm{\sigma}_3$.

From this, we compute the inelastic neutron cross section which is proportional to
\begin{equation}
S(\mathbf{k},\omega)= \sum_{\alpha\beta} (\delta_{\alpha\beta}-\hat{k}_{\alpha}\hat{k}_{\beta})S^{\alpha\beta}(\mathbf{k},\omega),
\label{eq:FW_DSF}
\end{equation}
with
\[
S^{\alpha\beta}(\mathbf{k},\omega) = \int dt e^{i\omega t} \langle  \mu^{\alpha}_{-\mathbf{k}}(t)\mu^{\beta}_{\mathbf{k}}  \rangle
\]
where $\bm{\upmu}= g_l \mathbf{l} + g_{\fs{S}} \bm{\mathsf{S}}$ and the $g$-factor for the effective orbital moment is $g_l = -3/2$ and $g_{\fs{S}}\approx 2$.

\subsection{Spin-Orbital Exchange}

In the context of the effective spin one-half model used to fit the data, we considered various types of anisotropic exchange. To nearest neighbor, for example, we saw that six couplings are allowed by symmetry including Kitaev and $\Gamma$ couplings. In this section, we briefly review how the effective spin one-half model can be understood through a superexchange calculation mediated by cobalt-oxygen-cobalt bonds and we use these results to carry out a mean field calculation of the ground states of coupled spin-$3/2$ effective orbital moments with spin-orbital exchange couplings.

First we consider the geometry of the principal exchange pathways in \cto.
The crystal structure consists of edge-sharing CoO$_6$ octahedra forming a honeycomb arrangement in the $ab$ plane as illustrated in Supplementary Figure~\ref{fig:twins} left panel. The cobalt honeycomb network is buckled like cyclohexane in its chair conformation ($\pm$ signs on the blue balls indicate alternating cobalt positions above/below the plane) so the only rotational symmetry at the cobalt sites is 3-fold and the oxygen geometry breaks the cobalt sublattice out-of-plane mirror symmetries. Exchange interactions between neighboring cobalt ions is primarily mediated by a pair of oxygen ions forming a planar unit with two near $90^{\circ}$ Co-O-Co bonds (at least within the errors of the crystal structure refinement).

For the purpose of identifying the most relevant exchange mechanisms, in the following we consider an idealized version of the actual crystal structure, where the cobalt honeycomb is planar (no buckling) and the oxygen octahedra are regular, as illustrated schematically in Supplementary Figure~\ref{fig:FW}d). The spin-orbital exchange interactions for this idealized Co-O-Co bonding geometry have been considered by Liu and Khaliullin \cite{PhysRevB.97.014407} and Sano, Kato and Motome \cite{PhysRevB.97.014408}. The calculation of the exchange in this paper was carried out within the set of effective orbital $l=1$ and spin $3/2$ degrees of freedom on the cobalt ions. The geometry of the low-lying triplet of $d$ orbitals is central to this calculation as hopping through the mediating oxygens with right-angle bonds allows the orbital character to change. For example, in the natural Cartesian frame $\fs{xyz}$ shown in Supplementary Figure~\ref{fig:FW}d) the $d_{\fs{yz}}$ orbital on one site is connected via the oxygen to $d_{\fs{zx}}$ on the other site.

There are several possible exchange processes and hence several distinct spin-orbital exchange couplings. These processes can be grouped into a pair of classes. The first class is set of d orbitals involved in the exchange (I) $t_{2g}-t_{2g}$ (II) $t_{2g}-e_{g}$ and (III) $e_{g}-e_{g}$. The second class is the type of intermediate state (A) $d^7 d^7 \rightarrow d^6 d^8$ in which the on-site Coulomb energy $U$ enters, (B) charge transfer where two holes are created on one of the oxygen ions $d^7 p^6 d^7 \rightarrow d^8 p^4 d^8$ and (C) cyclic exchange where, again, holes are created on oxygen ions but one hole on each oxygen. There are then, naively, nine possible exchange processes obtained from pairs with one from each class. However, processes IIIA and IIIC vanish by symmetry leaving seven routes. The resulting interactions are the product of isotropic exchange in spin space and angular-momentum violating exchange in orbital space. We consider the following four distinct spin-orbital couplings:
\begin{widetext}
\begin{align}
{\cal{H}}_{\rm SOexchange}  = & \fs{t}_1 \left( \bm{\mathsf{S}}_i\cdot \bm{\mathsf{S}}_j + \fs{S}^2  \right) \left[ n_{ia}n_{jb}  + (a \leftrightarrow b) \right] \nonumber \\
+ &  \fs{t}_2 \left(\bm{\mathsf{S}}_i\cdot \bm{\mathsf{S}}_j + \fs{S}^2  \right) \left[  a_i^\dagger b_i a^\dagger_j b_j + (a \leftrightarrow b) \right] \nonumber \\
+ &  \fs{t}_3 \left( \bm{\mathsf{S}}_i\cdot \bm{\mathsf{S}}_j + S^2  \right) \left[  a_i^\dagger c_i c^\dagger_j b_j + c_i^\dagger a_i b^\dagger_j c_j + (a \leftrightarrow b) \right] \nonumber \\
+ &  \fs{t}_4 \left( \bm{\mathsf{S}}_i\cdot \bm{\mathsf{S}}_j - \fs{S}^2  \right) n_{ic} n_{jc}
\label{eq:SOexchange}
\end{align}
\end{widetext}
on a single honeycomb nearest neighbor bond in the frame illustrated in Supplementary Figure~\ref{fig:FW}d) where, following Liu and Khaliullin, we use the notation $a=d_{\fs{yz}}$, $b=d_{\fs{zx}}$, $c=d_{\fs{xy}}$ (where $\fs{xyz}$ are the axes of the Cartesian 111 frame) and $n_{a}=a^\dagger a$ etc. Coupling $\fs{t}_1$ originates from IA and IC, $\fs{t}_2$ from IA, $\fs{t}_3$ from IA  and $\fs{t}_4$ from IIB, IB and IA. While the couplings are fixed by microscopic terms, in the ensuing calculations we treat them as free parameters. Pure Heisenberg spin exchange may also arise microscopically, through process IIIB, and we also include these $\fs{J}_n$ couplings for $n=1$ to 6 in ${\cal H}_{\rm Heisenberg}=\sum_{\langle i,j \rangle_n} \fs{J}_n \bm{\mathsf{S}}_i\cdot \bm{\mathsf{S}}_j $.

The above calculations assume an idealized cobalt-oxide structural bonding and below we find that the spin-orbital exchange terms derived in this case in Supplementary Equation~(\ref{eq:SOexchange}) are sufficient to account for the main features of the ground state selection and excitation spectrum. The actual crystal structure has additional distortions, in particular the cobalt network is buckled, which amounts to a roughly $12^{\circ}$ tilt of the cobalt-oxygen-cobalt unit about an axis through the pair of oxygens mediating that bond. Further refinement of the ground state selection and fine structure of the spin wave spectrum may necessitate including the local rotation of the exchange coming from this buckling or perhaps including higher order contributions to the exchange coupling beyond the terms in Supplementary Equation~(\ref{eq:SOexchange}).

So far we have not discussed the spin-orbit coupling and trigonal distortion. As the hopping and Coulomb scales are the dominant energy scales the superexchange calculation is carried out without them and they are then included on an equal footing with the large $U$ or Hund coupling spin-orbital exchange. To obtain the effective spin one-half exchange model of previous sections of this paper, one may project the spin-orbital exchange onto the spin-orbit coupled trigonally distorted doublet; the anisotropy in the effective spin model is inherited from the angular momentum violating orbital couplings.

\subsection{Mean Field - Flavor Wave Results}

As described in the previous section, the microscopic exchange to nearest neighbor acting on the spin $3/2$ and effective orbital angular momentum $1$ states will couple these degrees of freedom leading to an effective anisotropic exchange within the effective spin one-half model obtained by projecting the exchange onto the single-ion ground state doublet.

We consider a mean field theory including the microscopic single-ion physics, the spin-orbital exchange $\fs{t}_n$ ($n=1,2,3,4$) to nearest neighbor and pure spin isotropic exchange coupling the $n$th  neighbor $\fs{J}_n$ for $n=1,\ldots,6$, together with single-ion spin-orbit coupling $\lambda$ and trigonal distortion parameter $\delta$ as given in~\ref{sec:cef}. The collinear, easy plane ordered state of \cto\ with antiferromagnetically coupled layers is obtained straightforwardly by setting $\fs{J}_1<0$, $\fs{J}_2>0$. In Supplementary Figure~\ref{fig:FW}a-b) we illustrate the magnon spectrum obtained via the flavor-wave approach for a representative set of exchange parameters chosen such as to approximately reproduce the in-plane and out-of-plane spin wave bandwidths seen in experiments. In the calculation of the dynamical structure factor, as discussed above, we use idealized spin-orbital exchange that omits the effects of buckling of the cobalt honeycombs, but we {do} include the effects of the buckling on the spin wave intensities using this simplified exchange, i.e. we use the actual cobalt positions in the crystal structure in the calculation of dynamical structure factor. Panel a) shows the case for spin-only exchange, when the ground state energy is independent of the in-plane moments' orientation angle $\phi$ and consequently the magnon spectrum has a gapless Goldstone mode, emerging out of the magnetic Bragg peak position $(1,1,3/2)$ {
and the magnon spectrum has double-helix nodal lines as illustrated for the XXZ model in Fig.~3b).} Supplementary Figure~\ref{fig:FW}b) shows the case when a finite spin-orbital exchange perturbation $\fs{t}_3$ is switched on. This selects the family of $\phi=0$ modulo $\pi/3$ ground state moment orientations and consequently opens a gap in the magnon spectrum. In addition to capturing the discrete ground state selection and spectral gap, a further advantage of the flavor wave picture is that it also gives the spectrum of exciton modes. Supplementary Figure~\ref{fig:FW}c) shows the obtained dynamical structure factor for the lowest-energy exciton modes, the calculated spectrum bears strong resemblance to the data in Fig.~4c). For the parameters used in the above calculations, the largest effect of the finite spin-orbital exchange is in opening of a magnon spectral gap, the magnon spectrum still displays nodal lines, and the effect on the exciton modes is relatively small.

We note that in the present treatment of the spin-orbital exchange, each of the four $\fs{t}_{1-4}$ terms in Supplementary Equation~(\ref{eq:SOexchange}), irrespective of their sign, selects the family of $\phi=0$ modulo $\pi/3$ ground states at least when their magnitude is compatible with the experimentally observed spin wave gap. We leave it as subject for future research to investigate whether this is true in general for these couplings and, if so, how to understand this perturbatively in the spin-orbital exchange coupling over the crystal field gap. We also leave for the future the question of whether other symmetry-allowed spin-orbital exchange terms not explicitly listed in Supplementary Equation~(\ref{eq:varphi_k2}), could select the alternative set of $\phi=\pi/2$ modulo $\pi/3$ ground states.

We established in~\ref{sec:gap} that the effective spin one-half model at the mean field level has an accidental $U(1)$ degeneracy that is lifted by quantum fluctuations. In contrast, the spin-orbital exchange model discussed in this section can be viewed as an example of order arising from virtual crystal field fluctuations first discussed in the context of Er$_2$Ti$_2$O$_7$ \cite{McClarty_2009,PhysRevB.93.184408}. The theory developed in this section is based around a mean field theory that omits the order-by-disorder corrections discussed in~\ref{sec:gap} that act within the effective spin one-half set of states. Yet the accidental degeneracy, that is present in the spin-orbital model when projected down to the low-energy doublets on each site, is lifted within the full mean field theory leading to a discrete set of ground states. The discrete symmetry breaking in this case originates from the enlarged Hilbert space and the admixing of excited crystal field levels into the ground state and is therefore suppressed in powers of the inverse crystal field gap. In CoTiO$_3$ one expects that both order-by-disorder mechanisms are operative. While disentangling the relative contributions of the two effects is non-trivial, the fact that the exchange scale is a significant proportion of the crystal field gap in the material strongly suggests that order by virtual crystal field fluctuations is an important factor in the ground state selection in the system.

\begin{figure*}[!htbp]%
    \includegraphics[width=\textwidth]{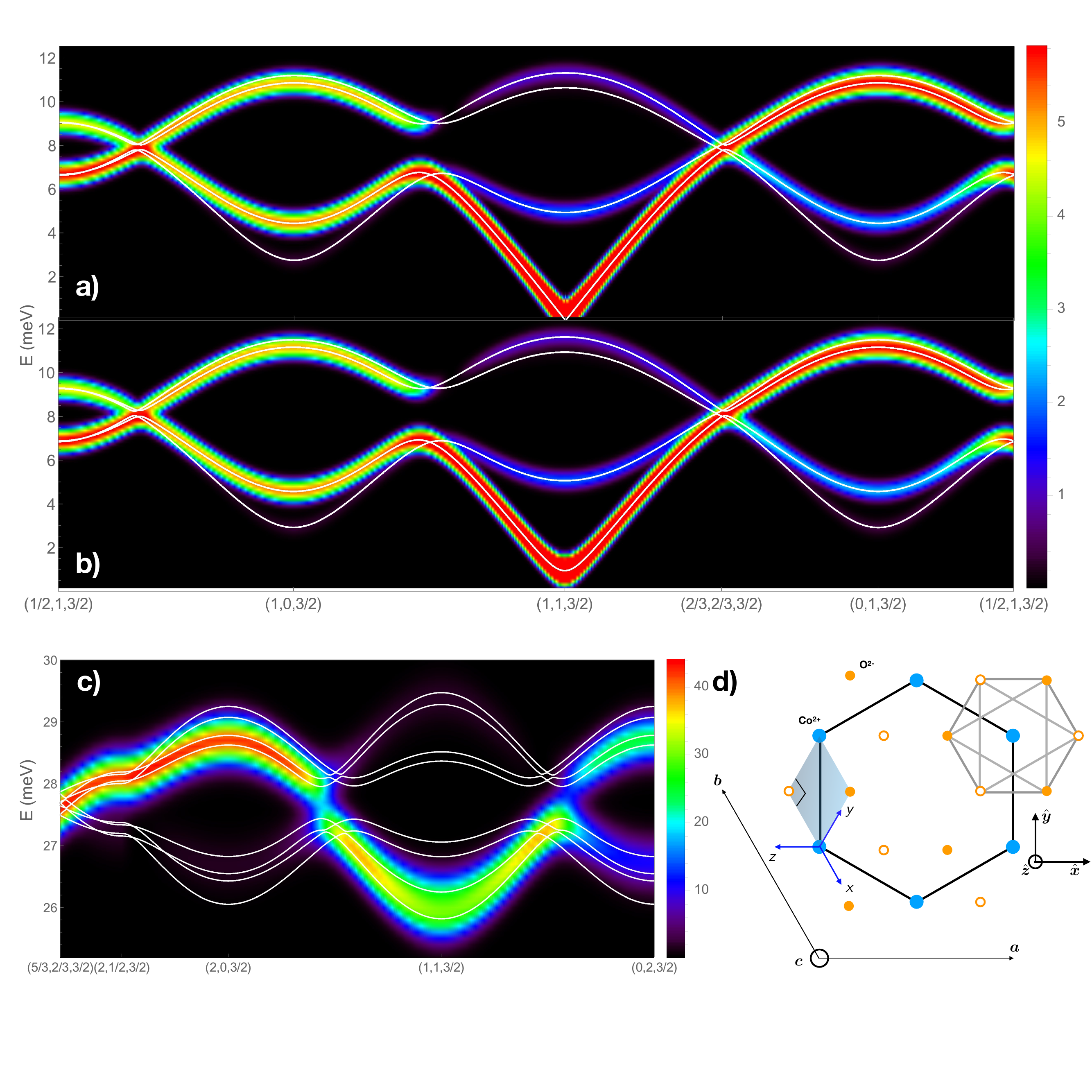}
    \figcaption{Flavor wave model} {Calculated dynamical structure factor in Supplementary Equation~(\ref{eq:FW_DSF}) for a-b) magnon and c) lowest-energy exciton modes along the same reciprocal-space paths as in Figs.~2a) and 3c), respectively. The exchange parameters in ${\cal H}_{\rm Heisenberg}$ are $\fs{J}_1=-0.9$, $\fs{J}_2=0.25$, and $\fs{J}_6=0.25$ (all in meV) and in ${\cal H}_{\rm SOexchange}$ $\fs{t}_3=0$ in a) and $\fs{t}_3=-0.15$~meV in b-c), with  $\fs{J}_3=\fs{J}_4=\fs{J}_5=\fs{t}_{1,2,4}=0$ in all panels. The calculations have been convolved with a Gaussian energy lineshape of standard deviation $\sigma= $0.17~meV. Colorbars show the dynamical structure factor, on a linear scale in c), and a log scale in a-b) $\ln(S({\bf k},\omega)+1)$ in order to highlight weak features in the magnon spectrum. All three panels correspond to the magnetic domain with $\phi=0$. The thin solid lines show the calculated dispersion relations (4 magnon modes in a-b) and 8 exciton modes in c)). d) Local geometry of an idealized cobalt oxide layer projected onto the $ab$ plane assuming regular oxygen octahedra, planar cobalt layers and hence 90$^{\circ}$ Co-O-Co bonds [Co at (0,0,1/3), O at (1/3,0,1/4) and $c/a=\sqrt{8}$ in Supplementary Table~\ref{TAB::crystal_structure}]. Solid blue dots are cobalt ions and orange dots are oxygens - filled/open for above/below the nearest cobalt ions. The blue arrows show the orientations of the $\fs{xyz}$ axes in the 111 coordinate frame. We also indicate the hexagonal primitive lattice vectors $\mathbf{a}$ and $\mathbf{b}$ and the Cartesian $xyz$ frame used to specify the nearest neighbor anisotropic exchange in~\ref{sec:eta}. Note the orientation of this figure is rotated around $\mathbf{c}$ by $-60^{\circ}$ compared to the orientation of Supplementary Figures~\ref{fig:BZ}c-d) and \ref{fig:twins}.}%
    \label{fig:FW}%
\end{figure*}

\end{document}